\documentclass[preprint2]{aastex}

\def\sun{\hbox{$\odot$}}
\def\farcm{\hbox{$.\mkern-4mu^\prime$}}
\def\farcs{\hbox{$.\!\!^{\prime\prime}$}}

\defcitealias{smi05}{Paper I}

\shorttitle{Hinge Clumps}
\shortauthors{Smith et al.}

\begin{document}

\title{Extra-Nuclear Starbursts: 
Young Luminous Hinge Clumps in Interacting Galaxies}

\author{
Beverly J. Smith\altaffilmark{1},
Roberto Soria\altaffilmark{2},
Curtis Struck\altaffilmark{3},
Mark L. Giroux\altaffilmark{1},
Douglas A. Swartz\altaffilmark{4},
and Mihoko Yukita\altaffilmark{5}
}

\altaffiltext{1}{Department of
Physics and Astronomy, East Tennessee
State University, Johnson City TN  37614;
smithbj@etsu.edu; girouxm@etsu.edu}
\altaffiltext{2}{Curtin Institute of Radio Astronomy, Curtin University, 1 
Turner Avenue, Bentley, WA 6102, Australia; 
rsoria@physics.usyd.edu.au} 
\altaffiltext{3}{Department of Physics and Astronomy, Iowa State University, Ames IA  50011; curt@iastate.edu}
\altaffiltext{4}{University Space Research
Association, NASA Marshall Space Flight Center, ZP 12, Huntsville AL}
\altaffiltext{5}{Department of Physics
and Astronomy, Johns Hopkins University, Baltimore
MD 21218; 
myukita1@pha.jhu.edu}

\begin{abstract}

Hinge clumps are 
luminous knots of star formation near the base
of tidal features in some interacting galaxies.   
We use archival Hubble Space Telescope UV/optical/IR
images
and Chandra X-ray maps along with GALEX UV, Spitzer IR, and ground-based
optical/near-IR images to investigate the star forming properties
in a sample of 12 hinge clumps in five interacting
galaxies.
The most extreme of these hinge clumps have star formation
rates of 1 $-$ 9 M$_{\sun}$~yr$^{-1}$, comparable to or larger than the `overlap'
region of intense star formation between the two disks 
of the colliding galaxy system the Antennae. 
In the HST
images, we have found remarkably
large and luminous sources at the centers of
these hinge clumps. 
These objects are much larger and more luminous than typical
`super-star clusters' in interacting galaxies,
and are sometimes embedded in a linear ridge of fainter
star clusters, consistent with star formation along
a narrow caustic.
These central sources have
diameters of $\sim$70 pc, compared to $\sim$3 pc in 
`ordinary' super-star clusters.
Their absolute I magnitudes range from
M$_I$ $\sim$ $-$12.2 to $-$16.5, thus if they are individual
star clusters they would lie near the top of 
the `super star cluster' luminosity function
of star clusters.
These sources may not be individual star clusters, but instead
may be 
tightly packed groups of clusters that are blended together in the HST 
images.
Comparison to population synthesis modeling indicates
that the hinge clumps contain a range of stellar ages.
This is consistent with expectations based on models
of galaxy interactions, which suggest that star formation may be prolonged
in these regions.
In the Chandra images, we have found
strong X-ray emission from several of these hinge clumps.
In most cases, this emission is well-resolved with Chandra
and has a thermal X-ray spectrum, thus it is likely due to  
hot gas associated with the star formation.  
The ratio of the extinction-corrected diffuse X-ray luminosity to the
mechanical energy rate (the X-ray production efficiency)
for the hinge clumps is similar to that in the Antennae galaxies,
but higher than those for regions in the normal spiral galaxy
NGC 2403.
Two of the hinge clumps have point-like X-ray emission much brighter
than expected for hot gas; 
these sources are likely
`ultra-luminous X-ray sources' (ULXs)
due to accretion disks around black holes.
The most extreme of these sources, in Arp 240, has a hard
X-ray spectrum and an absorbed X-ray luminosity
of $\sim$2 $\times$ 10$^{41}$ erg~s$^{-1}$; this is above the luminosity
expected by single high mass X-ray binaries (HMXBs), thus it may be either
a collection of HMXBs or an intermediate-mass black hole ($\ge$80 M$_{\sun}$).

\end{abstract}


\section{Introduction}

Some of the most luminous extra-nuclear star forming regions known
in the local Universe lie near the base of tidal structures in 
interacting galaxies.  We have dubbed these regions `hinge clumps'
as they lie near the `hinge' of the tail (\citealp{hancock09, smith10}).
Hinge clumps can have L$_{H{\alpha}}$ $>$ 10$^{40}$
erg/s \citep{hancock07}, earning them the label of
extra-nuclear starbursts.  
Their location on the outskirts of galaxy disks means 
they may be important in enriching the intergalactic
medium via winds.
Hinge clumps may be more extreme sites of star formation
than `tidal dwarf galaxies', larger but lower density star forming
regions found 
near the ends of tidal features.
Hinge clumps may be as common
as tidal dwarfs, but are not nearly as well-studied.
Hinge clumps bear an intriguing resemblance to the massive
star forming regions formed by gravitational instabilities
in high redshift disks (e.g., \citealp{cowie95, elmegreen09,
forster11, elmegreen14}),
thus the study of these regions 
broadens our perspective
on high redshift disk evolution.

Our recent analytical models of galaxy interactions
\citep{struck12}
have provided insights into
why star formation is so intense in these regions.
These models sometimes produce
intersecting caustics near the base of tidal
tails, where a caustic is a narrow pile-up zone produced by
orbit crowding.   Intersecting gas flows within or between
caustics may trigger star formation at these locations, producing
hinge clumps.   
In retrospect,
evidence for overlapping waves or compressions in the hinge
regions of tidal tails can also 
be seen in many classical papers that modeled galaxy
flybys (e.g., \citealp{toomre72, elmegreen91, gerber94}).
More generally, the higher velocity dispersions in the interstellar gas in interacting galaxies
may lead to more massive self-gravitating clouds \citep{elmegreen93,
bournaud08, teyssier10}, 
and therefore more efficient star formation and more luminous star forming regions
in interacting systems compared to more isolated galaxies.

Numerical and analytical models show that
intersecting caustics can be produced in both the inner regions
of interacting galaxies as well as in tidal features.
Strong prograde encounters can produce `ocular'
structures in disks: eye-shaped ovals formed from intersecting
spiral arms \citep{elmegreen91}. Luminous knots of star
formation are sometimes observed along the `eyelids' and in the
`points' of oculars \citep{elmegreen06, hancock07, hancock09}.
In tidal features, models reveal a wide variety of caustic morphologies.
A single strong caustic is sometimes produced along the leading edge of
tidal features; in other cases, two approximately parallel caustics
or two diverging or branching caustics are seen in tidal tails
\citep{elmegreen91, donner91, gerber94, struck12}.
In some tidal features, a diverging caustic branches back towards the main disk,
creating a loop-like structure, while in other cases two
caustics diverge to produce a double-tail morphology
\citep{struck12}.
Models also sometimes show a `narrowing'
of a tidal feature where two caustics converge 
\citep{gerber94, struck12}.
At such intersections between caustics,
gas build-up and gravitational collapse is
expected to occur, triggering star formation.
Converging caustics can also occur in the outer portions
of tidal features, potentially producing knots of star formation
\citep{struck12},
but
most tidal dwarfs may form
via gas accumulation and gravitational collapse near the end of a tidal feature
\citep{duc04, bournaud06, wetzstein07}.

Continued
gas inflow into hinge clumps along caustics may produce
sustained star formation in these regions, rather than
instantaneous bursts.  
High spatial resolution imaging is 
one way to test this hypothesis.
At high spatial resolution,
knots of star formation in nearby galaxies resolve into
multiple young star clusters \citep{zhang01, larsen04, 
bastian05, mullan11}.
The most luminous of these young clusters are termed `super star clusters'
or `young massive clusters' 
(e.g., \citealp{larsen00}).  
For clumps of star formation in 
the disk of Arp 284W, 
the average age of the stars in the 
clump derived from single-burst stellar population
synthesis models 
is 
older than that found
for the observed star clusters in that clump seen in Hubble Telescope
images \citep{peterson09}.
This indicates that more than one generation of stars is present 
in the region, with the older clusters fading and possibly dissolving
with time but still contributing to the total light of the clump.

Regions with sustained star formation may have strong
X-ray emission, as very young stars may exist simultaneously
with the end products of stellar evolution.
Given the high star formation rates in hinge clumps, 
strong diffuse x-ray emission 
may be present due to 
stellar winds, supernovae, and shocks.
In spiral and irregular galaxies, the diffuse X-ray luminosity
correlates with the star formation rate \citep{strickland04, owen09, mineo12b}.

Hinge clumps may also host bright X-ray point sources, as
the number of high mass X-ray binaries (HMXBs) is correlated
with the star formation rate in spiral galaxies 
\citep{grimm03, gilfanov04, persic04, mineo12a}.
The number of `ultra-luminous' X-ray point sources (ULXs; L$_{\rm X}$ $\ge$ 
10$^{39}$ erg~s$^{-1}$)
is also correlated with 
the star formation
rate in both normal spirals and pre-merger interacting
galaxies \citep{swartz04, swartz11, liu06, smith12}.
This correlation suggests that most ULXs are X-ray luminous HMXBs.
This conclusion is supported by studies that show that the local
environments around ULXs tend have blue optical colors
\citep{swartz09, smith12}.
The most extreme ULXs, however, with L$_{\rm X}$ $\ge$ 10$^{41}$ erg~s$^{-1}$
\citep{farrell09, sutton11, sutton12}, 
may be intermediate mass black holes, since black holes with masses less than
80 M$_{\sun}$ are not expected to produce such high luminosities
\citep{zampieri09, belczynski10, swartz11, ohsuga11}.

In order to 
better understand star formation 
in hinge clumps, we have searched the archives
of the Hubble Space Telescope (HST) and the Chandra X-ray telescope
for available images 
of a sample of hinge clumps. 
In Section 2 of this paper, we define our sample of hinge clumps
and describe the available datasets.
Along with the HST and Chandra data, we also include UV images from
the Galaxy Evolution Explorer (GALEX) telescope, infrared images
from the Spitzer telescope and the 
Two Micron
All Sky Survey (2MASS),
ground-based broadband optical images
from the Sloan Digitized Sky Survey (SDSS), and published
H$\alpha$ maps.
In Section 3, we describe the photometry.
We provide
approximate star formation rates for the clumps
in Section 4.
In Section 5, we discuss UV/IR colors for the clumps, and in
Section 6, we present the X-ray/UV/IR colors of the clumps.
In Section 7 we compare with
population synthesis models, and in Section 8
we 
compare the X-ray luminosities of the clumps with their 
star formation properties. 
Conclusions are presented in Section 9.
In an Appendix, we describe the individual galaxies in the sample and
their morphologies, and provide a detailed 
discussion of the intense star formation
regions within the Antennae galaxies.

\section{Sample and Data}

\subsection{Sample Selection}

To study extra-nuclear star formation in interacting galaxies,
we have obtained 
GALEX UV and Spitzer 
infrared images 
of three dozen nearby pre-merger 
interacting galaxy pairs selected from
the \citet{arp66} Atlas 
(the
`Spirals, Bridges, and Tails' (SB\&T)
survey;
Smith et al. 2007; 2010).  
We 
expanded this search to include other Arp pre-merger pairs with
available
archival 
GALEX and Spitzer images 
(Struck \& Smith 2012).
We also searched the literature for non-Arp systems
with similar morphologies
and suitable Spitzer and GALEX images.
We identified knots of star formation in these systems
by visual inspection of the 8 $\mu$m Spitzer images,
classifying them as either tidal or disk regions.
In a later paper, we will investigate the 
statistical properties
of the full sample of star forming regions.
In the current paper, we focus on several systems out of this larger
sample that contain prominent hinge clumps, with ample data to allow
their detailed study.

We define a hinge clump as a discrete knot of
star formation near the base of a tidal tail.
To define the `hinge region', we draw a line from the center
of the galaxy through the base of the tail, out to approximately
twice the radius of the disk.
At right angles to this line in the direction of the spiral of the tail,
we draw a second line out from the center of the galaxy,
again extending out to approximately
twice the radius of the disk.  The portion of the tail that
lies within this pie slice defines the hinge region.
The choice of 90 degrees here is motivated by the
analytical models. In tidally perturbed systems,
in alternate quadrants of the galaxy material
either gains or loses orbital angular momentum. In the former
case the material is flung out in tails,
which can commonly form sharp outer/inner edge caustics.
In the latter case, disk material compresses, making the
light falloff much steeper, if not actually a sharp edge.
The meeting point of these two 'sharpened' edges often
appears as a cusp point in tailed galaxies.
In the models, most of the strong wave overlaps occur within
90 degrees of the base of the tail.

In the current paper, we focus on five systems hosting 
hinge clumps with
high quality multi-wavelength data available (Table 1).
To be included in this study, we 
required the system 
to have both 
Spitzer 24 $\mu$m and GALEX near-UV images available, as well
as either HST or Chandra images.
The total number of hinge clumps in the sample is 12, as 
some of these systems contain more than one
hinge clump.
All of these hinge clumps are bright, 
exceeding a S/N of 25 above the surrounding light in either 24 $\mu$m
or the NUV.
Some of the hinge clumps in our sample 
show up brightly in both the UV and the mid-infrared.
Others are bright in the mid-infrared but with little observed
UV or with the UV peak offset from that in the mid-infrared.
In these cases, we use the mid-infrared peak as the assumed
center of the hinge
clump.
These selection criteria limit the current
sample to hinge clumps
that are relatively young and luminous; older and fainter 
regions may be missed by these requirements.
Other candidates will be discussed in a later paper.

We provide detailed descriptions 
of the morphologies of these galaxies in the Appendix to this paper.
In the Appendix, we also provide a discussion
of numerical and analytical models of the interactions, when available.
For four of the five systems, numerical simulations that approximately
reproduce the large-scale structure of the galaxies have been published. 
In addition,
for three systems (Arp 82, Arp 240, and Arp 256) we
have identified analytical models from \citet{struck12}
that approximately match the morphologies of the galaxies.
These models are displayed in the Appendix, and compared with
the available images.   
These models suggest that 
the hinge mechanism, i.e., intersecting caustics, 
may be responsible for the strong star formation in these regions.
The morphologies of the two remaining systems (Arp 270 and NGC 2207) 
are more complex, and 
do not match any of the analytical models in \citet{struck12}.
We include them in our sample because of the availability
of high quality data and the existence of high S/N clumps of star formation
in the hinge region, however, it is unclear whether the star formation
in these clumps is triggered by the same mechanism as the other regions.
This issue is discussed further in the Appendix.

Throughout this paper we use 
distances from the NASA Extragalactic
Database (NED\footnote{http://ned.ipac.caltech.edu}), 
assuming
H$_0$ = 73 km~s$^{-1}$~Mpc$^{-1}$ and accounting for peculiar velocities
due to the Virgo
Cluster, the Great Attractor, and the Shapley Supercluster.
These distances range from 29.0 Mpc for Arp 270 to 109.6 Mpc
for Arp 256 (Table 1).

We compare these hinge clumps to star forming regions within 
other interacting galaxies, including the Antennae galaxies,
as well as the normal spiral galaxy NGC 2403.  The available data
for these galaxies
are discussed further
in Section 3.4.

\subsection{Spitzer/GALEX/SDSS/H$\alpha$/2MASS Datasets}

Of the five systems in Table 1, 
all have GALEX near-UV (NUV) images available
and 
two also have GALEX 
far-UV (FUV) images.
All five have both Spitzer near-infrared
(3.6 $\mu$m and 4.5 $\mu$m) and mid-infrared (5.8 $\mu$m, 8.0 $\mu$m, and 24 $\mu$m)
images
available. 
All but one have ground-based SDSS ugriz images available.

The GALEX FUV band has an effective wavelength
of 1516 \AA~ with a full width half maximum (FWHM) of 269 \AA, 
while the NUV band has an effective
wavelength of 2267 \AA ~and FWHM 616 \AA.
The GALEX images have 1\farcs5 pixels, and the point spread
function has a FWHM of $\sim$ 5$''$.
The Spitzer FWHM spatial resolution is
1\farcs5 $-$ 2$''$ for
the 3.6 $\mu$m $-$ 8 $\mu$m bands, 
and $\sim$6$''$ at 24 $\mu$m. 
The 3.6 $\mu$m $-$ 8 $\mu$m images have 0\farcs6~pixel$^{-1}$,
while the 24 $\mu$m images have 2\farcs45~pixel$^{-1}$.
The SDSS pixels are 0\farcs4, and the SDSS FWHM spatial 
resolution is typically about 1\farcs3.
The SDSS u, g, r, i, and z filters have effective 
wavelengths
of 3560 \AA, 4680 \AA, 
6180 \AA, 7500 \AA, and 8870 \AA, respectively.
For more details about the Spitzer, GALEX, or SDSS 
observations see \citet{smith07}, \citet{smith10}, and
\citet{elmegreen06}. 

We also obtained copies of published H$\alpha$ maps for all 
of the hinge clump systems.
These include H$\alpha$ images of Arp 240 and Arp 256
\citep{bushouse87}, Arp 82 \citep{hancock07}, Arp 270 \citep{zaragoza13},
and NGC 2207 \citep{elmegreen01, elmegreen06}.
The pixel sizes in these H$\alpha$ images range from 0\farcs2 
to 0\farcs595.
When necessary, we registered these images to match those at other
wavelengths.

We also utilize near-infrared J, H, and K$_{\rm S}$
(1.235 $\mu$m, 1.662 $\mu$m, and 2.159 $\mu$m)
images
of these galaxies 
from the 2MASS survey 
\citep{skrutskie06}.
These images have 1\farcs0 per pixel.
Some of the hinge clumps are visible as discrete sources on the 2MASS images.
The 2MASS seeing was typically FWHM 2\farcs5 $-$ 3\farcs4
\citep{cutri06}.

\subsection{Hubble Space Telescope Datasets}

To study the morphology of these star forming regions
at higher resolution, we have searched the HST archives 
for suitable images.  Of our sample, four systems have suitable HST images
available that cover
the hinge clumps.  These galaxies and their HST datasets are listed 
in Table 2. 
Together, these four systems host a total of eight hinge clumps.

As can be seen in Table 2, a variety of instruments and filters were used
for the HST observations.   The data includes images obtained with the 
Wide Field Planetary Camera (WFPC2),  the Advanced Camera for Surveys (ACS),
the Wide Field Camera 3 (WFC3), and the Near-Infrared Camera and Multi-Object
Spectrometer (NICMOS).   Bandpasses available include the far-UV F140LP filter,
the near-infrared F160W filter (H band), as well as a range of optical
bands.   
The data that we used was obtained from the Hubble 
Legacy Archive\footnote{http://hla.stsci.edu/}.
The WFPC2 images have a pixel size of 0\farcs1, while the
pixel size for the ACS F140LP FUV images is 0\farcs025.  The ACS optical
and
the 
NICMOS F160W images have 0\farcs05 pixel
sizes, while the WFC3 F160W image has 0\farcs09 pixels.

\subsection{Chandra Telescope Data}

To investigate stellar feedback and evolution in these regions, we searched
the Chandra archives for suitable datasets.
We found archival Chandra images of
four
interacting systems containing a total of eleven hinge clumps 
(Table 3).

\section{Photometry}

\subsection{Spitzer, GALEX, SDSS, H$\alpha$, and 2MASS Photometry}

From the Spitzer, GALEX, SDSS, H$\alpha$, and 2MASS images,
we extracted aperture photometry of the sample hinge 
clumps 
using the 
Image Reduction and Analysis Facility
(IRAF\footnote{http:$//$iraf.noao.edu})  
{\it phot} routine.
We used a 5$''$ radius aperture, which 
is a compromise between our desire to 
study detailed 
regions within the galaxy and the limiting spatial 
resolution of the GALEX and Spitzer 24 $\mu$m images.
This aperture 
corresponds to 0.70 kpc to 2.7 kpc
at the distances of these galaxies.
We used a sky annulus with the mode sky
fitting algorithm, an inner
radius of 6$''$ and an outer radius of 12$''$.

Aperture corrections were determined for each of the GALEX images by 
determining the 
counts within 5$''$ and 17$''$ radii for three to eight moderately bright 
isolated point sources
in the field.  These values are tabulated in Table 4. 
For the Spitzer images, we used aperture corrections from
the IRAC and MIPS Data Handbooks.
No aperture corrections are needed for the SDSS images.
We also did not do aperture corrections for the 
2MASS data, as these corrections are expected to be small for photometry
with 5$''$ radii \citep{cutri06}.
They are also expected to be small 
for the H$\alpha$ images.

When necessary,
the H$\alpha$ luminosities have been approximately
corrected for the nearby [N~II] lines
in the filter.
We assume a 30\% calibration uncertainty
for the H$\alpha$ images in addition to 
the statistical uncertainties.
We calculated H$\alpha$ equivalent widths for the clumps using
the SDSS r band flux for continuum, approximately correcting this 
flux for contamination by H$\alpha$.  For most of the clumps,
this correction is only 7$\%$ $-$ 10$\%$; it is $\sim$30\% for 
Arp 82-1, 
Arp 240-3,
and Arp 270-4. 
For NGC 2207, for the H$\alpha$ continuum we extrapolated between the 
HST F555W and F814W bands.

We corrected the 
SDSS photometry for Galactic reddening 
as 
in \citet{schlafly11}, as provided by NED.
The GALEX photometry was corrected
for Galactic reddening
using the \citet{cardelli89} attenuation law.
The final photometry is given in Tables 5 and 6.  

\subsection{HST Photometry}

The HST morphologies of these hinge clumps are often quite remarkable,
with a luminous central source embedded in a row of fainter clusters.
These linear structures are quite different from the 
more clustered and amorphous
groupings of star clusters seen in HST images of many other galaxies,
for example, in the disk of M51 \citep{bastian05}.
The morphologies of the individual systems
are discussed in more detail in the Appendix.

Using the IRAF {\it phot}
routine we extracted magnitudes for the central sources in the various
HST bands using apertures of 0\farcs15 radii, and sky annuli with inner
radii of 0\farcs15 and outer radii of 0\farcs30.
When necessary, we adjusted the registration of the images 
obtained with different
filters to match.
We applied aperture corrections as in \citet{holtzman95},
\citet{sirianna2005}, \citet{dieball07},
the NICMOS Instrument Handbook, and the WFC3 Instrument Handbook.
We corrected the 
HST
photometry for Galactic extinction using the same method
as for the GALEX and SDSS photometry.
The final photometry is given in Table 7.  

The absolute I magnitudes of the central sources 
in the hinge clumps\footnote{These magnitudes 
and those given in the Appendix
are in the Johnson system,
using the approximate
conversions given in the WFPC2 Photometry Cookbook.}, 
uncorrected for internal extinction,
range from 
M$_{\rm I}$ = 
$-$12.2 to $-$16.5. 
Young unobscured clusters typically have V $-$ I colors
of $-$0.4 to 0.7 (e.g., \citealp{chandar10}), 
so most of these sources are more 
luminous than typical super-star clusters
in interacting galaxies, defined to have 
absolute V magnitudes 
M$_V$ $\le$ $-$11 (e.g., \citealp{larsen00}).
If these sources are individual star clusters, they would 
lie
near the top of the
luminosity function of super-star star clusters (e.g., \citealp{gieles10}).
The most luminous
of these objects are comparable in luminosity to the most massive star cluster
in the Antennae galaxies, which lies in an intense
starburst in the `overlap' region
between the two galaxian disks.  This Antennae cluster has
M$_{\rm V}$
$\sim$ $-$15.4 after correction for
internal extinction \citep{whitmore10}.  

The sources in these hinge clumps are more luminous than many
nuclear star clusters, for example, the nuclear star clusters in low luminosity
spirals have M$_{\rm B}$ $\sim$ $-$10 and M$_{\rm I}$ $\sim$ 
$-$11 \citep{matthews99}.  Most nuclear star
clusters in late-type spirals
have absolute I magnitudes between 
$-$10 to $-$14 \citep{boker04}, thus the central objects in our
hinge clumps are as luminous as or more luminous than many 
nuclear star clusters.

The sizes of the sources at the centers of the hinge
clumps are also quite large,
with diameters of 40 pc to 100 pc (see the Appendix).  
This is much larger than
the typical sizes of $\sim$3 pc for 
most luminous young star clusters (e.g., \citealp{larsen04}).
In the Appendix of this paper, we compare 
the hinge clumps to HST images
of knots of star formation in the Antennae galaxies, smoothed
to the same effective resolution as the hinge clumps.
We conclude that, if the Antennae were at the distance of the hinge
clumps, multiple star clusters would appear blended together,
creating larger sources similar in size to those seen in the hinge clumps.   
Thus we conclude that the 
large sizes we measure for the central regions of the hinge
clumps are likely 
the consequence of the blending of multiple clusters
closely packed together in a complex, rather than a single large cluster. 

We also extracted larger-aperture photometry of these hinge clumps
from the HST data (Table 8), using 5$''$ apertures
to match those used for the GALEX, Spitzer, SDSS, H$\alpha$, and 2MASS photometry
given in Tables 1 and 2.  
In most cases, we used sky annuli with 6$''$ inner radii and 12$''$ outer radius,
with the mode sky fitting algorithm.  
For the Arp 82 images, where the clump is near the edge of the field of view,
we used a 9$''$ outer radius.  For the ACS F140LP (FUV) images and the NICMOS F160W
(H band) images, which have smaller fields of view, 
we used an annulus with an inner radius of 5$''$ and an outer radius
of 6$''$.
No aperture corrections were
needed for the large aperture HST photometry.

Comparison of Table 2 to Table 3 shows that the central
source within these hinge clumps provides only
a small percentage of the total light 
coming from the hinge clump.  
The percentage coming from the central source 
increases with decreasing wavelength.
In the F140LP FUV filter, the 
percentages range from 2.6\% to 6.0\%, with a median of 3.5\%.
For the F435W filter, 
the percentages range from 1.1\% to 5.2\%, with a median of 2.2\%.
For the F814 filter, 
the percentages range from 0.4\% to 3.6\%, with a median of 1.0\%.
The lowest percentages are found for the near-infrared F160W filter,
ranging from 0.15\% to 1.8\%, with a median value of 0.7\%.
This trend suggests that the stars in the central source
are younger 
on average 
than the stars in the clump as a whole. 
This suggest that, as with the regions within the disk of Arp 284
studied by \citet{peterson09}, the hinge clumps
contain luminous `jewels' of young stars 
embedded in a more extended `crown' of older stars.

\subsection{Chandra X-Ray Spectra and Luminosities }

We have detected copious X-rays from 
eight of the eleven
hinge clumps with archival Chandra data available.
In five of these clumps, the X-ray emission is well-resolved
with Chandra. 
In one clump 
(clump 3 in Arp 240W), 
the x-ray emission is point-like,
in a second it appears point-like but may be marginally resolved
(clump 1 in Arp 270),
while in a third
(clump 4 in Arp 240W), there is a point-like hard component
superimposed on an extended soft component.
The X-ray morphologies of the individual sources are discussed
further in the Appendix of this paper.

We extracted X-ray fluxes for these clumps, 
and used Cash statistics to analyze their X-ray spectra.
No background flares were present in the data, so no filtering was needed.
The X-ray spectra are generally 
soft and in most cases appear dominated by thermal
emission (Figure 1). 
In Table 9, we provide 
the observed 0.3 $-$ 8 keV flux for the central X-ray source, along
with a rough estimate of its size and its coordinates.

Clump 1 in Arp 240W has an X-ray spectrum consistent with
a thermal plasma.  
Clump 2 is also probably thermal plasma, but has too few counts
to strongly constrain the parameters.
Although clump 3 in Arp 240W is unresolved with Chandra, its X-ray spectrum
is intrinsically soft but appears highly absorbed, 
consistent with a hot plasma (for example, a compact starburst
region with diameter $\le$ 300 pc or a disk blackbody), 
rather than a power-law
spectrum.
The point source in clump 4 in Arp 240W has a very hard
spectrum, with
80 counts between 1.2 and 7 keV, and only 2 counts $\le$ 1.2 keV,
and may be very absorbed (see
Figure 1).
The spectrum can be fit with 
a power law 
with a photon index $\gamma$ = 2.7 $\pm$ $^{1.0}_{0.8}$.
Clump 5, at the base of the southern tail in Arp 240E, is extended
with a thermal spectrum.  
The point source in clump 1 of Arp 270 may also be a highly absorbed thermal
spectrum, but this is not
well-constrained.

We are not able to provide strong constraints
on the absorbing hydrogen column density of any of these
sources based on the X-ray spectra itself. 
We therefore estimated the absorption 
using the 
L$_{24}$/L$_{H\alpha}$ ratios.
These corrections and the corrected luminosities
are discussed at length in Section 7.3.
To match the
large-aperture GALEX/SDSS/Spitzer/HST photometry,
we also obtained X-ray luminosities 
within a 5$''$ radius aperture.
In most cases, these are less than twice that of the central source.

\subsection{Photometry of Comparison Systems}

To put these hinge clumps into perspective, 
we compare them to star forming regions in other galaxies.
We focus in particular on two comparison systems which have high quality 
X-ray data available in addition to UV/optical/IR images:
the Antennae galaxies and the normal spiral galaxy NGC 2403.

For the Antennae, we 
use the 4\farcs5 radius (530 pc)
Spitzer, GALEX, and ground-based H$\alpha$ 
photometry of 34 positions provided by \citet{zhang10}.
Most of these positions were selected based on peaks
in the 24 $\mu$m map, although three were selected
based on GALEX UV sources.

To augment this UV/optical/IR
photometry, we used archival Chandra data
for the Antennae galaxies.
We combined six Chandra observations of the Antennae
(ObsID 700479/80/81/82/83) taken
in the FAINT mode, removed flares with a sigma clipping method
(2.5 $\sigma$), giving a total of 329 ksec of observations.
This Chandra data has been used in numerous
earlier studies of the Antennae
(e.g., \citealp{fabbiano01, fabbiano04, zezas02, zezas02a, zezas02b,
zezas06}),
with the diffuse 
X-ray emission in the
Antennae previously
being
studied by \citet{metz04} and \citet{brassington07}. 
We extracted X-ray fluxes
for the 34 \citet{zhang10} regions,
using 4\farcs5 (530 pc) radius apertures.
In the Appendix of this paper, we provide a detailed
description of the overlap region of the Antennae,
including its X-ray spectra, for comparison to the
hinge clumps.

We also compare the hinge clumps to star forming
regions within the normal spiral galaxy NGC 2403.
For
the star forming regions in NGC 2403, 
we used IRAF to extract Spitzer and GALEX photometry 
from archival data 
using the same positions and apertures
as used by \citet{yukita10} to measure the diffuse X-ray emission.
These apertures have 
radii of 7\farcs3 $-$ 13\farcs8, 
corresponding to 110 pc to 210 pc.
For this photometry, we used background annuli with inner and outer radii
of 13\farcs8 and 20\farcs0, respectively, and the mode sky fitting algorithm.
We repeated this process for narrowband H$\alpha$ and off-H$\alpha$ red continuum
images of NGC 2403 from van Zee et al.\ (2013, in preparation).

In Section 5 of this paper, 
we also compare the UV/optical/IR colors of
the hinge clumps with published values for
star forming regions within other strongly 
interacting galaxies as well as regions within the Magellanic Clouds.

\section{Star Formation Rates }

The H$\alpha$ luminosities for some of the hinge clumps
are very high (Table 6).
Of our twelve hinge clumps,
nine
have observed 
L$_{H\alpha}$ $>$ 10$^{40}$ erg~s$^{-1}$ and six have L$_{H\alpha}$
$>$ 10$^{41}$ erg~s$^{-1}$.  For comparison, 30 Doradus in the Large Magellanic
Cloud has an observed 
L$_{H\alpha}$ of only 5 $\times$ 10$^{39}$ erg~s$^{-1}$,
and the giant H~II region complex NGC 5461 in M101
has an observed L$_{H\alpha}$ $\sim$ 1.5 $\times$ 10$^{40}$ erg~s$^{-1}$
\citep{kennicutt84}.

To complement the H$\alpha$ luminosities,
in Table 10 we provide monochromatic 
luminosities
($\nu$L$_{\nu}$) of the hinge clumps
in the NUV and 24 $\mu$m bands.
Six of our 12 hinge clumps have L$_{NUV}$ larger than all ten
of the TDGs in the SB\&T sample \citep{smith10}.
Seven of the clumps have 
L$_{24}$ $\ge$ 6 $\times$ 10$^{42}$ erg~s$^{-1}$.
For context, 
out of a sample of 26 normal spirals,
only four have {\it total} 24 $\mu$m luminosities
in this range \citep{smith07}.  The entire 24 $\mu$m
luminosity for the Antennae is 2 $\times$ 10$^{43}$ erg~s$^{-1}$,
and for the starburst galaxy Arp 284 it is 3 $\times$ 10$^{43}$
erg~s$^{-1}$ \citep{smith07}.

In Table 10, we also provide two estimates
of the star formation rates (SFRs)
of the clumps.   First, we
use the 
equation SFR (M$_{\sun}$~yr$^{-1}$) 
= 5.5 $\times$ 10$^{-42}$[L$_{H\alpha}$ + 0.031~L$_{24}$] (erg/s),
where the 24 $\mu$m luminosity
L$_{24}$ is defined as $\nu$L$_{\nu}$.
This relationship was 
found for H~II regions in nearby galaxies
assuming a Kroupa initial mass function
\citep{calzetti07, kennicutt09}.
We make a second estimate of the SFR 
from the NUV luminosity first correcting for extinction using
L$_{NUV}$(corr) = L$_{NUV}$ + 2.26L$_{24}$ 
and then using the relation
log(SFR) = log(L$_{NUV}$(corr)) $-$ 43.17,
where L$_{NUV}$ =
$\nu$L$_{\nu}$
\citep{hao11, kennicutt12}.
The two methods 
agree reasonably well (Table 10).
There is a large range in the inferred SFR for the hinge clumps 
in our sample, from fairly low values (0.02 M$_{\sun}$~yr$^{-1}$ for
Arp 270-1) to very high values.
Seven of the 12 hinge clumps have estimated
SFRs greater than 1 
M$_{\sun}$~yr$^{-1}$, with the highest
being Arp 240-5, with a rate of 9
M$_{\sun}$~yr$^{-1}$.
For the two methods of estimating SFR, using global fluxes we calculated
the total SFR for the parent galaxies of the target hinge clumps and
compared with the SFR for the hinge clump itself.  The percent
of the total SFR from the galaxy due to the hinge clump ranges from
1\% in Arp 270-1 and Arp 270-2 to 56\% in Arp 240-5 (Table 10).

These
estimates of SFR are very
approximate, as
these formulae
were derived assuming 
constant star formation rates over the last $\sim$10 $-$ 100 Myrs
(see \citealp{kennicutt12}), while the clumps have probably undergone recent
bursts of star formation.
In spite of the uncertainties,
however, these numbers aid comparison to other studies,
which frequently quote SFRs for both individual knots of star formation
within galaxies
(e.g., \citealp{boquien07, boquien09b, boquien11, kennicutt07,
cao07, beirao09, pancoast10})
as well as galaxies as a whole (e.g., \citealp{kennicutt12}).

For comparison to the hinge clumps, we calculated
SFRs for the Antennae regions using the two methods
described above.   
In these calculations, we assume a distance of 24.1 Mpc
to the Antennae\footnote{As for the hinge clumps, in calculating
this distance we assume
H$_0$ = 73 km~s$^{-1}$~Mpc$^{-1}$ and account for peculiar velocities
due to the Virgo
Cluster, the Great Attractor, and the Shapley Supercluster.
A smaller distance of 13.3 Mpc was found by \citet{saviane08}
based on the apparent tip of the red
giant branch. This latter result was questioned
by 
\citet{schweizer08}, 
who find 22.3 Mpc
based on a type Ia supernova.   They suggest that the 13.3 Mpc
estimate was 
probably due to a mis-identification
of the red giant branch.
For more information, see \citet{tammann13}.
}.
The most intense star formation in the Antennae is occuring in 
regions 3 and 4 from the 
\citet{zhang10} study, which lie in the overlap region.
For these regions, the implied SFRs are
1 $-$ 2 M$_{\sun}$~yr$^{-1}$.  Several of our hinge clumps exceed these rates
(see Table 10).  Summing over all of the regions in the Antennae,
we find a total SFR for the Antennae
of 9 $-$ 12 M$_{\sun}$~yr$^{-1}$ for the two methods.

Although several of these hinge clumps have very high star formation
rates, their H$\alpha$ equivalent widths (Table 6) are not particularly high,
compared to those of H~II regions within other galaxies (e.g.,
\citealp{cedres05, boquien09a, popping10, sanchez12}).  
This suggests that the hinge clumps contain a wide range of stellar ages
and possibly considerable underlying older stellar population.
The areas covered by our hinge clump apertures are 
larger than typically used for H$\alpha$ photometry of individual H~II regions
in nearby galaxies, thus we are likely adding light from surrounding older stars
to the continuum flux.
Other factors that may affect the H$\alpha$ equivalent width include
extinction variations within the clump, metallicity effects, 
stellar absorption of H$\alpha$, 
and 
differences in the 
initial mass function (IMF).

\section{UV/IR Colors }

In Figure 2, 
we plot the [3.6 $\mu$m] $-$ [24 $\mu$m] colors of the
hinge clumps
vs.\ the [8.0 $\mu$m] $-$ [24 $\mu$m] and NUV $-$ [24 $\mu$m] colors.  
We compare these values with published photometry for 
star forming regions in several other
interacting galaxies.
These include 
knots of star formation within the disks of the Antennae
galaxies 
(Arp 244;
\citealp{zhang10}), 
additional star formation regions in Arp 82 besides
the hinge clump (from \citealp{hancock07}),
as well 
as 
tidal and disk clumps in 
the interacting galaxies
Arp 107 \citep{smith05, lapham13},
Arp 24 \citep{cao07}, 
and 
Arp 143 \citep{beirao09}, 
along with Arp 105 and Arp 245
\citep{boquien10}.
In this Figure, we excluded sources that
are likely to be foreground/background objects 
not associated with the galaxies (see \citealp{lapham13}).
In the left panel of
Figure 2, we also include Spitzer colors for H~II regions
in the Small and Large Magellanic Clouds \citep{lawton10}.
In Figure 2, we also compare with 
star forming regions within the normal spiral
galaxy NGC 2403, as discussed in Section 3.4.

As can been seen in Figure 2, there is a range in colors for
the hinge clumps, however, on average, the hinge clumps have 
redder [3.6] $-$ [24] colors
(i.e., high 
L$_{24{\mu}m}$/L$_{3.6{\mu}m}$
ratios)
than the other clumps. 
The 3.6 $\mu$m
emission from galaxies is generally assumed to be dominated by light
from the older stellar population/underlying stellar mass 
(e.g., \citealp{helou04}),
while the 24 $\mu$m light is emitted from `very small interstellar dust grains'
(VSGs),
heated mainly by UV light from massive young stars 
(e.g., \citealp{li01}).  
Thus
a red [3.6] $-$ [24] color
implies
very obscured star formation,
a very young burst, and/or a high young/old stellar mass ratio.

The [3.6] $-$ [24] colors of some of the hinge clumps
approach that of the `overlap' region in the Antennae galaxies
(Figure 2).  
The overlap region corresponds to regions 3 and 4 in the 
\citet{zhang10} Antennae study, with region 4 being more obscured.
In particular, 
the [3.6] $-$ [24] color of the NGC 2207 region is similar to
that of regions 3 and 4, and is much redder than the 
other hinge clumps.
The Arp 82 hinge clump and Arp 240 clump 3
are also quite red in this color. 
This supports the idea that these are young regions.

Figure 2 shows that 
the [3.6] $-$ [24] color is correlated with [8.0] $-$ [24],
with some of the hinge clumps having very red [8.0] $-$ [24] colors.
The [8.0] $-$ [24] color is a function of the ultraviolet
interstellar radiation field (ISRF), with a 
stronger ISRF producing
a higher 24 $\mu$m flux
compared to that at 8 $\mu$m (i.e., a redder [8.0] $-$ [24] color)
\citep{li01, peeters04, lebouteiller11}.
The 8 $\mu$m Spitzer band contains emission from both
very small dust grains
and polycyclic aromatic
hydrocarbons (PAHs).  PAHs
can be excited
by non-ionizing photons
as well as UV, so the 8 $\mu$m emission
may be powered in part by lower mass
stars \citep{peeters04, calzetti07, lebouteiller11}.
The Small Magellanic Clouds regions have redder [8.0] $-$ [24]
colors for a given [3.6] $-$ [24] colors, likely caused by
less PAHs in the 8 $\mu$m band.  Low metallicity
dwarfs tend to be deficient in the 8 $\mu$m Spitzer band compared to 
normal spirals \citep{engelbracht05, rosenberg06, rosenberg08, draine07,
wu07}.

The FUV to infrared ratio has sometimes been
used as an indicator of dust reddening (e.g., \citealp{boquien09a}),
with redder colors meaning more absorption of the UV by dust. 
Since some 
of our hinge clumps do not have FUV images available,
we use NUV $-$ [24] as an alternative indicator of extinction.
Figure 2 shows that there is a range in 
extinction for the hinge clumps.
In Figure 2, 
the NUV $-$ [24] color correlates to some
extent with [3.6] $-$ [24], but there is considerable
scatter.  In general, the regions within the Antennae
are more obscured than many of the other
clumps for the same [3.6] $-$ [24] color, particularly
those in the normal spiral NGC 2403.
\citet{zhang10} region 4, in the 
overlap region of the Antennae galaxies, is particularly
red in NUV $-$ [24].

In the left panel of Figure 3, 
we plot [3.6] $-$ [24] vs.\ [3.6] $-$ [4.5] for the hinge clumps and
the comparison samples.  
For spiral galaxies as a whole,
both the 3.6 $\mu$m and 4.5 $\mu$m Spitzer bands are generally assumed to be
dominated by starlight, as the 
global [3.6] $-$ [4.5] colors of spirals are usually close to zero 
(within 0.1 magnitudes) (e.g., \citealp{pahre04, smith07}).
However,
low metallicity starburst galaxies sometimes show large
excesses in these bands above the stellar continuum, with red
[3.6 $\mu$m] $-$ [4.5 $\mu$m] colors
\citep{smith09}.
These excesses may be due to 
hot dust and/or nebular emission (Br$\alpha$ line emission
and/or the nebular continuum).  All of these factors 
are expected to be stronger
in younger clumps, while the nebular components will be enhanced
in low metallicity systems.

Within interacting galaxies, localized knots
of intense star formation
sometimes have large excesses above
the stellar continuum in the 3.6 $\mu$m and 4.5 $\mu$m bands, particularly
in the 4.5 $\mu$m filter
\citep{smith08, zhang10, boquien10}.
In Figure 3 (left), 
a trend is seen, such that the regions with redder [3.6] $-$ [24] colors
are redder in [3.6] $-$ [4.5].   This is consistent with increasing
4.5 $\mu$m non-stellar excess with 
younger stars or higher young/old star ratios.
However, there is considerable scatter
in this plot. In particular, the Magellanic Cloud regions 
are redder in [3.6] $-$ [4.5] 
for the same [3.6] $-$ [24] color compared to 
the regions within the interacting galaxies (Figure 3).
This may be a consequence of lower metallicity.
Such an offset is also seen in the global colors of 
low metallicity dwarfs compared
to higher metallicity galaxies \citep{smith09}.

Interestingly, some of the NGC 2403 regions also lie to the right of the
correlation marked by the clumps in the interacting galaxies, in the regime
populated by the Magellanic Cloud regions.  
This may also be a metallicity effect.
The red NGC 2403 regions are preferentially found at larger galactic radii.
NGC 2403 is a moderate luminosity (M$_{\rm B}$ = $-$18.9; 
\citealp{moustakas10}) 
late-type (Scd) spiral.
The NGC 2403 region with the largest [3.6] $-$ [4.5] color, 
J073628.6+653349, has a deprojected distance from the nucleus
of 6.21 kpc \citep{garnett97}.  Its oxygen abundance 
has been determined by the
direct electron temperature method to be 
log(O/H)~+~12 = 8.10 $\pm$ 0.03
\citep{garnett97} or
log(O/H)~+~12 = 8.28 $\pm$ 0.04
\citep{berg13}.
This is in the range where red [3.6] $-$ [4.5] 
colors become more common for dwarf galaxies \citep{smith09}.
No metallicities are available for our hinge clumps at present.

In the right panel of Figure 3, we plot the [3.6] $-$ [24] color against
the 24 $\mu$m luminosity.
A rough correlation is seen in this plot, such that the clumps
with the highest luminosity tend to be redder.  The large scatter
in this plot is likely due in part to the fact that the apertures
cover different physical sizes in the different galaxies.
Note that the NGC 2403 and Magellanic Cloud luminosities are lower on average, due 
to the smaller areas covered.
The LMC and SMC apertures range from 105 pc to 450 pc \citep{lawton10}.

\section{X-Ray vs. UV/IR Colors}

In Figure 4 
we plot the observed (uncorrected for
dust attenuation) log L$_X$/L$_{24{\mu}m}$ against 
[3.6] $-$ [24] and against NUV $-$ [24], for the hinge clumps with Chandra
data available.
In this Figure we compare
with both the 
star forming knots in the Antennae as well as the star forming
regions within the normal spiral NGC 2403.
For the regions in NGC 2403, we used the diffuse X-ray 
fluxes from \citet{yukita10},
which were obtained 
from the Chandra map after bright point sources were removed.
This plot shows that the NGC 2403 regions have lower 
observed L$_X$/L$_{24{\mu}m}$ for a given [3.6] $-$ [24] color
than either the hinge clumps
or the positions in the Antennae.  
They also have lower [NUV] $-$ [24] values, indicating lower
extinction. 
Some of the hinge clumps lie in the region populated by the 
Antennae points, while some are lower on this plot.   
The NGC 2207
clump 
lies between Antennae overlap regions 3 and 4 in these plots, with large
[3.6] $-$ [24] and low observed 
L$_X$/L$_{24{\mu}m}$. 

Correcting these X-ray fluxes 
for dust attenuation is critical, 
given the large absorbing columns 
towards some of these regions implied by 
their red NUV $-$ [24] colors.
The correction for dust attenuation is discussed further in Section 7.3.

Two of the hinge clumps
with X-ray point sources 
(Arp 270-1 and Arp 240-4)
likely host ULXs, as their X-ray luminosities
are very high compared to other regions with similar [3.6] $-$ [24]
colors (Figure 4).  
Alternatively, these regions may contain two or more 
lower luminosity HMXBs very close together.
The 
third X-ray point source,
Arp 240-3, has a location on this plot similar to that
of the clumps resolved by Chandra, thus
its X-ray emission may be due to a compact young star formation
region rather than a ULX.
This is consistent with its intrinsically soft X-ray spectrum
(Section 3.3).

\section{Stellar Population Synthesis Models}

\subsection{Overview}

We compared the large aperture
GALEX/SDSS
broadband 
UV/optical colors of these clumps with
population synthesis models
to determine the ages of their stellar populations and their
dust attenuations.
We used the large aperture HST photometry as a check on these fits.
We did not use 
the 2MASS near-infrared or Spitzer 3.6 and 4.5 $\mu$m fluxes in these fits,
although
this light may also be dominated by starlight.   This point
is discussed further in Section 7.2.
We obtained a second independent estimate of the stellar age from the H$\alpha$
equivalent width.

For this analysis we assume a single instantaneous burst.
As discussed
below, for some of these clumps more than one generation of 
stars may be present.
In these cases, the ages we derive from the 
single-burst models
are a luminosity-weighted average age for the stars in the clump.
As in our earlier studies
\citep{smith08, hancock09, lapham13},
we use the Starburst99 population synthesis code
\citep{leitherer99} and include the Padova asymptotic giant
 branch stellar models \citep{vaz05}.
We assume
a Kroupa initial mass function
and solar metallicity.
We integrated the model spectra
over the GALEX and SDSS bandpasses,
including
the
H$\alpha$ line from Starburst99
in the model spectra
as well as other optical emission lines,
derived using the prescription given by \citet{anders03} for
solar metallicity star forming regions.
We used the \citet{calzetti94}
starburst dust attenuation law.

The best-fit ages and dust reddening
E(B$-$V) values from the broadband photometry
are given in Table 11, along with uncertainties
on these values.
These were computed using a chi squared ($\chi$$^2$) minimization
calculation as in our earlier studies.
For these fits, we only used the filters with reliable detections; 
upper limits were ignored.  When available, we used the FUV $-$ NUV,
NUV $-$ g, u $-$ g, g $-$ r, r $-$ i, and i $-$ z colors for fitting.
In calculating the $\chi$$^2$ values, in addition to the statistical
errors, we included additional uncertainties in the colors
due to uncertainties in the GALEX aperture corrections (see Table 4) and 
in
background subtraction. 
To estimate the uncertainties due to background subtraction,
we calculated colors using
an alternative second sky annulus with an inner
radius of 8$''$ and an outer radius of 14$''$.
To estimate the uncertainties in the best-fit parameters, we used
the $\Delta$$\chi^2$ method \citep{press92} to determine
68.3\% confidence levels for the parameters.
In Table 11, we provide the reduced $\chi$$^2$, equal to $\chi$$^2$/(N$-$2), 
where N is the number of 
colors used in the fit.

As can be seen in Table 11, for some of the clumps
the fits are quite good.  For a few, however, the $\chi$$^2$ values
indicate a very poor fit (e.g., Arp 82, Arp 240-1, Arp 240-3,
and especially Arp 256).
This suggests that more than one age
of stars are present.  Two or more bursts of star formation
may have occured in the region, or the star formation continued over an extended time
period.    

We obtain a second independent estimate of the age of the stellar
population using the H$\alpha$ equivalent
width and assuming an instantaneous burst.  
These ages are tabulated in Table 12.
Except for regions Arp 240-1, Arp 256-1, Arp 270-1, and Arp 270-3,
these ages 
are younger than the ages
derived from the broadband UV/optical fluxes (Table 11).
Such differences in derived ages
have been found earlier for some knots of star formation
in Arp 284 and Arp 107 \citep{peterson09, lapham13},
and 
provide additional support for the idea that these hinge clumps
host a range of stellar ages.   
Ages from H$\alpha$ equivalent widths tend to be biased
towards the younger populations in a region, while
the broadband ages are weighted towards the somewhat
older stars that sometimes dominate the UV/blue light from
star forming regions.
The comparison of these two 
age estimates illustrates the uncertainties in 
age determinations via
population synthesis. 

\subsection{SED Plots and Near-IR Excesses}

In Figures 5 to 7, we plot the large-aperture spectral energy
distributions (SEDs) for these clumps, including the
GALEX, SDSS, Spitzer, 2MASS, and HST large aperture photometry.
On these plots, we overlay
the best-fit Starburst99 single-population
models from the broadband data.  As an indication of the uncertainties
in these models, in these figures we also plot models with the
best-fit reddening and the best-fit age $\pm$ the 1$\sigma$ uncertainty
in the age.  As can be seen from these plots, in some cases a single-burst
model does not provide a good match to the SED.
For NGC 2207, which lacks SDSS data, we do not provide 
a best-fit model, however, comparison to the other
SEDs shows that it has quite red colors, consistent with
high obscuration (see Section 7.3). 

As shown in Figures 5 $-$ 7, for some of the clumps the Spitzer 
3.6 $\mu$m and 4.5 $\mu$m
fluxes are higher than predicted by the best-fit single-burst models, which
do not include dust emission.
In some cases, the HST or 2MASS near-infrared fluxes are also higher than expected
from our single-burst models.
This is additional evidence for two or more
generations of stars or on-going star formation.
Alternatively, as noted in Section 5, there may
be excess flux in these bands above the stellar continuum due to hot dust
or to interstellar gas emissions (see \citealp{smith09}).

Instead of instantaneous bursts, a better model of the star formation history
in these clumps
might be prolonged star formation
(continuous for an extended period, then shut off), or 
an
exponentially-decaying star formation rate (e.g., \citealp{boquien10}).
As discussed in the Appendix of this paper, regions 3 and 4
in the Antennae galaxy each host stars with a range of
ages; our hinge clumps, which cover even larger physical
sizes than the Antennae photometry, may also contain a range
of stellar ages.
As an approximation, in
\citet{lapham13} we
explored models with two instantaneous bursts for fitting clumps
of star formation within Arp 107.
These provide better fits to the SEDs than
single bursts, but we were unable to constrain
the parameters of the fit well, 
as we were able to find multiple models with very
different parameters that fit the data equally well.  
In general, adding more parameters to the fitting routine
by allowing more than one stellar age increases the uncertainties
in the derived parameters.   Thus in this work, we focus on single-burst
models, and emphasize that the derived ages are luminosity-weighted
stellar ages, averaged over the timescale of the burst, and 
the star formation was 
likely prolonged rather than truly instantaneous.

It is also possible to include the mid-infrared photometry in
the population synthesis modeling (e.g., \citealp{noll09, boquien10}).
However, this also adds additional parameters to the model,
including the dust properties and the location of the dust relative
to the stars.   Thus in this work we derive ages from only the UV/optical
data.

\subsection{Attenuation}

In Table 11, we provide the 
E(B$-$V) estimates 
determined 
from the broadband Starburst99 modeling.
We obtained a second independent estimate of the absorption using the ratio
of the 24 $\mu$m luminosity to H$\alpha$ luminosity
(Table 12).
For this calculation, we used the empirically-determined 
relationship between
the
absorption in the H$\alpha$ emission
line and the 24 $\mu$m emission obtained 
by \citet{kennicutt07} for star forming regions within M51:
A$_{H\alpha}$ = 2.5~log[1 + 0.038L$_{24}$/L$_{H\alpha}$]. 
As noted by these authors,
there is considerable scatter in this relation, and for a given
star forming region the true relation likely depends upon the geometry,
the stellar types, and the age.
In the hinge clumps,
we expect an even larger scatter in the relationship,
as 
we are likely averaging over a range of extinctions within
the beam.
We converted from
A$_{H\alpha}$ to
hydrogen column density 
using A$_{H\alpha}$ = 0.82A$_{\rm V}$
and the \citet{calzetti00} starburst total to selective attenuation 
ratio R$_{\rm V}$
= A$_{\rm V}$/E(B $-$ V) = 4.05.

For some of the clumps, the absorption derived from 
A$_{H\alpha}$ agrees reasonably well with that obtained from
the broadband photometry.  For others,
the H$\alpha$-derived estimates
are higher than that inferred
from the broadband photometry.
This difference may be caused by a range of ages and extinctions
in the clumps, with the younger stars being more obscured.
In starburst galaxies, the ionized gas tends to be more obscured
than the starlight on average (e.g., \citealp{calzetti01}).
For the region in NGC 2207, our estimate of A$_{\rm V}$ = 4.1 from
L$_{24}$/L$_{H\alpha}$
(Table 12) is consistent with the \citet{kaufman12} limit of A$_{\rm V}$
$\le$ 4.9 determined from the ratio of the 6 cm flux to that in H$\alpha$.

As mentioned in Section 3.3, we are not able to
determine X-ray absorptions directly from the Chandra data itself.
We therefore correct the X-ray data for absorption using column densities
obtained from the 
L$_{24}$/L$_{H\alpha}$ ratio, as it is likely more valid for the ionized
gas than values obtained from broadband photometry.
We first converted from E(B$-$V) to hydrogen column density using
N$_{\rm H}$(cm$^{-2}$) = 5.8 $\times$ 10$^{21}$E(B$-$V) 
\citep{bohlin78}.
These estimates range from N$_{\rm H}$ = 1 $-$ 6 $\times$ 10$^{21}$ cm$^{-2}$
(see Table 12).
We emphasize that these estimates are quite uncertain, since 
the X-ray emitting gas may be less obscured than
the H$\alpha$-emitting ionized gas within H~II regions.
However, these estimates 
are similar to 
the absorbing column densities
found by 
\citet{mineo12b}, who fit X-ray spectra of the global diffuse
X-ray emission of a sample of nearby 
late-type star forming galaxies.
Obtaining higher quality X-ray data 
for our hinge clumps 
would be very valuable to better constrain the X-ray absorption.

Using these estimates of the absorbing column,
we next determined the ratio of the
attenuation in the 0.3 $-$ 8 keV X-ray band to the hydrogen column density
A$_X$/N$_{\rm H}$
using 
the 
{\it wabs} routine 
in the 
XSPEC\footnote{http://heasarc.gsfc.nasa.gov/docs/xanadu/xspec/index.html}
software (this uses 
the Wisconsin absorption cross-sections, e.g.,
\citealp{morrison83}).
The 
A$_X$/N$_{\rm H}$
ratio depends upon both the X-ray spectrum and the hydrogen column density.
For this calculation, we assumed a 0.3 keV thermal spectrum, thus
A$_X$/N$_{\rm H}$ varies between 7.19 $\times$ 10$^{-22}$ mag~cm$^{2}$ 
for N$_{\rm H}$ = 3 $\times$ 10$^{20}$ cm$^{-2}$ and 
A$_X$/N$_{\rm H}$ = 2.73 $\times$ 10$^{-22}$
mag~cm$^2$
for N$_{\rm H}$ = 2 $\times$ 10$^{22}$ cm$^{-2}$.

These corrections have been applied to the observed X-ray fluxes.
In Table 9, we provide 
absorption-corrected 
X-ray luminosities for each hinge clump.  We provide both the luminosity
of the central source, as well as luminosities
within the larger 5$''$ radius aperture, 
assuming the same correction factor for both.
These corrections are generally multiplicative
factors of 2 $-$ 5, with larger
correction factors for the two most obscured regions Arp 240-5 and NGC 2207
(factors of 9 and 13, respectively).
We emphasize that these corrected luminosities are quite uncertain
due to uncertainties in both the correction method and in the fluxes
themselves.

The absorption-corrected
X-ray luminosities L$_{\rm X}$ 
(0.3 $-$ 8 keV)
of the eight detected hinge clumps (Table 9) are all
greater than 10$^{39}$ erg/s, with seven greater than 10$^{40}$ erg/s.
For the three hinge clumps that were undetected by Chandra,
we provide upper limits in Table 9 
assuming 
the reddening from the 
L$_{24}$/L$_{H\alpha}$ ratio.

The most X-ray luminous region in Table 9 
is 
the point source in clump 4 of Arp 240W, which has a  
corrected 0.3 $-$ 8 keV luminosity of
$\sim$2 $\times$ 
10$^{41}$ erg~s$^{-1}$.  
The absorption correction for this source may be overestimated by
the above method, if the source has a intrinsically hard spectrum.
However,
even without any correction at all, the luminosity of this source is high, 
5 $\times$ 10$^{40}$ erg~s$^{-1}$. 
Obtaining higher S/N X-ray spectroscopy of this source would be very valuable,
to better determine the absorbing column and therefore the intrinsic
luminosity. 
At a luminosity of $\sim$2 $\times$ 
10$^{41}$ erg~s$^{-1}$, 
this would be one of the most luminous ULXs known, and therefore a possible
intermediate mass black hole.

\subsection{Stellar Masses}

For both methods of determining 
ages and extinctions,
we estimated stellar masses for the
clumps (Tables 11 and 12).
As expected, the masses derived using the H$\alpha$ equivalent widths
are sometimes lower than those obtained from the broadband photometry,
since the inferred ages are younger.
We note that stellar masses
derived by 
scaling directly from the 
K and 3.6 $\mu$m
photometry (e.g., \citealp{bell01, bell03, into13}) 
are considerably larger (factors of 2 $-$ 25 times larger) than
the masses obtained from individual
fitting of the UV/optical broadband photometry.
These scaling factors are intended
for galaxies as a whole rather than individual star forming regions
within galaxies, and, 
as discussed by \citet{gallazzi09}, 
they may over-estimate the stellar mass in systems undergoing recent
bursts.
Even when population synthesis fitting is done, the stellar masses
are quite uncertain
(up to a factor of 5; \citealp{smith08}).

\section{L$_{\rm X}$ vs.\ Star Formation Properties}

\subsection{L$_{\rm X}$/SFR Ratios:
The Hinge Clumps, Antennae, and NGC 2403 }

In Table 12 we provide 
L$_{\rm X}$/L$_{H\alpha}$ and 
L$_{\rm X}$/SFR ratios for each
hinge clump, corrected for extinction 
using the 
H$\alpha$/24 $\mu$m-derived values.
For these calculations, we use the X-ray fluxes within a 5$''$ radius.
Two of the hinge clumps (Arp 240-4 and Arp 270-1) have
much higher values than the rest, supporting
the idea that they are ULXs.
For the remaining clumps,
this table shows that 
the X-ray luminosity is not perfectly correlated
with SFR, since there is a scatter of about a factor of four in these ratios.

For the Antennae and NGC 2403
we calculated 
L$_{\rm X}$/SFR ratios using the same procedure.
As with the hinge clumps, 
for both the Antennae and NGC 2403 it is difficult
to constraint the internal extinction from the X-ray spectra alone
(e.g., \citealp{metz04, yukita10}).
We therefore estimate the dust absorption 
within these regions  
using the 24~$\mu$m
and H$\alpha$ fluxes within our apertures, 
as we did for the hinge clumps.
For the Antennae
these range from A$_{\rm V}$ = 0.67 to 3.23, with region 4 having the highest value.
For the NGC 2403 regions, the implied absorption is much smaller than for the Antennae
and the hinge clumps,
with A$_{\rm V}$ between 0.08 and 0.50.

Excluding the regions with bright X-ray point sources,
the median
extinction-corrected
L$_{\rm X}$/SFR 
for the Antennae
areas is 
5.7 $\times$ 10$^{39}$ (erg/s)/(M$_{\sun}$~yr$^{-1}$) using 
the H$\alpha$+24 $\mu$m estimate
of SFR.
For the NUV+24 $\mu$m method, 
the median
L$_{\rm X}$/SFR 
is 8.3 $\times$ 10$^{39}$ (erg/s)/(M$_{\sun}$~yr$^{-1}$) for the Antennae regions.
These are similar to the ratios for the non-ULX hinge clumps
(Table 12).

For NGC 2403, the 
absorption-corrected
L$_{\rm X}$/SFR ratios have a 
median value of 6 $\times$ 10$^{38}$ 
(erg/s)/(M$_{\sun}$~yr$^{-1}$) 
for the H$\alpha$+24 $\mu$m method
and 4 $\times$ 10$^{39}$
(erg/s)/(M$_{\sun}$~yr$^{-1}$) 
for the NUV+24 $\mu$m method.
The latter value agrees reasonably well with those of the hinge clumps,
while
the ratio determined using the 
H$\alpha$+24 $\mu$m method 
is somewhat lower.  

\subsection{L$_{\rm X}$/SFR Ratios From Other Studies }

For a sample of nearby spirals and irregulars, \citet{mineo12b}
extracted the diffuse X-ray emission and compared with the SFR.
Their best-fit value for the ratio of 
the extinction-corrected 0.3 $-$ 10 keV X-ray luminosity
from hot gas 
to the SFR is 
L(0.3 $-$ 10 keV)/SFR = (7.3 $\pm$ 1.3) $\times$ 10$^{39}$ 
(erg/s)/(M$_{\sun}$~yr$^{-1}$).  
Given their somewhat different definition of SFR, their 
different energy range, and their different method of correcting
for internal extinction, their results are 
consistent with
our values for the hinge clumps (Table 12).
They also agree well with our values for the Antennae regions dominated
by diffuse X-ray emission (Section 8.1).
The \citet{mineo12b} ratios are somewhat higher than
the mean L$_{\rm X}$/SFR of
1.4 $\times$ 10$^{39}$ (erg/s)/(M$_{\sun}$~yr$^{-1}$)
found by \citet{li13}
for the coronal X-ray emission from 53 edge-on disk
galaxies.  

In \citet{smith05b}, we collated X-ray luminosities, 
extinction-corrected H$\alpha$ luminosities, and 
published ages for 20 H~II regions within Local
Group galaxies and the nearby galaxies 
M101 and NGC 4303.
Regions with ages $\le$3 Myrs (before supernovae
turn on) show dramatically lower 
L$_{\rm X}$/SFR ratios
than 
older regions.  The younger H~II regions have a median
L$_{\rm X}$/SFR of
3.4 $\times$ 10$^{36}$ (erg/s)/(M$_{\sun}$~yr$^{-1}$).
This contrasts sharply with the 
median value of the rest of the regions of 
2 $\times$ 10$^{39}$ (erg/s)/(M$_{\sun}$~yr$^{-1}$).
Our hinge clumps have
ratios similar to those of the older H~II regions, indicating that
supernova activity has begun in the hinge clumps.

In \citet{smith05b},
we found extended X-ray emission
from four 
H~II regions in the primary disk of the interacting pair 
Arp 284 using Chandra.
Using the same conversion as above, for these regions we find 
L$_{\rm X}$/SFR between 
8 $\times$ 10$^{39}$ 
(erg/s)/(M$_{\sun}$~yr$^{-1}$) and 
4 $\times$ 10$^{40}$ 
(erg/s)/(M$_{\sun}$~yr$^{-1}$).  These are similar to the values for our
hinge clumps.

From these comparisons, we conclude that the hinge clumps are
producing X-rays 
at a rate relative to the SFR similar to those of star forming regions
in other galaxies.  

\subsection{HMXBs }

The intrinsically soft X-ray spectra of most of the hinge clumps
argues that this emission is dominated by hot gas.
However, a fraction of the observed
X-ray emission may be due to the combined light of multiple
unresolved X-ray point sources.   
For star forming regions,
this additional component is likely dominated by HMXBs, 
as the number of HMXBs in a galaxy is correlated with the SFR
\citep{grimm03, gilfanov04, persic04, mineo12a}.
Other kinds of X-ray-emitting objects, such as 
YSOs, hot stars, and low mass X-ray binaries, are expected to be 
less important 
to the observed X-ray emission from strongly star forming
systems
(e.g., \citealp{bogdan11, mineo12b}).
HMXBs are 
associated with populations with ages between 20 and 70 
Myrs \citep{antoniou10, williams13}.  
These ages are consistent with 
our estimates of the average stellar ages in some of the hinge clumps
from the broadband UV/optical photometry.  Thus sufficient time may
have passed to
produce HMXBs in at least some of our hinge clumps.

To estimate the contributions from HMXBs to the X-ray luminosities
of the hinge clumps, we use the relation found by \citet{mineo12a}
for HMXBs in nearby star forming galaxies of 
L(0.5 $-$ 8 keV)(HMXBs) = SFR $\times$ (2.6 $\times$ 10$^{39}$ erg~s$^{-1}$).
As discussed at length by \citet{mineo12a}, this relation agrees well with
relations found by \citet{grimm03} and \citet{ranalli03}, after conversion
to the same energy range and SFR definitions.
Excluding the two candidate ULXs, 
this relation implies that about 15\% $-$ 30\% of the X-ray light from the hinge
clumps comes from HMXBs, with an estimated fraction of 40\% for clump 1
in Arp 240.   These estimates are very uncertain, 
as there is a lot of scatter in the SFR-HMXB relation \citep{mineo12a},
and this relation was derived by averaging over entire galaxies, 
thus it is not necessarily appropriate for individual star forming regions.
However, it provides a very approximate estimate, which supports
our conclusion based on the soft X-ray spectra that in most cases the 
X-ray emission from these clumps
is dominated by radiation from hot gas rather than HMXBs.

As noted earlier, the absorption-corrected X-ray luminosity of
clump 4 in Arp 240 (2 $\times$ 10$^{41}$ erg~s$^{-1}$)  
is higher than expected
for HMXBs, thus it may host an intermediate-mass black hole.
Alternatively, it may contain multiple luminous HMXBs.  Its
L$_{\rm X}$/SFR of 7 $\times$ 10$^{40}$ erg~s$^{-1}$~(M$_{\sun}$/yr)$^{-1}$ 
(Table 12) is more than an order
of magnitude higher than that found by \citet{mineo12a}
for HMXBs in nearby spiral galaxies.   If this X-ray emission
is due to a collection of HMXBs,  this region
is either unusually rich in HMXBs per SFR compared to
other systems, or its
star formation rate was considerably higher in the very recent past.
However, our H$\alpha$ and NUV estimates of star formation
rate for this clump agree well (2.8 M$_{\sun}$~year$^{-1}$ 
and 2.3 M$_{\sun}$~year$^{-1}$;
Table 10), although H$\alpha$ and NUV
are sensitive to star formation
on different timescales ($\sim$10 Myrs and $\sim$200 Myrs, 
respectively;
\citealp{kennicutt12}).
Thus there is no evidence for
a dramatic drop in SFR in this region
in the $\le$70 Myrs timescale for HMXB production.  
The total stellar mass of this clump
(2 $\times$ 10$^8$ to 1 $\times$ 10$^9$ M$_{\sun}$; 
Tables 11 and 12) could be produced by a steady rate
of $\sim$ 2 $-$ 3 M$_{\sun}$~year$^{-1}$ over $\sim$200 Myrs, 
without requiring a significantly
higher rate in the recent past,
but that steady rate would not be 
enough to produce such a high HMXB luminosity today.

\subsection{ULXs}

A related question is 
whether the hinge clumps host ULXs
at the rate expected from their SFRs, 
where a ULX is defined to be a point source with 
L$_{\rm X}$ $\ge$ 10$^{39}$ erg~s$^{-1}$.  
As noted earlier, most ULXs with luminosities between 10$^{39}$ erg~s$^{-1}$
and 10$^{41}$ erg~s$^{-1}$ are likely high luminosity HMXBs.
Out of our 12 hinge clumps, three host point sources
with 
L$_{\rm X}$ $\ge$ 10$^{39}$ erg~s$^{-1}$, 
however, 
the source in Arp 240-3 has a likely thermal spectrum, and
as discussed below, 
its X-ray emission is probably dominated by hot gas. 
The other two point sources, Arp 240-4 and Arp 270-1, 
are candidate ULXs.

To determine whether the frequency of ULXs in hinge clumps relative
to the SFR is consistent with those of galaxies as a whole, we compare to 
our statistical studies of ULXs in normal galaxies
\citep{swartz04, swartz11} and strongly interacting
galaxies from the Arp Atlas \citep{smith12}.   
In those studies, we determined
the number of ULXs per far-infrared luminosity L$_{\rm FIR}$ for various
samples of galaxies.   
To compare with these studies, 
we estimated L$_{\rm FIR}$ for the hinge clumps using the
\citet{calzetti05} relation between the 8 $\mu$m and 24 $\mu$m fluxes
and the total infrared luminosity L$_{\rm IR}$, and used the approximate
relation between L$_{\rm FIR}$ and L$_{\rm IR}$ of 
L$_{\rm FIR}$ = 0.55L$_{\rm IR}$ from \citet{helou88}.

For all but one of our sample galaxies, the Chandra sensitivity is insufficient
to detect all ULXs down to the limiting luminosity of 
L$_{\rm X}$ $\ge$ 10$^{39}$ erg~s$^{-1}$.  
However, the Chandra data for all of our systems have sufficient sensitivity
to detect ULXs more luminous than 
L$_{\rm X}$ $\ge$ 10$^{40}$ erg~s$^{-1}$.  
We can compare the number of ULXs above  
L$_{\rm X}$ $\ge$ 10$^{40}$ erg~s$^{-1}$
with
the combined estimated
far-infrared luminosity of all of the sample
hinge clumps, 2.2 $\times$ 10$^{44}$ erg~s$^{-1}$.
We find only one candidate ULX in our hinge clump sample above 
10$^{40}$ erg~s$^{-1}$, Arp 240-4.
This gives a ratio of the number of ULXs above 
10$^{40}$ erg~s$^{-1}$ per far-infrared luminosity of
N$_{\rm ULX}$/L$_{\rm FIR}$ 
= 4.6 $\times$ 10$^{-45}$ 
(erg~s$^{-1}$)$^{-1}$.
This is consistent within the uncertainties with the ratios found
for spiral galaxies of 
N$_{\rm ULX}$/L$_{\rm FIR}$ =
6.7 $\pm$ $^{2.9}_{2.1}$ $\times$ 10$^{-45}$ 
(erg~s$^{-1}$)$^{-1}$
\citep{swartz04}
or 9.4 $\pm$ 2.2 $\times$ 10$^{-45}$  
(erg~s$^{-1}$)$^{-1}$
\citep{swartz11}.  
Interestingly, the Arp sample appears to have a deficiency of ULXs
in this luminosity range, with 
N$_{\rm ULX}$/L$_{\rm FIR}$ =
9.6 $\times$ 10$^{-46}$ 
(erg~s$^{-1}$)$^{-1}$
\citep{smith12}.  This may be because the Arp sample includes
a number of ultra-luminous infrared galaxies (ULIRGs), which
dominate the combined far-infrared luminosity of the sample
yet have relatively 
few ULXs.  ULXs may be more obscured in ULIRGs, leading to a
deficiency in the observed number of ULXs in this sample;
alternatively, hidden AGN may be contributing to L$_{\rm FIR}$ in the
ULIRGs \citep{smith12}.
Hinge clumps appear to be forming 
L$_{\rm X}$ $\ge$ 10$^{40}$ erg~s$^{-1}$
ULXs at a rate relative to their SFR similar to spirals.
They do not
appear to be deficient in these ULXs relative to the SFR, in contrast to ULIRGs.

Only for Arp 270 is the Chandra sensitivity sufficient to detect all ULXs
down to 10$^{39}$ erg~s$^{-1}$.  The combined estimated L$_{\rm FIR}$ for the 
four hinge clumps in Arp 270 is 1.5 $\times$ 
10$^{42}$ erg~s$^{-1}$, giving
N$_{\rm ULX}$/L$_{\rm FIR}$ 
= 6.6 $\times$ 10$^{-43}$ (erg~s$^{-1}$)$^{-1}$
for L$_{\rm X}$ $\ge$ 10$^{39}$ erg~s$^{-1}$. 
This is considerably higher
than the
N$_{\rm ULX}$/L$_{\rm FIR}$ ratio for spirals
of 
6.5 $\pm$ 0.7 $\times$ 
10$^{-44}$ (erg~s$^{-1}$)$^{-1}$ \citep{swartz04}
or 5.3 $\pm$ 0.5 $\times$ 10$^{-44}$ (erg~s$^{-1}$)$^{-1}$ 
\citep{swartz11}.
This difference may simply be due to small number statistics;
alternatively, the candidate ULX in Arp 270-1 may be 
a background source rather than a true ULX.

In the ULX luminosity range
$\ge$ 10$^{39}$ erg~s$^{-1}$,
the N$_{\rm ULX}$/L$_{\rm FIR}$ ratio
for Arp systems is similar to that of spirals, 
7.6 $\pm$ 1.3 $\times$ 10$^{-44}$ (erg~s$^{-1}$)$^{-1}$ 
\citep{smith12}.
The subset of Arp systems in the \citet{smith12} sample that
have Chandra sensitivity to 
10$^{39}$ erg~s$^{-1}$
point sources 
contains fewer high L$_{\rm FIR}$ systems than the larger Arp sample
with Chandra sensitivity to only 10$^{40}$ erg~s$^{-1}$.  
The 10$^{39}$ erg~s$^{-1}$ sample of Arp systems appears similar to spirals 
in their ULX populations.

\subsection{Electron Densities in the Hot Gas}

To estimate the 
electron number density 
n$_{\rm e}$
in the hot gas within the hinge clumps,
we used the angular extents of the central X-ray source
(Table 9) to calculate the volume of X-ray-emitting gas, 
assuming an ellipsoidal 
shape with the third axis equal to the average of the other two
dimensions.
For the NGC 2403 regions, we estimated the volume using
the radii of 
the X-ray-emitting regions obtained by \citet{yukita10}, and assumed
a spherical shape. 
For the Antennae,
we estimated angular extents of the diffuse X-ray emission
for each region from the co-added Chandra map, 
assuming an ellipsoidal 
shape as for the hinge clumps.
For each of these regions,
we calculated 
n$_{\rm e}$$\sqrt{f}$, where f is the volume filling 
factor (Table 12).
We used 
standard cooling
functions \citep{mckee77, mccray87} and assumed 
thermal emission with
a temperature kT of 0.3 keV.
We used the L$_{\rm X}$ of the central
source for these calculations
rather than the large aperture luminosity.
We emphasize that these 
estimates of n$_{\rm e}$$\sqrt{f}$
are very uncertain, because of 
uncertainties in the measured
angular extent 
of the X-ray emission which is a function of sensitivity,
as well as  
lack of information about 
the line-of-sight path length through the ionized
region and the three-dimensional geometry of the regions.

\subsection{L$_{\rm X}$/L$_{H\alpha}$ vs.\ Other Parameters}

In Figures 8, 9, and 10, we plot 
the extinction-corrected L$_{\rm X}$/L$_{H\alpha}$ ratios against
the H$\alpha$ equivalent widths, the hydrogen column densities
(determined from the H$\alpha$/24 $\mu$m ratios), 
and 
n$_{\rm e}$$\sqrt{f}$.
We calculated H$\alpha$ equivalent widths for the Antennae
regions using the \citet{zhang10} photometry, 
interpolating between the HST F555W and F814W fluxes for the red continuum.
For the NGC 2403 regions, we used fluxes from
the van Zee et al. (2013, in preparation) H$\alpha$ and off-H$\alpha$
images to calculate equivalent widths.
These agree reasonably well with 
spectroscopic H$\alpha$ equivalent widths determined for a few of our
NGC 2403 regions by \citet{berg13}.

Figures 8, 9, and 10
show clear differences between
the NGC 2403 regions and the other regions,
while the values for
the Antennae and hinge clumps overlap.  
The NGC 2403 equivalent widths are generally larger than those for the Antennae
regions and the hinge clumps.  
In the hinge clumps and in the Antennae regions, 
the ionized gas may be more obscured relative
to the overall starlight, producing smaller observed equivalent widths.
Lower metallicity in some of the NGC 2403 regions may also contribute
to higher H$\alpha$ equivalent widths.
In addition,
the 
NGC 2403 apertures cover smaller physical areas on the galaxy, closer
to the central star formation;
more older stars 
may be included in the larger Antennae and hinge clump apertures.  
Since both the hinge clumps and the Antennae
regions likely contain a range of stellar ages (see Appendix),
their H$\alpha$ equivalent widths are a rough proxy for the 
luminosity-weighted mean
stellar age.  

The NGC 2403 regions 
have lower N$_{\rm H}$ and n$_{\rm e}$$\sqrt{f}$ values than
the hinge clumps and Antennae regions.   This is expected, as star forming
regions
in normal
galaxies are expected to be less obscured and less dense on average
than regions
in strongly interacting galaxies.

The NGC 2403 regions also have 
lower extinction-corrected L$_{\rm X}$/L$_{H\alpha}$ ratios
than the other regions.   
This may be due to 
younger ages on average in the NGC 2403 regions,
as suggested by their higher H$\alpha$ equivalent widths.
Very young star forming regions 
are not expected to host significant supernovae activity, while
hot gas may build up with time in
older regions with successive generations of star formation. 
Alternatively, the lower gas number densities and gas column densities 
towards the NGC 2403
regions 
may be responsible for their lower relative X-ray emission.
The higher the density of the region, the more X-ray production
is expected due to a higher particle collision rate.
This topic is discussed further in the next section.

\subsection{The X-Ray Production Efficiencies}

To further investigate these issues, in this section we compare 
the large-aperture X-ray luminosities with 
Starburst99 predictions 
of the rate of mechanical
energy injection into the region
from hot star winds (ignoring red giant stars)
and supernova.   The ratio of the absorption-corrected X-ray
luminosity to the mechanical luminosity is defined as 
the X-ray production
efficiency (XPE; 
the fraction of the mechanical luminosity
emitted in X-rays).
To calculate XPEs 
it is necessary to have estimates of the extinction and the average
stellar age.
For this calculation,
we use the ages and reddenings derived from the H$\alpha$ data
(Table 12), as these are likely
more relevant for the 
X-ray-emitting hot gas than values obtained from the broadband
photometry (Table 11).
Using these values and SB99, we derive mechanical
energy rates and thus XPEs for the clumps assuming
that all of the X-ray light comes from hot gas and not HMXBs.
These XPEs are tabulated in Table 12.
Most of the hinge clumps 
with extended X-ray emission have similar inferred XPEs of 
$\sim$0.6\%,
while Arp 240-1 has a lower value of $\sim$0.2\%.

We emphasize that 
these XPEs are uncertain for several
reasons. First, there 
is likely a range in both age and attenuation
in these clumps, making our estimates of both the
mechanical luminosity and the extinction uncertain.
Second, the cooling time for the X-ray emitting gas 
can be quite long, while
our determination
of the XPE uses an instantaneous estimate
of the mechanical luminosity.  Ideally,
in calculating XPE one should average 
the rate of mechanical energy injection 
over an extended time period, taking into account
the variation of the star formation rate with time
along with gas cooling rates.
Given these caveats, it is unclear how significant the 
variations in XPEs from one hinge clump to another
are. 

However, even taking these caveats into account, 
two of the X-ray point sources, Arp 270-1 and Arp 240-4,
have very high derived values of 
XPE (8.0\% and 3.2\%, respectively).  This 
again suggests that these two
regions host ULXs rather than hot gas,
in spite of the possible thermal X-ray spectrum for Arp 270-1.
The third X-ray point source, Arp 240-3, has an XPE similar to that
of the clumps with diffuse X-ray emission, suggesting 
that its X-ray emission is dominated by light from hot gas.

For many of the clumps,
the older ages obtained from the broadband photometry (Table 11)
are inconsistent with the X-ray fluxes, if the X-ray emission
is only due to hot gas, the burst is instantaneous, and the
hot gas cools quickly.
A dramatic drop-off in the mechanical luminosity
is expected for ages older than
40 Myrs, when supernovae Type II cease.  
However, as noted earlier, the cooling times for the hot
gas can be quite long, and hot gas can build up in the
regions over an extended period, thus our instantaneous
estimates of XPEs may not be very accurate for these regions.

We derived  
XPEs for the Antennae and NGC 2403 regions in the same way as we did the hinge clumps.
For the Antennae regions
with $\ge$3$\sigma$ detections
in the X-ray, we find XPEs ranging from 0.16\% to 0.91\%,
with a median of 0.38\%, after eliminating the regions with bright X-ray point sources.
These values are similar to the XPEs found for the hinge clumps.
For the NGC 2403 regions,
we find a lower median XPE of 0.16\%.
In all cases, we find a large range in the derived XPEs, perhaps reflected
the uncertainties in these estimates.

In Figures 11 though 14, 
we plot XPE against H$\alpha$ equivalent width,
Spitzer [3.6] $-$ [24] color, N$_{\rm H}$, and n$_{\rm e}$$\sqrt{f}$.
The XPE for the NGC 2207 clump is similiar to or higher than that of
the other hinge clumps and regions within the Antennae, in spite
of its low observed 
L$_{\rm X}$/L$_{24}$ ratio (see Figure 4).
This suggests that 
its low observed 
L$_{\rm X}$/L$_{24}$ ratio 
is primarily caused by high absorption,
rather than by extremely young age.  It has the highest inferred N$_{\rm H}$
of the hinge clumps (see Table 12) thus the largest
correction for extinction, 
but only a moderate H$\alpha$ equivalent width.

Antennae region 3  has an XPE similar to that of NGC 2207-1 but a lower 
inferred
column density and a 
higher H$\alpha$ equivalent width.   Thus it appears younger but
less obscured.
Antennae region 4 has a low inferred XPE and a high implied
column density.  
Region 4 in the Antennae does not have an extremely high
H$\alpha$ equivalent width (Figure 11), in spite of the 
very young estimated age of 1 Myrs for the most luminous cluster WS80 
in this region
\citep{whitmore10}.
This suggests that region 4 contains a wide range of stellar ages
and extinctions.
The extinction of the H$\alpha$
line towards the youngest stars may be significantly 
higher than that of the observed
red stellar continuum, which would artificially
bias the observed
H$\alpha$ equivalent width towards older ages.

The three hinge clumps that are undetected by 
Chandra (Arp 270-2, Arp 270-3, and Arp 270-4) have
XPE upper limits consistent with the X-ray detections of the other hinge 
clumps and the detected Antennae regions (Figures 11 $-$ 13). 
This suggests that their lack of X-ray emission 
is simply due to low SFRs.
They have the lowest SFRs in our
sample (Table 10) with the exception of Arp 270-1, which has
a candidate ULX.
Arp 270 is the closest system in our sample, thus our 
selection criteria reaches lower luminosities for clumps
in Arp 270 than for the other galaxies.

A weak correlation between XPE and 
H$\alpha$
equivalent width is visible in Figure 11, while
the XPEs do not correlated with the [3.6] $-$ [24] 
colors of these clumps (Figure 12).
Theoretical studies
suggest that the X-ray production efficiency in star forming 
regions should increase with time
\citep{silich05, anorve09, hopkins12}.
The large scatter in this plot may be because there is likely 
a range of
stellar ages and/or extinctions within a clump, thus
our simple estimate of XPE using a single age and single extinction
is quite uncertain.  In addition, there may be 
contributions from HMXBs to the extended X-ray emission
in these regions, which are not taken into account
in our calculation of the XPE.
Another factor is the timescale for cooling of the hot gas;
hot gas can build up in the region over an extended period,
thus the rate of mechanical energy injection into
the region from the star formation
should be averaged over an extended time period.

We also see a weak correlation between XPE and hydrogen column
density N$_{\rm H}$ (Figure 13),
and between XPE and 
n$_{\rm e}$$\sqrt{f}$ (Figure 14),
when excluding the sources that are likely ULXs. 
This is consistent with
theoretical expectations, which 
suggest that XPE should increase
with increasing gas density
(Silich et al.\ 2005; Anorve-Zeferino et al.\ 2009;
Hopkins et al.\ 2012).
Lower density gas 
allows 
stellar winds to escape more freely, while
higher density gas would incur more collisions
thus producing more X-ray emission.
The large scatter in XPE with 
n$_{\rm e}$$\sqrt{f}$ (Figure 14)
may be 
due in part to the large uncertainties on both of these 
quantities.  
In addition to
uncertainties in the derived stellar 
ages and the possibility that
there is a range of 
ages within the clumps, 
other contributing factors may be 
uncertainties in the measured
angular extent 
of the X-ray emission which is a function of sensitivity,
as well as  
uncertainties in  
the line-of-sight path length through the ionized
region.

\section{Summary and Discussion}

We have investigated the properties of a visually-selected sample of
 12 bright hinge clumps in five interacting galaxy systems.
We limited our sample to those with extensive multi-wavelength archival data available
 including GALEX UV, Spitzer IR, 2MASS, H$\alpha$, and SDSS imaging.
Most critically, we have also examined high-resolution HST images of the
 hinge clumps to investigate their morphology in detail and Chandra X-ray
 data to study their hot gas and luminous X-ray binary populations.
Comparisons have been made with star-forming regions in the
Antennae galaxies and in the normal spiral NGC~2403.
Some of these hinge clumps are forming stars at prodigous rates,
between 1 $-$ 9 M$_{\sun}$~yr$^{-1}$, higher than the global values for
many normal spiral galaxies, while others are more quiescent.

We find remarkably large ($\sim$70 pc) and luminous (M$_{\rm I}$
$\sim$ $-$12.2 to $-$16.5) UV/optical sources at the centers of these 
clumps.
These sources are sometimes embedded in long arcs or linear ridges containing fainter
star clusters, suggestive of star formation along a narrow caustic.
The sizes of these central sources are much larger than typical super star clusters,
and their luminosities are near the high luminosity end of the super star cluster
luminosity function.   These results suggest that they are
likely composed of close concentrations of multiple 
star clusters,  rather than individual clusters.

Comparison to stellar population synthesis models 
suggests that
the hinge clumps contain a range of stellar ages.
This is consistent
with expectations based on 
models of interacting galaxies which predict prolonged
inflow of gas into the hinge region, producing sustained
star formation or multiple bursts rather than a single instantaneous burst.
The central sources seen in the HST
images are bluer on average than the clump as a whole,
again suggesting a range of stellar ages within the clump.
A jewel-in-the-crown mode of star formation may 
operate within hinge clumps, in which luminous young
star clusters form within a `crown' of older
stars, dispersed from earlier star formation episodes.

These results all indicate that hinge clumps, as a population, are relatively 
long-lived structures.
How long hinge clumps persist and form stars is an unanswered question.
Our selection criteria 
for this study
selected targets that are bright in either UV or 24~$\mu$m light,
which restricts our sample clumps to be younger than $\sim$100~Myr,
 the lifetime of the UV-producing stellar population.
However, our morphological definition of hinge clumps, that they are 
discrete knots of
recent star formation in the 
inner half of a tidal feature, can include 
fainter
                                                                               and presumably older structures.  

Analytical and numerical models (Struck \& Smith 2012) suggest
 hinge clumps arise in regions of global compression formed by multiple 
converging
 density waves that produce ocular waves and caustics in the outer 
disks of interacting galaxies.
These structures may be long lived but are more subject to shear in the 
relatively
 flat rotation curve environment of outer disks compared to nuclear starbursts
 that reside near the bottom of the galaxy potential.
The models suggest much more intense interaction between
azimuthal and radial caustic waves in the outer disk, near the base of
the tails, than elsewhere in the system in the pre-merger stage of a 
tail-producing encounter.  

Determining how long hinge clumps may last and their ultimate 
fate depends on the
 specific details of the galaxy interaction that produced them.
While the dynamical explanation for hinge clump formation appears robust,
 current simulations lack sufficient resolution and the necessary 
microscale physics
 needed to make reliable predictions of their ultimate fate.
However, if the galaxy interaction is a simple flyby, then it is likely that the
larger
 hinge clumps will eventually spiral into the nucleus through dynamical friction
while the smaller
 ones may disperse through shear in a few rotation periods.
If the galaxy interaction results in a merger, the star clusters
within hinge clumps may be 
dispersed intact as
 self-gravitating bodies and may eventually become globular clusters.
It is uncertain if and how such globular clusters could be differentiated from the
 ordinary populations of old globular clusters.

Several of the hinge clumps studied in this work are prodigous X-ray emitters
(L$_{\rm X}$ $\sim$ 10$^{40}$ $-$ 10$^{41}$ erg~s$^{-1}$).
In most cases, this emission appears extended and is best interpreted as
 originating from a hot, $\sim$0.3~keV, thermal plasma.
Hot gas is expected in regions of intense star formation.
Where this gas is confined by dense surrounding cooler interstellar medium,
the hot gas cooling times are short and the energy input into the hot gas is
quickly radiated away giving rise to a high X-ray luminosity.
In lower-density regions, in contrast, hot gas bubbles 
do $P\,dV$ work on the
 surroundings giving rise to galactic winds but with little radiative losses
\citep{hopkins12}.
The converging flow of gas and stellar orbits that produce hinge clumps may be
 sufficient to confine the hot gas produced by supernovae and stellar winds 
and thus
produce the high X-ray luminosities 
and high XPEs
observed in several of our sample hinge clumps,
similar to those of star forming
regions within the merging Antennae galaxies.  In contrast,
the star forming regions in the normal spiral galaxy
NGC 2403 have lower XPEs, consistent with their lower inferred 
electron number densities and hydrogen column densities. 

In three of the hinge clumps, the X-ray emission is either 
point-like or point-like with an underlying extended
 soft component (Arp~240-4).
In one case (Arp~240-3), the X-ray spectrum and luminosity 
indicate a compact star forming region; the other two sources are likely ULXs.
Arp~240-4 is the most X-ray luminous hinge clump in the sample,
with a hard X-ray spectrum
and an extreme X-ray luminosity of $\sim$2 $\times$ 10$^{41}$ erg~s$^{-1}$.
This luminosity is difficult to explain by a single HMXB; it may
be a collection of HMXBs or alternatively,  
an intermediate-mass black hole. 
If this source is indeed an intermediate-mass black hole, its mass is
1600 $-$ 16,000 M$_{\sun}$, assuming it is radiating at 10\% $-$ 100\% 
of the Eddington luminosity.  The ratio
of the mass of this black hole to the stellar mass of this clump 
(Tables 11 and 12) would then be 1.5 $\times$ 10$^{-4}$
to 7.3 $\times$ 10$^{-6}$.
Massive clumps and clusters hosting 
intermediate mass black holes 
may eventually migrate
into the center of the galaxy, where the
black holes may merge into supermassive central black 
hole (e.g., \citealp{ebisuzaki01, elmegreen08}).

\acknowledgements

This research was supported by NASA Astrophysics 
Data Analysis Grant ADAP10-0005 and National Science Foundation
Extragalactic Astronomy grant AST-1311935.
We thank the anonymous referee for helpful suggestions.
We also thank Michele Kaufman, Debra Elmegreen, Hongxin Zhang, 
Yu Gao,
Howard Bushouse, Javier Zaragoza-Cardiel,
Marcel Clemens,
and Liese van Zee 
for providing copies of their data and helpful communications.
We thank Mark Hancock for the use of his scripts to determine
population ages.
This research has made use of the NASA/IPAC Extragalatic Database (NED), which is operated by the Jet Propulsion Laboratory, California Institute of Technology, 
under contract with NASA.
This work is based in part on observations made
with the Spitzer Space Telescope, which is operated by
the Jet Propulsion Laboratory (JPL), California Institute
of Technology under a contract with NASA.
This research is also based in part on
observations made with the NASA/ESA
Hubble Space Telescope, and obtained from the Hubble
Legacy Archive, which is a collaboration between the Space
Telescope Science Institute (STSci/NASA), the Space
Telescope European Coordinating Facility (ST-ECF/ESA) and 
the Canadian Astronomy Data Centre (CADC/NRC/CSA).
This study also uses data from the NASA Galaxy
Evolution Explorer (GALEX), which was operated for NASA
by the California Institute of Technology under
NASA contract NAS5-98034.
This research has also made use of data obtained from
the Chandra Data Archive and software provided by
the Chandra X-ray Center (CXC).
This publication makes use of data products from the Two Micron All Sky Survey, which is a joint project of the University of Massachusetts and the Infrared Processing and Analysis Center/California Institute of Technology, funded by the National Aeronautics and Space Administration and the National Science Foundation.
This research has made use of the NASA/IPAC Infrared Science
Archive, which is operated by JPL, the California
Institute of Technology, under contract with the
National Aeronautics and Space Administration.

\appendix{\bf Appendix I: Morphologies of Individual Galaxies}

\section{Arp 82 (NGC 2535/6)}

Arp 82 is an unequal mass pair with a long tail extending to the north
(Figure 15 left).  
The more massive galaxy in the north
NGC 2535 has an ocular structure (\citealp{elmegreen91, kaufman97, 
hancock07}).
In an earlier study \citep{hancock07}, we conducted
a detailed analysis of the 
GALEX/Spitzer/H$\alpha$ photometry of numerous clumps of
star formation in Arp 82.  
Luminous knots of star formation are present along the eyelids
and points of the ocular \citep{hancock07}.
In the current study, we target 
the most luminous knot of star formation in the tidal features,
which lies near the base
of the northern tail
(marked in Figure 15). 
We classify this region as a hinge clump. 
This knot shows up as a discrete source in the 20 cm radio continuum
image and the 21 cm HI
map of \citet{kaufman97}.

In Figure 16, we present a close-up view of the HST
F606W image of
the hinge clump. 
In the center of this clump, we find a very luminous
source with 
absolute I
magnitude
M$_{\rm I}$ $\sim$ $-$13.3.  This source
is unresolved with HST (diameter $\le$ 70 pc).
This object lies near the center of
a straight line of lower luminosity star clusters that extends 15$''$
(4 kpc) in length.
This linear structure suggests star formation along a caustic, perhaps
caused by local gravitational instability
and competitive cluster growth in a linear pile-up zone.

Arp 82 has not been observed with Chandra.

We presented a numerical simulation of the Arp 82 interaction
in \citet{hancock07}.
The model indicates that the long tail was produced
in a prograde planar encounter, with the companion in an elliptical orbit
around the primary.  
In the right panel of Figure 17, we display
an analytical model 
of a prolonged prograde interaction (approximated
by three tidal impulses) 
from \citet{struck12} (Figure 12 in that paper).
We compare with 
a close-up view of the \citet{arp66} Atlas
photograph of the hinge region of Arp 82 (left panel).
Note the resemblance between the model and the galaxy.
The inner disk of the model galaxy shows an ocular structure
similar to that in NGC 2535.  
A larger oval-shaped structure
is visible in the model outside of the inner ocular.
This structure is produced when two caustics diverge and 
one arcs back to the main galaxy.    Arp 82 shows a similiar morphology
in the northeast, 
where a spiral arm loops back towards
the main galaxy.   
Further along the model tail to the
north, two caustics converge, causing a narrowing of the tail.
In the Arp picture, 
a similar narrowing of the tail occurs at the location of the hinge clump,
and a faint arc is visible extending to the west.
This suggests 
that the star formation in the hinge region was triggered
by intersecting caustics.

\section{Arp 240 (NGC 5257/8)}

Figure 18 compares the HST F435W image of 
Arp 240 with the Spitzer 8 $\mu$m image.  
Arp 240 is a pair of disk galaxies with similar masses, with a connecting
bridge and two short tails extending from the two galaxies.
Both galaxies in the pair 
are independently
classified as `Luminous Infrared Galaxies' (LIRGs), with far-infrared
luminosities of 2 $\times$ 10$^{11}$ L$_{\sun}$ \citep{howell10}.
Arp 240 has a total of five clumps that we define as hinge clumps, one in
the southeastern galaxy of the pair (NGC 5258), and 
four in the northwestern galaxy (NGC 5257).
These are marked in Figure 18.
X-ray emission is detected from all five of these clumps.

Arp 240W has an inner spiral,
along with two
clumpy ridges of star formation to the west and east of this spiral,
at the base of the tails.
In the Spitzer 8 $\mu$m image of Arp 240W, three 
knots of star formation are seen along the base of the southern tail,
and a fourth at the base of the northern tail.
These four regions are all bright in the CO (3$-$2) map of \citet{wilson08}
as well as in the \citet{bushouse87} H$\alpha$ map.

A closer view by HST 
of Arp 240W 
(Figure 19 left) 
shows that each of the four hinge clumps
in this galaxy
contains a central luminous source.
The absolute B magnitudes
M$_{\rm B}$ of these sources range from
$-$14.4 for Arp 240-4 to $-$15.1 for Arp 240-1,
while M$_{\rm I}$ ranges from 
$-$14.4 for Arp 240-4 to $-$15.6 for Arp 240-2.
These clusters are resolved in the HST images, 
with FWHM between $\sim$40 pc for Arp 240-2 to $\sim$100 pc
for Arp 240-3.
In the HST images, the brightest optical source in clump 2
is about 2$''$ off from the 8 $\mu$m
peak, but well within the 5$''$ radius used for the Spitzer photometry.
For clump 3, the brightest optical source 
is about 1$''$ from the 8 $\mu$m peak.
For clump 1, the source 
that is the brightest in the HST UV and optical images is
not the brightest in the near-infrared F160W image.
Both of these
sources are about 3$''$ from the 8 $\mu$m peak.

In Figure 19 (right), 
we overlay a smoothed version of the Chandra 
0.3 $-$ 8 keV map 
on the HST F814W image of Arp 240W.
Diffuse X-ray emission is present along the 
base of the tidal features; this extended emission
is predominantly soft X-rays.
Clump 4,
at the base of
the northern tail, has a bright compact source with 
a hard X-ray spectrum.
Clump 3 is also compact in the X-ray, however, its
spectrum is intrinsically soft (see Section 3.3).

In Figure 20, we compare
the HST F814W image of Arp 240W
with an analytical model from
\citet{struck12} 
(Figure 10 from that paper, inclined to match 
the observed inclination of Arp 240W).
In the model, two embedded oculars
are seen.
In Arp 240W, the inner spiral
is not completely closed into an ocular,
unlike in the model.  
In a ridge along the base of the southern tail lies 
three regions we
classify as hinge clumps (see Figure 19).
The extreme star formation
in this region may be due to overlapping caustics,
as in the model.
A similar overlap of caustics may be responsible
for clump 4 near the base of the northern tail.

Arp 240E has a bright mid-infrared source at the base of
the southern tail (Figure 18).
This region is the brightest 
CO (3$-$2) source in the galaxy \citep{wilson08}.
In the HST images (Figure 21), 
this clump has an extremely luminous
optical source 
near its center, with M$_{\rm B}$ $\sim$ $-$15.8
and M$_{\rm I}$ $\sim$ $-$16.5.
This source is unresolved in the HST images (FWHM $\le$ 75 pc).
It lies along a distorted arm-like structure, 
and is flanked by two fainter sources $\sim$170 pc (0\farcs35)
away along the arm.
The X-ray map shows strong extended X-ray emission from
this region (Figure 21 right).
Because of the high inclination of this galaxy and its very
distorted structure, unlike Arp 240W we are not able to match it
to one of our analytical models.
However, the location and luminosity of the star formation
region 
suggests it may be a hinge clump.

\section{Arp 256}

Arp 256 is a widely separated pair of spirals (Figure 22).  
The southern galaxy
has a far-infrared luminosity of 2.8 $\times$ 10$^{11}$ L$_{\sun}$,
thus it is classified as a
LIRG, while the
northern galaxy has L$_{FIR}$ = 2.3 $\times$ 10$^{10}$ L$_{\sun}$
\citep{howell10}.
According to an archival 70 $\mu$m Herschel image,
most of the far-infrared
luminosity from the southern galaxy comes from the inner disk, while
the far-infrared light from the northern 
galaxy is dominated by 
a luminous knot of star formation at the base of its long northern tail.
We classify this knot as a hinge clump.

In Figure 22, we compare the HST F435W image of Arp 256
with the Spitzer 8 $\mu$m image.
In addition to the hinge clump in the northern tail, the
northern galaxy in Arp 256
has a tidal dwarf galaxy near the end of its southern tail.
A 21 cm HI map of this system has been presented by \citet{chen02},
which shows that these tails are gas-rich.
The hinge clump is very bright in the 20 cm radio continuum 
\citep{chen02}.

In Figure 23,
we display a close-up view of the HST F814W image of
the hinge clump. 
This region is resolved in the HST
red and near-infrared
images into a flattened
structure with a FWHM $\sim$ 70 pc, much larger than the typical
size of the super-star clusters found in interacting
galaxies ($\sim$3 pc; Larsen 2004).
This source
is extremely luminous (M$_{\rm B}$ $\sim$ $-$14.6 and
M$_{\rm I}$ $\sim$ $-$14.8).
In the HST UV image, this source resolves into two
peaks separated by $\sim$0\farcs15 (80 pc)
north/south; this double structure may
be due to 
dust absorption; alternatively, the slightly higher spatial resolution
in the UV may be resolving a pair of clusters.
As in Arp 82 and 240, this
source is embedded in a linear ridge of star clusters.
This ridge lies
along the leading edge of the tidal tail.
The X-ray emission from this region (Figure 21, right
panel) is extended, with two possible peaks.

As with Arp 82, the long northern tail of Arp 256N signals a 
prograde planar encounter.   In an earlier study, we produced
a numerical simulation of the similar system Arp 305 \citep{hancock09}.
This same simulation can be applied to Arp 256, but at an earlier
timestep (see the second panel in Figure 4 in \citealp{hancock09}).
A closer look at the northern tail of Arp 256N shows a double structure
(Figure 24).
We find an approximate match to
this tidal morphology using the same 
analytical
model as for Arp 82, but at an earlier timestep.
This is illustrated in Figure 25, where we compare the HST images
of Arp 256N with this analytical model (3rd panel from
Figure 12 in \citealp{struck12}).
Note the two approximately parallel caustics along the model tail.
In the Arp 256 tail, the hinge clump lies near where the two main filaments
in the tail intersect (see Figure 25). 

In the HST images (Figure 25), two UV-bright knots of star formation 
are visible out of the main disk of the galaxy,
above and below the bulge of the galaxies.   The numerical
model suggests that that
star formation was triggered by material pulled out from the galaxy
by the interaction, which is now falling back in on the disk.
These star forming regions produce the characteristic `X' shape seen in the 
inner region of Arp 256 in the \citet{arp66} Atlas photograph (Figure 24).

\section{Arp 270 (NGC 3395/6)}

The GALEX NUV and Spitzer 8 $\mu$m images
of Arp 270 are displayed in Figure 26.
Arp 270 is a close pair of equal-mass spiral galaxies with a 10$'$ (79 kpc)
long
HI tail stretching from the eastern edge of the pair 
and extending to the south and west
\citep{clemens99}.
We include four tidal star forming regions in Arp 270
in our hinge clump sample.
All four of these are associated
with the western galaxy NGC 3395.
The first three Arp 270 clumps have the lowest H$\alpha$ luminosities
($<$10$^{40}$ erg/s) of the hinge clumps in our sample
(see Table 6).
Only one of the four hinge clumps
in Arp 270 was
detected in the X-ray, clump 1 (Figure 27 right).
This 
source was included in our earlier survey of X-ray point
sources in Arp galaxies \citep{smith12}.
As noted in Section 3.3, this source has a soft X-ray spectrum.
This source is coincident with the Spitzer
mid-infrared position, which is a bit offset from
the UV peak.  
A number of other X-ray point sources are
detected in Arp 270, but none of these are associated
with the other hinge clumps.
Diffuse X-ray emission is seen in the inner region of
the galaxy, but not in the hinge clumps.

Based on a numerical simulation, \citet{clemens99}
conclude that the Arp 270 encounter is prograde with respect to
the western galaxy NGC 3395 and retrograde with
respect to the edge-on galaxy in the east NGC 3396,
and the two galaxies are on their second approach.
They conclude that the long HI tail originated from the 
western galaxy NGC 3395, and its apparent connection
to NGC 3396 is a projection effect.
In this model, the tail originates from the western side of NGC 3395
and arcs around behind NGC 3396 \citep{clemens98}.  
In optical images (Figure 27), NGC 3395 has two tail-like structures
extending to the west: the outer tail containing our target clumps,
and an inner spiral.   
Whether the long HI tail connects to the outer tail
or the inner spiral
is uncertain.    

None of the analytical models in \citet{struck12} resemble
NGC 3395 in any detail, perhaps because of its
warped and distorted structure.   
The closeness of the two galaxies in this pair argues
that this system is in a later stage of evolution 
than the other systems in our sample
and closer to merger, thus it is harder to model analytically.
We include the four marked regions 
in our sample as possible hinge clumps due to their apparent location
near the base of tidal features and the availability
of high sensitivity Chandra data, however,
whether or not their formation was triggered by intersecting caustics
is uncertain.  
The only HST images of Arp 270 do not cover
the hinge clumps \citep{hancock03}, thus we do not have high resolution
optical images available to study the morphology of these regions
in more detail. 

\section{NGC 2207}

IC 2163 and NGC 2207 are two spiral galaxies with similar
masses in a very close interacting pair.
In Figure 28, we display the GALEX NUV and Spitzer 8 $\mu$m
images of this pair.
The Spitzer images of this system 
were previously presented by \citet{elmegreen06}, who noted
an extremely mid-infrared-bright knot of star formation
on the western edge of NGC 2207.
This source, which they call `feature i', is our 
hinge clump candidate, and lies in a distorted
spiral arm at the base of a short tail-like feature visible in
optical images.
Feature i lies near a 
massive concentration of HI gas ($\sim$10$^{9}$
M$_{\sun}$); a second similar 
concentration (8 $\times$ 10$^{8}$ M$_{\sun}$)
is seen 1$'$ (11 kpc)
to the north of knot i
\citep{elmegreen95}.
This second HI cloud does not have an optical or infrared counterpart,
and may be a gas-rich extension of the tail-like feature containing
feature i.
The companion galaxy IC 2163 to the east
has an ocular structure \citep{elmegreen91,
elmegreen06}, with an HI tail extending to the east \citep{elmegreen95}.
Strong star formation is seen in the `eyelids' and `points' of this
ocular \citep{elmegreen95, elmegreen06}.

The Spitzer mid-infrared peak for feature i
is offset slightly from the GALEX NUV peak, likely
because of obscuration.
Feature i is the brightest radio continuum source
in the galaxy \citep{vila90, kaufman12}.
The 6 cm/20 cm spectral index is consistent with non-thermal emission
in the radio \citep{vila90}.   \citet{kaufman12} noticed a possible
increase in 6 cm flux between 1986 and 2001, and suggested a possible
radio supernova. 

The HST images of NGC 2207 were previously analyzed
by \citet{elmegreen01},
before the Spitzer images were obtained.  
In that study, they only considered the optically brightest clusters,
which did not include the cluster in the center of 
the mid-infrared object feature i.
The HST morphology of the NGC 2207 hinge clump (Figure 29) 
differs somewhat
from that of the other hinge clumps in our sample.
The dominant source
in the NGC 2207 hinge clump is at the center
of a more extended grouping of fainter clusters, rather than in a
straight line of clusters.
However, this group of clusters is part of an extended
ridge of star formation along a spiral arm/short tail-like structure,
thus it may also have been produced by star formation along a caustic,
albeit a more distorted caustic.

The source at the center of 
this clump is 
extended in the HST images, with FWHM $\sim$ 70 pc.  
Its observed
absolute I magnitude is only about $-$12.2, 
however, its intrinsic
luminosity might be much higher if it is strongly obscured.
As noted by \citet{kaufman12}, there is a dust lane visible in
optical images that runs
across feature i.
The extinction to this clump is discussed further in Section 7.3.

The unsmoothed X-ray map of the NGC 2207 hinge clump is shown in 
Figure 29 (right).
This source is quite extended with Chandra but with a low
surface brightness, and has a soft X-ray spectrum.
The Chandra data for this source was previously analyzed by 
\citet{mineo13}, who 
also concluded that the source is 
soft and extended with Chandra, and found an observed flux
similar to ours.

We classify 
feature i as a possible 
hinge clump based on its location near the base of a short tail-like 
structure 
and its strong mid-infrared
emission.  However, this classification is uncertain because of the shortness
of the tail and the peculiar morphology, which differs from that of the other
systems in our sample. 
Feature i may instead be a tidally-disturbed disk region
rather than a true tidal structure.  

The NGC 2207/IC 2163 encounter has 
been modeled numerically by \citet{struck05},
who conclude that the collision was prograde with respect
to IC 2163, and retrograde for NGC 2207.  
In contrast, the other hinge clumps in our sample all lie
in longer tidal features produced by prograde encounters.
However,
the basic mechanism of star formation triggering due to intersecting
caustics may still hold for feature i.
Models suggest that the eastern HI tail of IC 2163 may eventually
develop hinge clumps, once sufficient time has elapsed for 
strong waves to develop and intersect in that region.

\vspace{0.3in}

\appendix{\bf Appendix II: Comparison to the Antennae}

\section{Overview}

To put our hinge clumps in perspective, 
we compare them to the well-studied `overlap'
region of intense star formation
between the two disks of the Antennae galaxies.
Like the hinge clumps, this region is not in a galactic nucleus,
thus it is easier to study and less confused than nuclear starbursts.
At a distance of only 24.1 Mpc,
the Antennae provides much better spatial resolution than the hinge clump
galaxies, thus it provides a good comparison system.
The star clusters in the Antennae have been investigated in a number
of studies, including a detailed population synthesis study by
\citet{whitmore10}.    

In Figure 30, we provide 
a multi-wavelength mosaic of the Antennae overlap region,
including the co-added Chandra 0.3 $-$ 0.8 keV map (upper left panel),
an archival HST ACS F814W (I band) image from the Hubble
Legacy Archive (upper right panel),
an archival HST WFC3 F160W image (near-infrared H band) (middle left
panel), and the Spitzer 8 $\mu$m map (middle right panel).  
We also compare to a 
230 GHz millimeter continuum map and a CO(2$-$1) map, respectively, 
from the 
Atacama Large Millimeter/submillimeter Array (ALMA) telescope,
previously presented by \citet{espada12} and 
obtained from
the ALMA science 
portal\footnote{http://almascience.nao.ac.jp/almadata/sciver/AntennaeBand6}
(lower left and low right panels in Figure 30).
On these images we superimpose circles marking two of the 4\farcs5
radius regions studied by \citet{zhang10}:
their
region 3 (right) and region 4 (left).
Region 4 lies near the peak of the 8 $\mu$m and 24 $\mu$m
emission, and has been called the mid-infrared `hot spot'.  It has the
reddest [3.6] $-$ [24] color of all the Antennae regions studied by
\citet{zhang10} (see Section 5).
Region 3 is also quite bright in the mid-infrared, and is the second
reddest region in [3.6] $-$ [24] in their survey.
These red colors imply intense and obscured star formation in
these regions.

\section{Region 4}

From the L$_{24}$/L$_{H\alpha}$ ratio we determine
an extinction of
A$_{\rm V}$ = 3.23 towards region 4.
Probably because we are averaging
over the entire region, this 
value is significantly less than that found 
by \citet{whitmore10}
towards the
brightest cluster 
visible in Region 4 in the HST F814W image,
for which they estimate
E(B $-$ V)
= 2.44.  
This cluster 
is also extremely massive, 8.2 $\times$ 10$^6$ M$_{\sun}$,
and has an extinction-corrected M$_{\rm V}$ = $-$15.4
\citep{whitmore10}.
This object is sometimes called `WS80', being object
number 80 in the \citet{whitmore95} study.
\citet{whitmore10} derive an age of only 1 Myrs for this cluster.
\citet{whitmore02} associate WS80 with the brightest
6 cm radio continuum source in the Antennae, as seen
in the \citet{neff00} maps.  The radio spectrum of this source
indicates that this emission is predominantly thermal.
The WS80 cluster 
is associated with the brightest 
CO(1-0) emission in the Antennae \citep{wilson00}.
As seen in the ALMA maps in Figure 30, this source is also very bright
in CO(2-1) and in the 230 GHz continuum.

In the HST images, WS80 is surrounded by fainter emission, probably
from nearby lower luminosity clusters, but no other clusters with
similar luminosities.
In subarcsecond resolution mid-infrared images, \citet{snijders06}
detect a possible second mid-infrared source 0\farcs54 to the northeast
of WS80, which is not seen in optical or near-infrared images.
They suggest that this may be a highly obscured cluster with A$_{\rm V}$
$\ge$ 72, which implies N$_{\rm H}$ $\ge$ 10$^{23}$ cm$^{-2}$.
In very high resolution 3.6 cm radio continuum maps, a second fainter
source is seen at this location, along with an H$_2$O radio maser 
\citep{brogan10}.

As seen in Figure 30, Region 4 in the Antennae is considerably 
fainter in the X-rays than Region 3.  
Region 4 hosts a moderately bright X-ray point source which lies $\sim$0\farcs5
southwest of WS80.  This point source is embedded in extended X-ray emission
that extends to the north towards WS80.   
From the archival Chandra data, 
we extracted the X-ray spectrum for both region 4 as a whole (using
a 4\farcs5 radius
aperture),
and for the point source (with a 1\farcs4 radius aperture).
Region 4 as a whole shows 
a composite X-ray spectrum, with both a power law and a thermal component
with kT = 0.68 $\pm$ 0.13 keV.
More than half of the power law flux arises from the point source.
The point source has a pure power-law spectrum with L$_X$ $\sim$ 3 $\times$
10$^{38}$ erg/s.

If the \citet{whitmore10} age estimate for WS80 of 1 Myrs is accurate,
WS80 itself is too young to host HMXBs or to have
developed supernovae. 
The point source may be a HMXB associated
with a nearby slightly older stellar population, while
the diffuse X-ray emission
in this region
may be powered by supernovae associated with
an earlier generation of stars or hot star winds 
rather than supernovae.   
The spatial offset between
the X-ray, the optical, and the mid-infrared sources in Region 4 suggest
an age sequence from older in the southwest to younger in the northeast.

\section{Region 3}

In contrast to WS80 in region 4, 
the brightest cluster in \citet{zhang10} region 3 is
closely surrounded by other bright clusters.   This complex
of clusters is known as `Knot B' \citep{rubin70, whitmore10}.
\citet{whitmore10} find a range of ages for the clusters
in this complex from 2.5 Myr to 50 Myr, with the youngest ages
found preferentially to the east.  
They suggest that this may be a case of sequential
triggered star formation.  
The clusters in Knot B are 
significantly
less obscured than WS80 in Region 4 but still quite reddened,
with E(B $-$ V) $\sim$ 0.12 $-$ 1.0
\citep{whitmore10}.
The most massive cluster in Knot B is almost as massive as
WS80, with 5.0 $\times$ 10$^6$ M$_{\sun}$.
This region
is much fainter in CO than WS80.

Most of the diffuse X-ray emission from Region 3 is located near Knot B.
In addition, an X-ray point source lies 
approximately 3$''$ to the south of knot B 
near an optical point source.
The extracted X-ray spectrum
of this source appears very absorbed,
implying a very large intrinsic L$_X$.
Its spectrum has 
a peak near 3 keV,
and can be fit with a disk blackbody model with kT $\sim$ 0.75 $-$ 1 keV.

\section{Comparison to Our Hinge Clumps}

The large apparent sizes and the high luminosities
of the objects seen in the centers of the 
hinge clumps by HST may be caused by
the blending of multiple clusters
very close together.
We can make use of the relative proximity of the Antennae
star forming regions to explore some of the results
of this blending.
We investigate what Regions 3 and 4 would look like in HST images
if they were at the distance of our hinge clumps.  
We smoothed
the HST F814W image with a Gaussian to the effective resolution
it would have if it were 
at a distance of 102 Mpc (the distance of Arp 240).  
The clusters in Knot B of Region 3 become blended together into a single
extended object with FWHM $\sim$ 110 pc $\times$ 120 pc.
In Region 4, the WS80 region is only marginally resolved 
in the smoothed image, with FWHM $\sim$ 80 pc.   
This exercise suggests that the extended sources we are 
seeing in
the HST images of the 
hinge clumps could be blended complexes of clusters
that are very closely packed together.   

We note that 
larger than normal clusters have also been found in the tidal features
of the merger remnant NGC 3256 and Stephan's Quintet
by \citet{trancho07, trancho12}.
These authors suggest that the large sizes may be 
a consequence of less tidal stripping of the outer stars in these clusters
or 
weak compression 
at the time of formation.  
It is possible, however, that the NGC 3256 and Stephan's Quintet sources
are 
also multiple clusters blended together, however, at 37 Mpc and 88 Mpc,
respectively, they are not as
distant as some of the hinge clumps in our sample.

At the distance of Arp 82
(53 Mpc),
Antennae
regions 3 and 4 would be included in a single 5$''$ Spitzer and GALEX aperture.
The 
[3.6] $-$ [24] color would still be quite red,
although these regions contain a range of 
stellar ages.
An area on the Antennae equivalent to 
that covered by our 5$''$ aperture at Arp 82 would be observed to have 
a [3.6] $-$ [24] color of 8.7, similar to the
value for region 3 alone.   
If the Antennae were further away, at the distance
of Arp 240, 
the southern nucleus would be included in a 5$''$ beam along with
regions 3 and 4.   Thus the Antennae overlap region is not
a perfect analogy 
to the hinge clumps, as it
is closer to its nucleus than the hinge clumps are to their galactic
nuclei.

In the Chandra map of the Antennae shown in Figure 30,
outside
of region 3 to the southeast is a ridge of diffuse
X-ray emission.  This
ridge continues several arcseconds to the southeast beyond
the field of view of the figure.   This ridge
of 
emission 
was studied by \citet{metz04}, who concluded
that it was likely caused by faint older stars ($\sim$100 Myrs).
In the HST F814W and F160W images, a few faint star clusters
are seen along this ridge.
If the Antennae were at the distance of Arp 240, this ridge
of emission would be included in our 5$''$ apertures.
The X-ray luminosity of this ridge is larger than that of knot B itself,
thus its inclusion in the beam would more than double the observed
X-ray flux.

This comparison to the Antennae 
emphasizes that care must be taken in our interpretation
of the constituents of hinge clumps, especially
those at greater distances.  Within our hinge clumps
there is likely
a range of stellar ages and extinctions, making detailed population 
synthesis difficult. 
As with Antennae regions 3 and 4, the X-ray emission
from the hinge clumps may 
be due to a combination of 
diffuse gas and HMXBs, again making interpretation difficult.
These points underline the cautions we give
in Sections 7 and 8, where we compare
colors and fluxes of the hinge clumps to simple population synthesis
models.

We note that 
linear structures made of star clusters are also seen in the Antennae.
\citet{whitmore10}
suggest that these structures
were produced by either sequential triggered star formation or large-scale
processes such as a density wave or a collision between two large clouds.

\begin{figure}
\plottwo{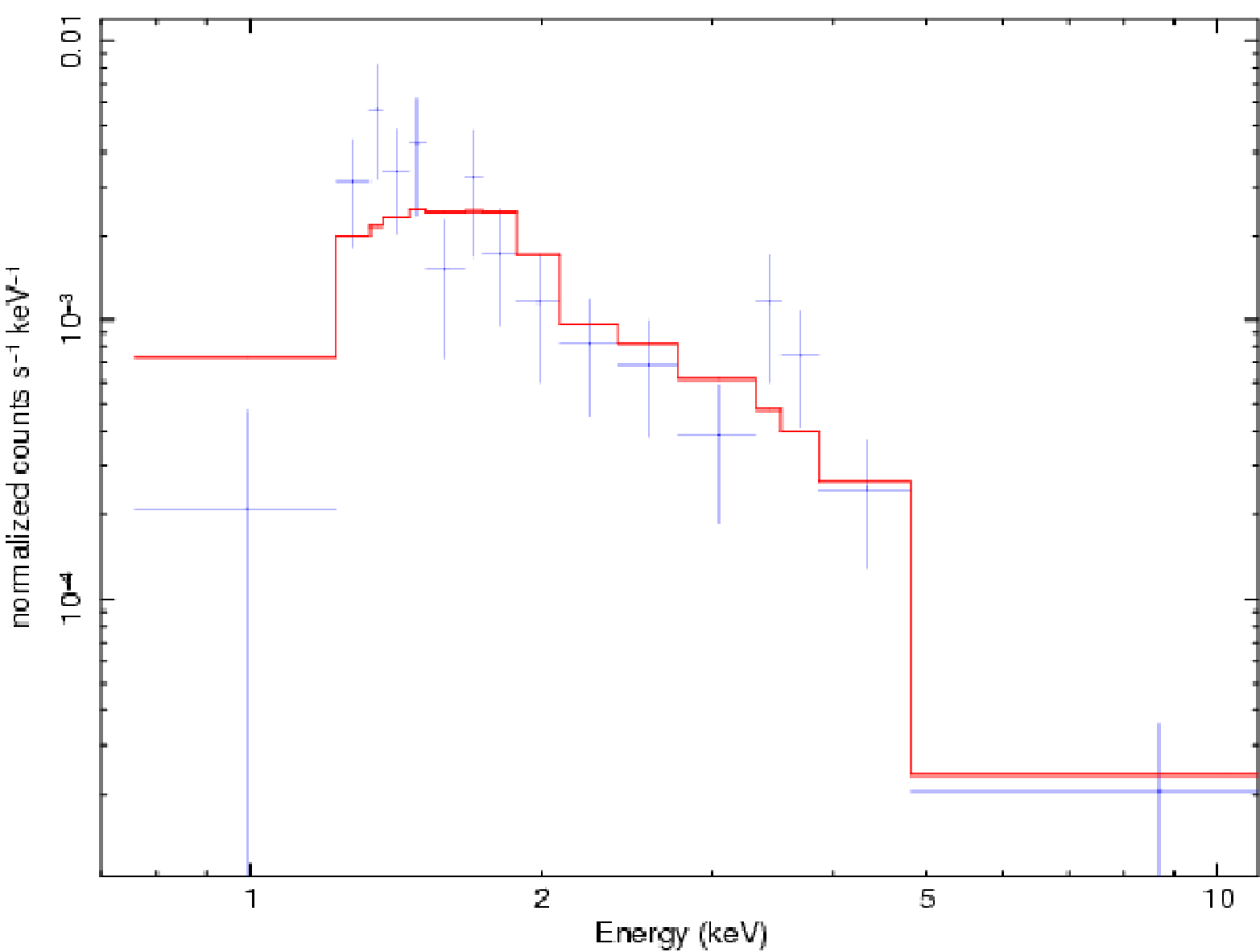}{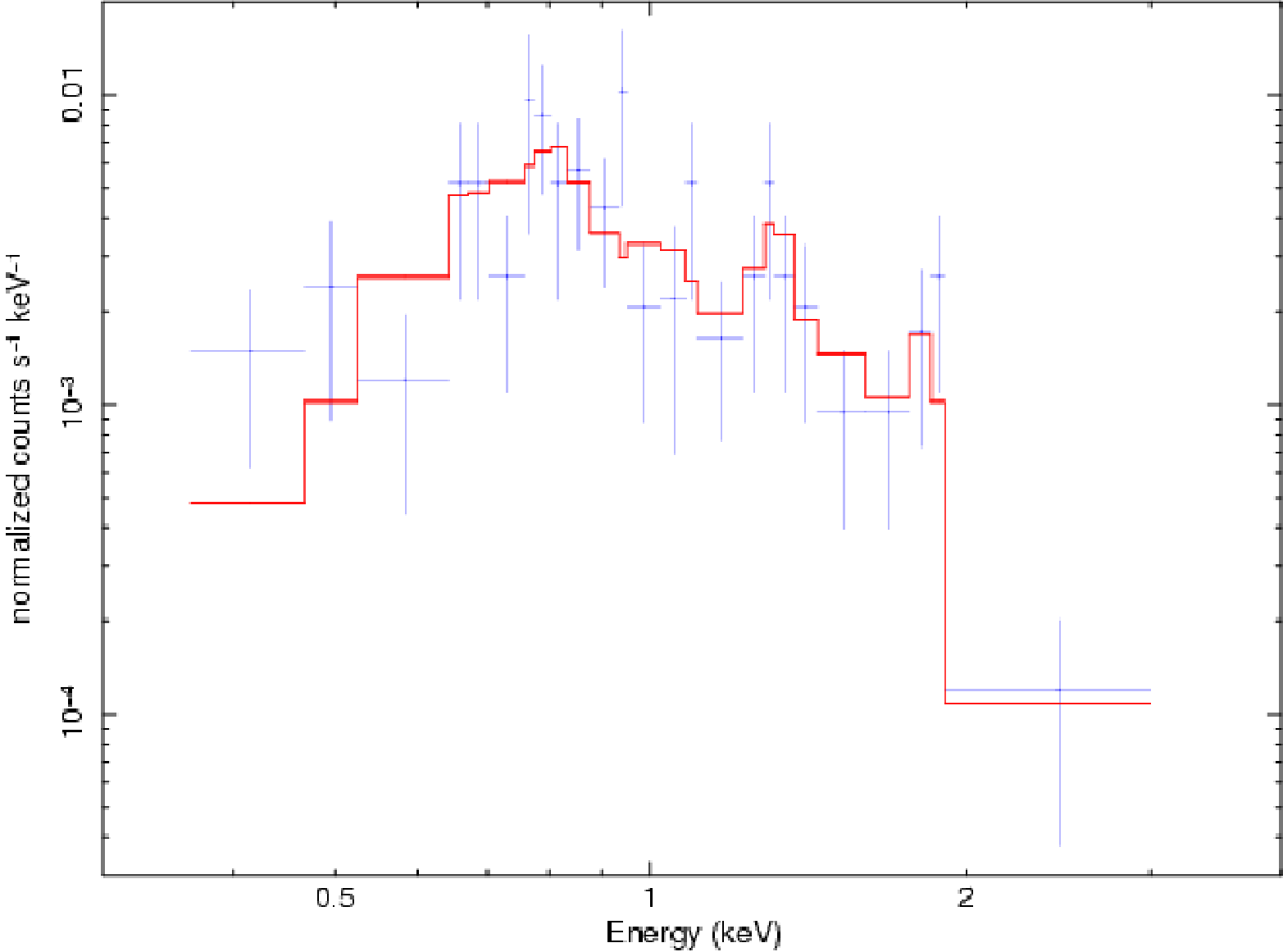}
\caption{
Left: The X-ray spectrum of the ULX in Arp 240W (clump 4).
The spectrum is well-fitted by a relatively steep 
power law with 
a photon index $\gamma$ = 2.7 $\pm$ $^{1.0}_{0.8}$
and an absorbing column N$_{\rm H}$ $\sim$ 1.5 $\times$ 10$^{22}$ cm$^{-2}$.
The small number of counts and the small energy range available
for spectral fitting do not allow us to constrain
the absorbing column or distinguish between a
power-law model and a curved model (eg, a Comptonized disk-blackbody). 
Right: The X-ray spectrum of the 
source in Arp 240E.
The spectrum is fit with a two-component model, in which the
column densities of the two components differ, but the temperatures
are the same.
The best-fit temperature is kT = 0.30 $\pm$ $^{0.09}_{0.07}$ keV.
The dominant component is highly absorbed (N$_{\rm H}$ = 1.6 $\times$
10$^{22}$ cm$^{-2}$, while the second is 2 $\times$ 10$^{20}$ cm$^{-2}$,
near the Galactic value.
}
\end{figure}

\begin{figure}
\plotone{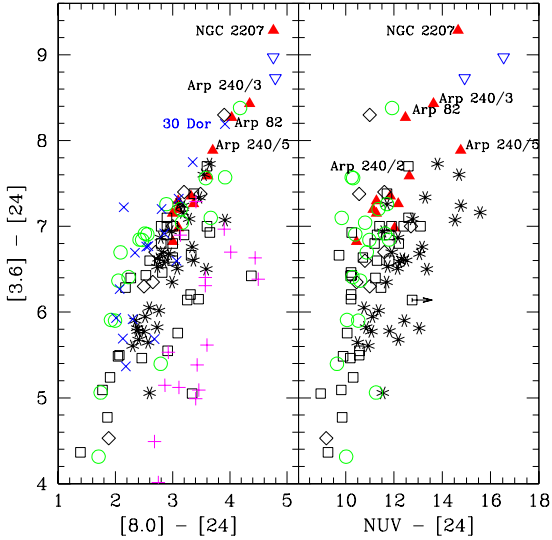}
\caption{UV/IR colors for the hinge clumps in the sample (red filled triangles),
compared to knots of star formation in other galaxies.  These include:
\citet{zhang10} regions 3 and 4 
in the overlap region of the Antennae
(blue upside down open triangles), 
other star forming knots in 
the Antennae (black asterisks),
regions in NGC 2403 (open green circles),
LMC regions (blue crosses), SMC regions (magenta plus signs), 
and clumps in other interacting galaxies, where 
the open black squares are disk clumps and the open
black diamonds are tail clumps.
For clarity, errorbars are omitted.   These are generally about the size of
the data points or slightly larger.
Some of the clumps are labeled.
The Spitzer 24 $\mu$m data for 30 Doradus in the LMC is saturated,
thus the [3.6] $-$ [24] and [8.0] $-$ [24] colors are lower limits.
}
\end{figure}

\begin{figure}
\plotone{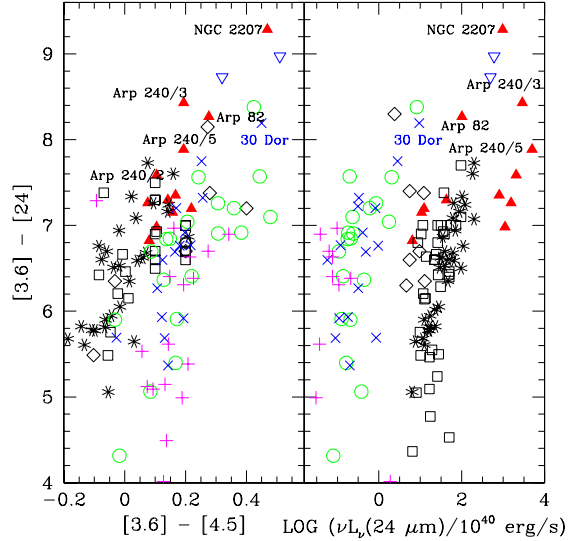}
\caption{
Left: The [3.6] $-$ [24] color plotted against [3.6] $-$ [4.5].
Right: The [3.6] $-$ [24] color plotted 
against L$_{24}$ = $\nu$L$_{\nu}$.
The hinge clumps are plotted as filled red triangles,
regions 3 and 4 in the overlap region of the Antennae
are shown as 
blue upside down open triangles, other star forming knots in 
the Antennae are black asterisks, regions in NGC 2403 are given
as open green circles,
LMC regions are blue crosses, SMC regions are magenta plus signs, 
disk clumps in other interacting galaxies are
open black squares, and tail clumps in other systems are 
black diamonds.
Some of the clumps are labeled.
The Spitzer 24 $\mu$m data for 30 Doradus in the LMC is saturated,
thus the [3.6] $-$ [24] color and the 24 $\mu$m luminosities are lower limits.
}
\end{figure}

\begin{figure}
\plotone{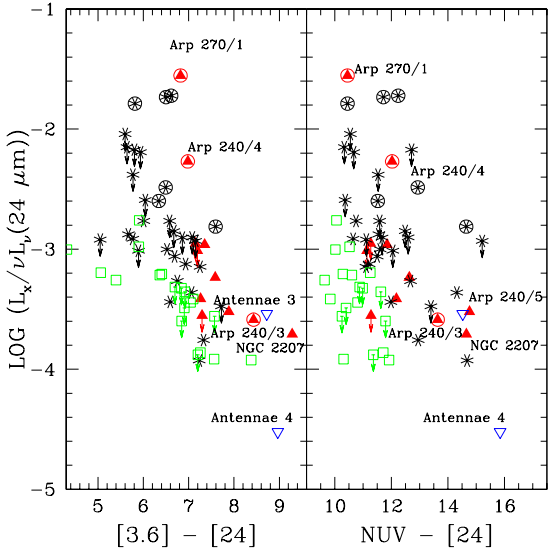}
\caption{
Plot of the observed L$_X$/L$_{24}$ vs.\ [3.6] $-$ [24] (left panel)
and vs.\ NUV $-$ [24] (right panel) for
the eight hinge clumps with Chandra data
(red filled triangles), 
regions 3 and 4 in the Antennae (upside down open blue triangles),
other locations in the Antennae (black asterisks), and
star forming regions within NGC 2403 (open green squares).
These values are {\it not} corrected for dust attenuation.
Points surrounded by circles contain bright X-ray point sources.  
}
\end{figure}

\newpage
\newpage

\begin{figure}
\plotone{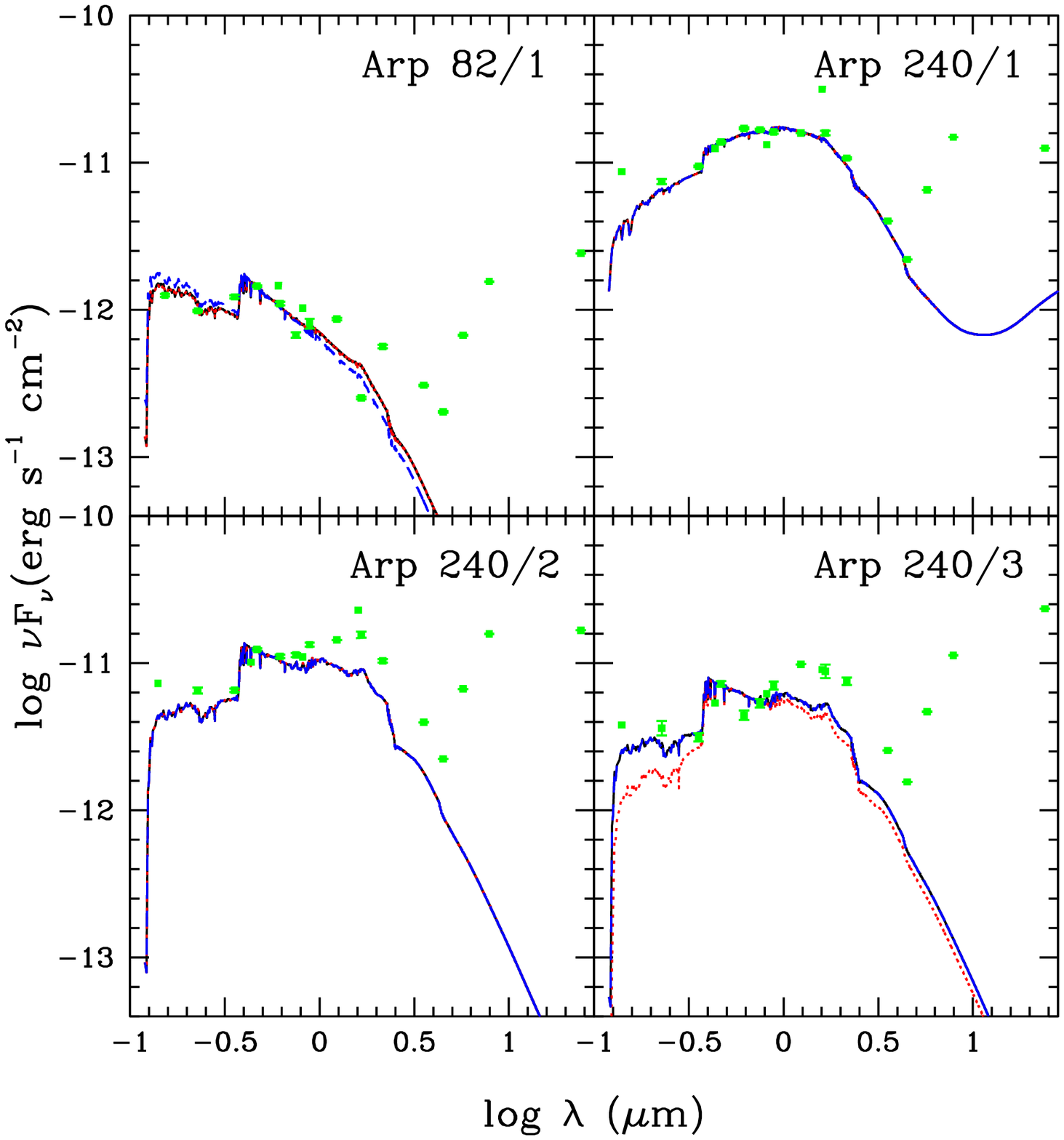}
\caption{
Broadband large aperture 
GALEX/SDSS/Spitzer/HST/2MASS
UV/optical/IR spectral energy distributions (SEDs) of some of
the clumps in the sample (green filled squares).
The black solid curve is the best-fit single-age instantaneous burst model.
The blue dashed curve is the burst model with the
best-fit reddening, and an age 1$\sigma$ less than
the best fit age.
The red dotted curve is the burst model with the
best-fit reddening, and an age 1$\sigma$ more than
the best fit age.
The plotted error bars only include statistical errors.
Note that the fits only include the GALEX/SDSS UV/optical photometry;
we extend the model stellar component to longer wavelengths to compare with
the Spitzer and near-infrared data 
to search for excesses in those bands (see Section 7.2).
}
\end{figure}

\newpage

\begin{figure}
\plotone{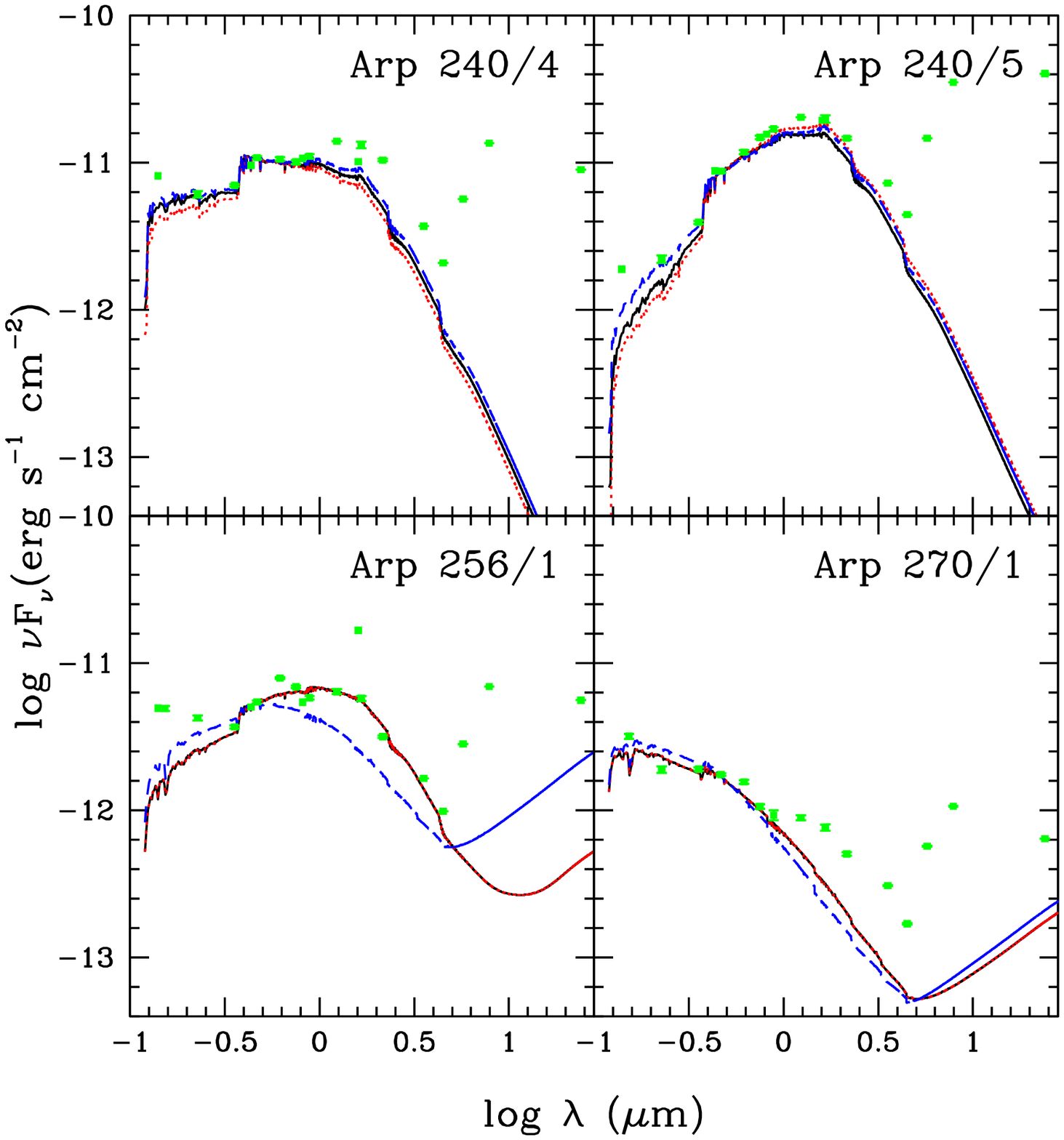}
\caption{
Broadband large aperture 
GALEX/SDSS/Spitzer/HST/2MASS
UV/optical/IR spectral energy distributions (SEDs) of some of
the clumps in the sample (green filled squares).
The black solid curve is the best-fit single-age instantaneous burst model.
The blue dashed curve is the burst model with the
best-fit reddening, and an age 1$\sigma$ less than
the best fit age.
The red dotted curve is the burst model with the
best-fit reddening, and an age 1$\sigma$ more than
the best fit age.
The plotted error bars only include statistical errors.
Note that the fits only include the GALEX/SDSS UV/optical photometry;
we extend the model stellar component to longer wavelengths to compare with
the Spitzer and near-infrared data to 
search for excesses in those bands (see Section 7.2).
}
\end{figure}

\begin{figure}
\plotone{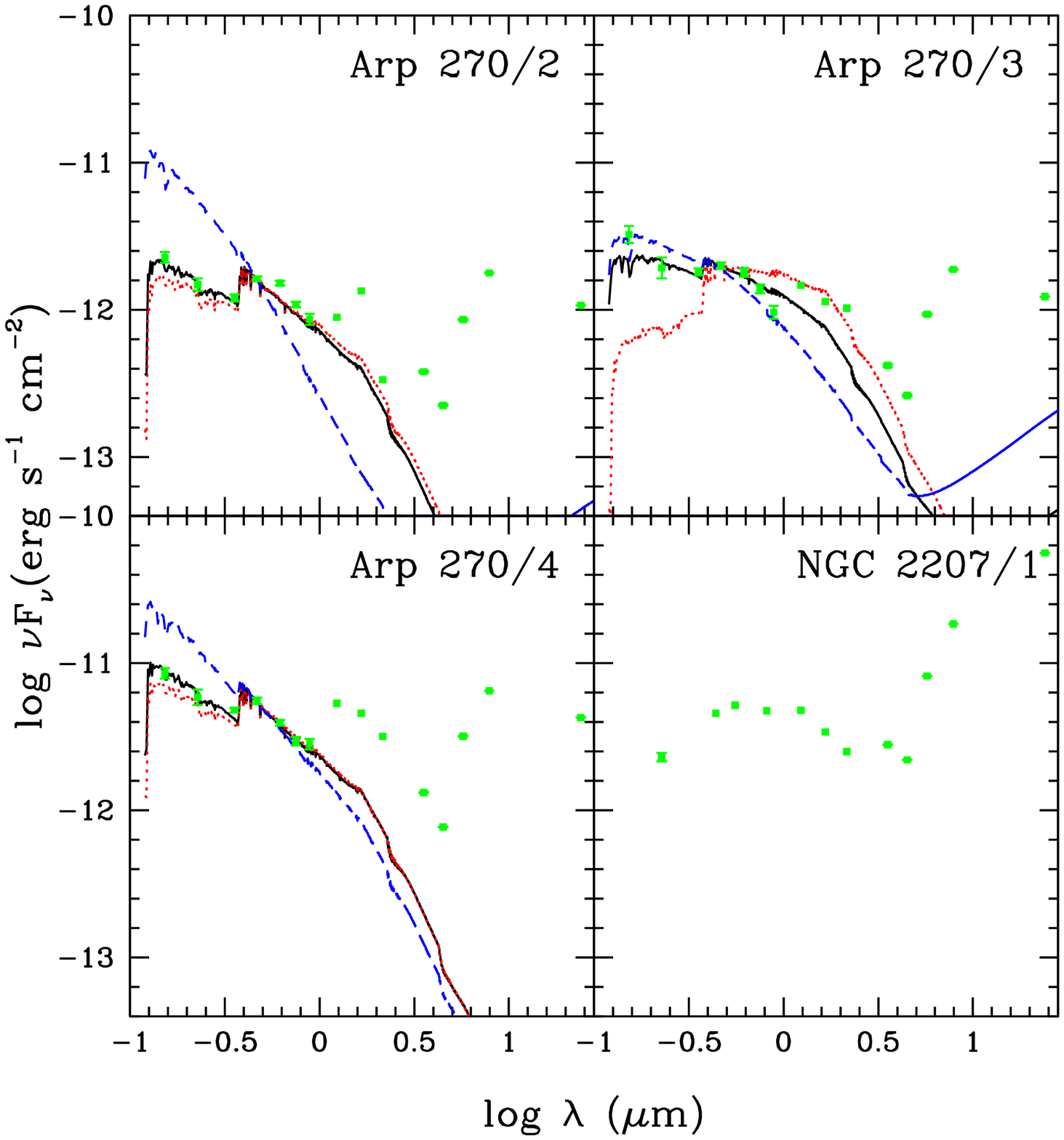}
\caption{
Broadband large aperture 
GALEX/SDSS/Spitzer/HST/2MASS
UV/optical/IR spectral energy distributions (SEDs) of some of
the clumps in the sample (green filled squares).
The black solid curve is the best-fit single-age instantaneous burst model.
The blue dashed curve is the burst model with the
best-fit reddening, and an age 1$\sigma$ less than
the best fit age.
The red dotted curve is the burst model with the
best-fit reddening, and an age 1$\sigma$ more than
the best fit age.
The plotted error bars only include statistical errors.
Note that the fits only include the GALEX/SDSS UV/optical photometry;
we extend the model stellar component to longer wavelengths to compare with
the Spitzer and near-infrared data to search for 
excesses in those bands (see Section 7.2).
Since no SDSS data is available 
for NGC 2207, 
we do not provide a best-fit model for that region.
}
\end{figure}

\clearpage

\begin{figure}
\plotone{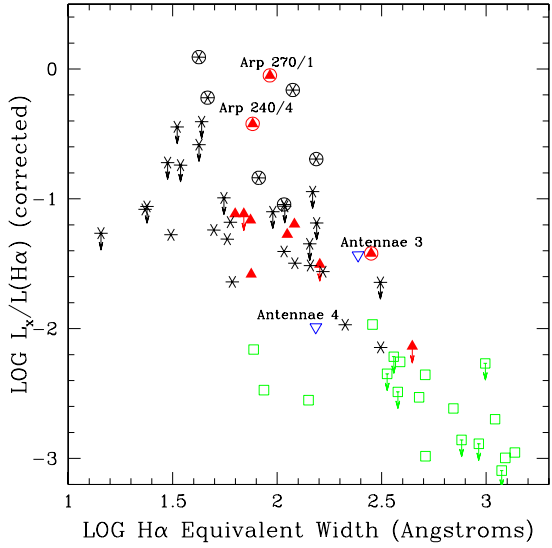}
\caption{
A plot of the reddening-corrected L$_{X}$/L$_{H\alpha}$ ratio against
H$\alpha$ equivalent width.   
The hinge clumps are the filled red triangles, the upside down open blue triangles
are regions 3 and 4 in the Antennae, the black asterisks
are other positions in the Antennae, and 
the open green squares are regions within NGC 2403.
For this plot, the reddening correction was determined
from the L$_{24}$/L$_{H\alpha}$
ratio as described in the text.
The symbols enclosed in circles represent regions
with strong X-ray point sources.
}
\end{figure}

\begin{figure}
\plotone{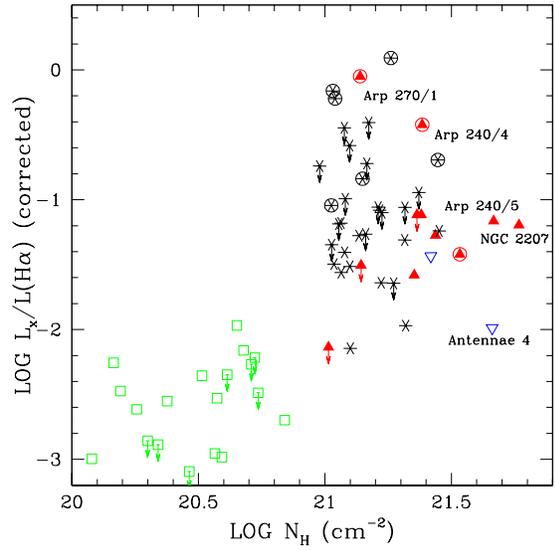}
\caption{
A plot of the reddening-corrected L$_{X}$/L$_{H\alpha}$ ratio against
hydrogen column density, both determined using the 
L$_{24}$/L$_{H\alpha}$ ratio as described in the text.
The hinge clumps are the filled red triangles, the upside down open blue triangles
are regions 3 and 4 in the Antennae, the black asterisks
are other positions in the Antennae, and 
the open green squares are regions within NGC 2403.
The symbols enclosed in circles represent regions
with strong X-ray point sources.
}
\end{figure}

\begin{figure}
\plotone{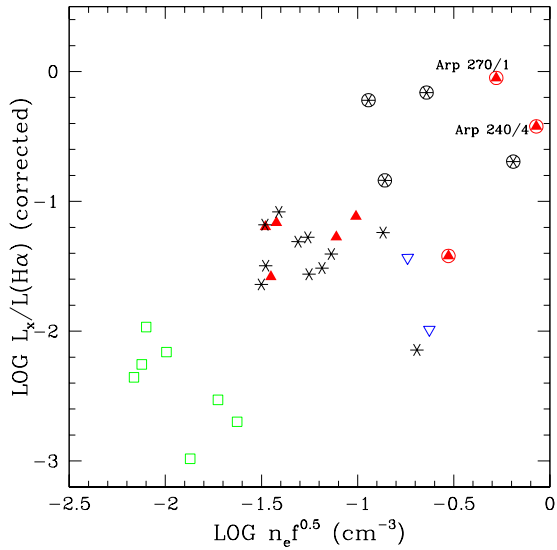}
\caption{
A plot of the reddening-corrected L$_{X}$/L$_{H\alpha}$ ratio against
electron number density n$_{\rm e}$ $\times$ $\sqrt{f}$, where f is the
volume filling factor.
The hinge clumps are the filled red triangles, the upside down open blue triangles
are regions 3 and 4 in the Antennae, the black asterisks
are other positions in the Antennae, and 
the open green squares are regions within NGC 2403.
The symbols enclosed in circles represent regions
with strong X-ray point sources; if these are ULXs,
the derived n$_{\rm e}$ are not relevant.
}
\end{figure}

\begin{figure}
\plotone{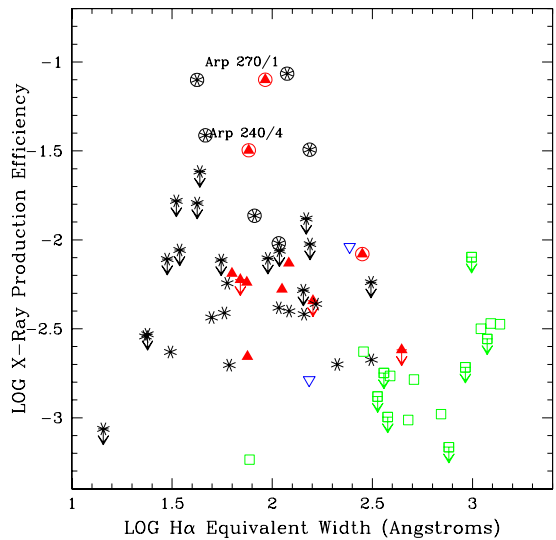}
\caption{
A plot of X-ray production efficiency = L$_X$/L$_{mech}$ vs.\ 
H$\alpha$ equivalent width.
The hinge clumps are the filled red triangles, the upside down open blue
triangles are regions 3 and 4 in the Antennae, the black asterisks
are other positions in the Antennae, and the open green squares are regions
within NGC 2403.
The symbols enclosed in circles represent regions containing strong
X-ray point sources.
}
\end{figure}

\begin{figure}
\plotone{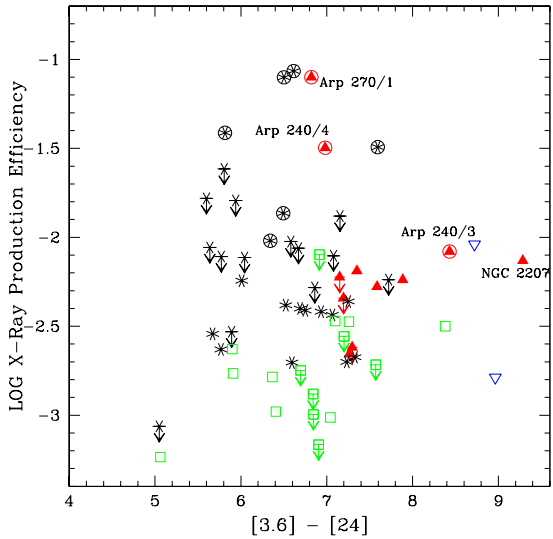}
\caption{
A plot of X-ray production efficiency = L$_X$/L$_{mech}$ against
the [3.6] $-$ [24] color.
The hinge clumps are the filled red triangles, while open blue upside down
triangles are regions 3 and 4 in the Antennae, the black asterisks
are other positions in the Antennae, and the open green squares are regions
within NGC 2403.
The symbols enclosed in circles represent regions containing strong
X-ray point sources.
The
reddest hinge clump in [3.6] - [24] is NGC 2207; the second reddest
is Arp 240-3.
}
\end{figure}

\begin{figure}
\plotone{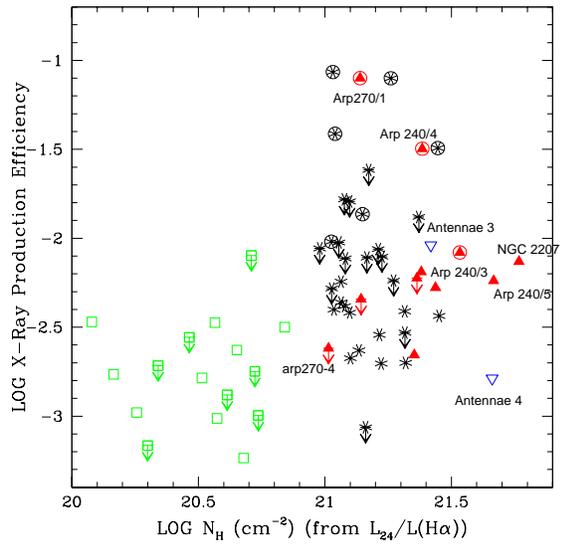}
\caption{
A plot of X-ray production efficiency = L$_X$/L$_{mech}$ vs.\ N$_H$,
where 
N$_H$ was calculated 
from the L$_{24}$/L$_{H\alpha}$ ratio, with the assumptions given in the text.
The hinge clumps are the filled red triangles, while open blue upside down
triangles are regions 3 and 4 in the Antennae, the black asterisks
are other positions in the Antennae, and the open green squares are 
regions within NGC 2403.
The symbols enclosed in circles represent regions containing strong
X-ray point sources.
}
\end{figure}

\begin{figure}
\plotone{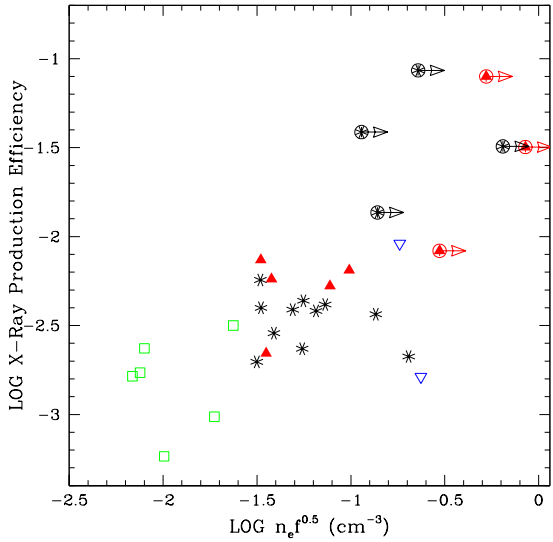}
\caption{
A plot of X-ray production efficiency = L$_X$/L$_{mech}$ vs.\ 
electron density n$_{\rm e}$ $\times$ $\sqrt(f)$,
where f is the volume filling factor.
The hinge clumps are the filled red triangles, while open blue upside down
triangles are regions 3 and 4 in the Antennae, the black asterisks
are other positions in the Antennae, and the open green squares are 
regions within NGC 2403.
The symbols enclosed in circles and marked by lower limit symbols represent regions containing strong
X-ray point sources.  These may be ULXs, for which the 
estimate of n$_{\rm e}$ is not valid.
}
\end{figure}

\begin{figure}
\plotone{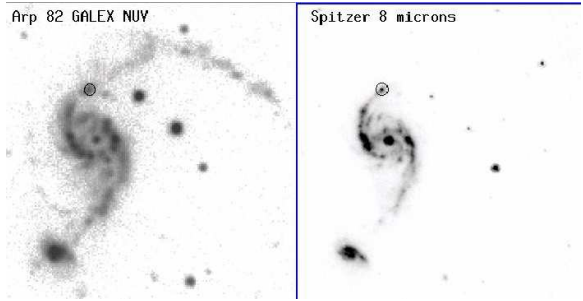}
\caption{
Left panel: GALEX NUV image of Arp 82.
Right panel: Spitzer 8 $\mu$m image of 
Arp 82.
The field of view is 4\farcm5 $\times$ 4\farcm5.
North is up and east to the left. 
The northern hinge clump
is marked by a circle with a radius of 5$''$.   Note the ocular (`eye-shaped') morphology of the northern galaxy.
The northern tail is much more evident in the NUV image than in the mid-infrared.
}
\end{figure}

\begin{figure}
\plotone{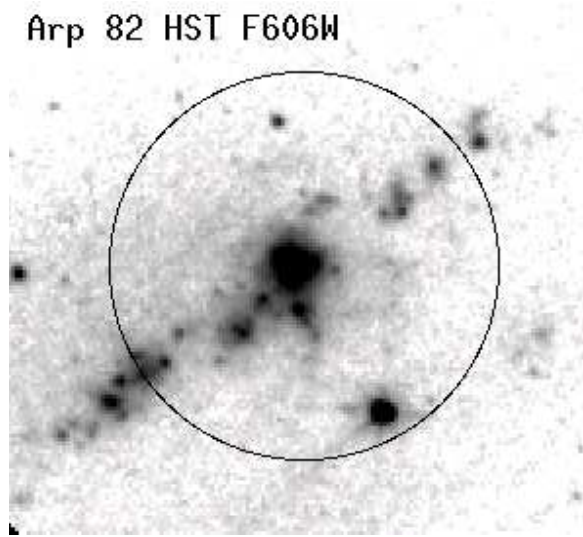}
\caption{
The archival HST F606W image of the region
around the hinge clump.   A 5$''$ radius circle is plotted
marking the hinge clump. Note the luminous source in the center
of a straight line of fainter star clusters.
The HST archival images of Arp 82 were
included in the Mullan et al.\ (2011) large statistical study of
the frequency of clusters in
tidal features, however, their study did not present
any analysis of individual clusters or cluster
associations.
No Chandra data is available for Arp 82.
}
\end{figure}

\begin{figure}
\plotone{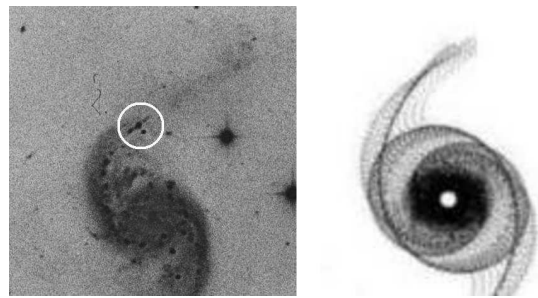}
\caption{
Left: a close-up view of the \citet{arp66} Atlas optical photograph
of the hinge region in NGC 2535 (the northern galaxy in Arp 82).
The hinge clump is marked.
Right: an analytical model of a prolonged prograde interaction
from
\citet{struck12}.   See text for more details.
}
\end{figure}

\begin{figure}
\plotone{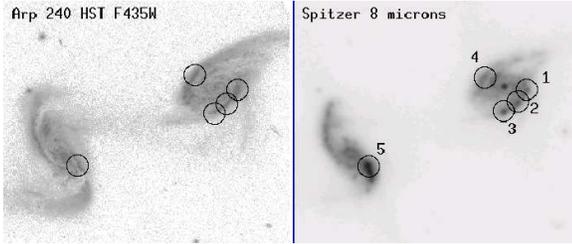}
\caption{ Left panel:
The HST F435W image of Arp 240.
Right panel: The Spitzer 8 $\mu$m image of Arp 240.
North is up and east to the left.
The field of view is 1\farcm7 $\times$ 2\farcm1.
The five hinge clumps are 
marked with 5$''$ radii and labeled.
The HST images of Arp 240 were previously used 
by \citet{haan11} to 
extract
nuclear profiles,
but no analysis of the non-nuclear
clusters was done.
}
\end{figure}

\begin{figure}
\plotone{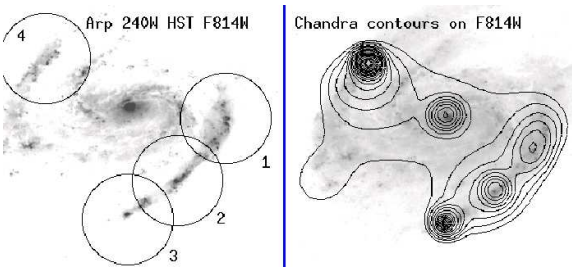}
\caption{ 
Left: 
A closer view of 
the HST F814W image of Arp 240W, with the greyscale
set to emphasize the small-scale structure.
Right: 
A smoothed Chandra 0.3 $-$ 8 keV map of Arp 240W 
plotted as contours on the HST F814W map.
In the left panel, 
the sample hinge clumps are marked by 
circles with radii of 5$''$ and are labeled.
The field of view is 30$''$ $\times$ 33$''$.
North is up and east to the left.
}
\end{figure}

\clearpage

\begin{figure}
\plotone{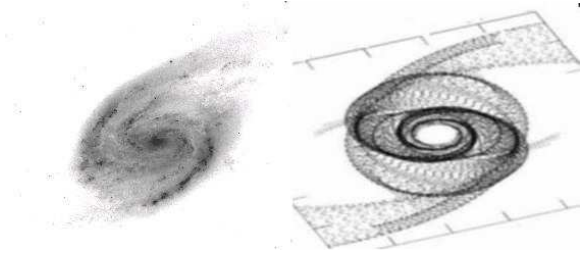}
\caption{
Left: The HST F814W image of Arp240W.
Right: An analytical model from \citet{struck12}
that approximates the structure of
Arp 240W (see text for details).
}
\end{figure}

\begin{figure}
\plotone{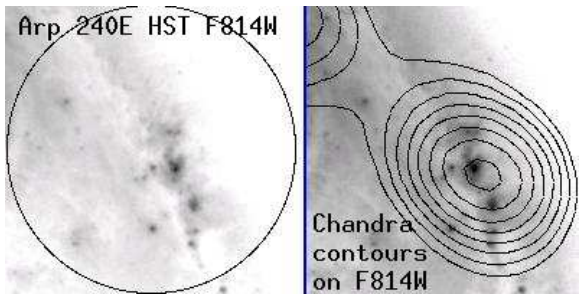}
\caption{
Left: A closer view of 
the HST F814W image of the hinge clump in Arp 240E (clump 5).
The hinge clump is marked by a 5$''$ radius circle.
Right: 
The smoothed 0.3 $-$ 8 keV Chandra map of Arp 240E.
North is up and east to the left.
Notice
the strong and spatially extended X-ray emission from this source (L$_X$ = 
1.4
$\times$ 10$^{41}$ erg~s$^{-1}$ after correction for
internal extinction).
The HST image shows an extremely luminous source near this 
location
(M$_{\rm I}$ = $-$16.5 with $\le$75 pc diameter.) 
}
\end{figure}

\clearpage

\begin{figure}
\plotone{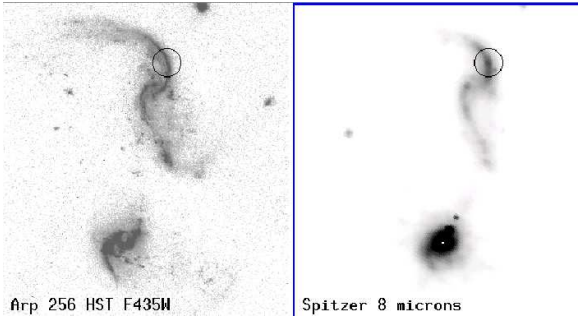}
\caption{
Arp 256. 
Left: HST F435W image.   Right: Spitzer 8 $\mu$m
image.  The field of view is 1\farcm67 $\times$ 1\farcm8.
North is up and east to the left. 
The hinge clump in the northern tail is marked.
Note the tidal dwarf galaxy in the southern tail
of the northern galaxy.
}
\end{figure}

\begin{figure}
\plotone{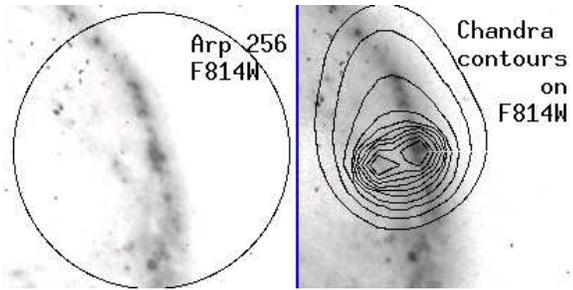}
\caption{
Left: A close-up view of the HST F814W image of Arp 256N.
A 5$''$ radius encircling the hinge clump is marked.
Note the flattened structure of the bright source near the
center of the hinge clump.
Right:
The smoothed 0.3 $-$ 8 keV Chandra image of the same region,
plotted as contours on the F814W image.
}
\end{figure}

\begin{figure}
\plotone{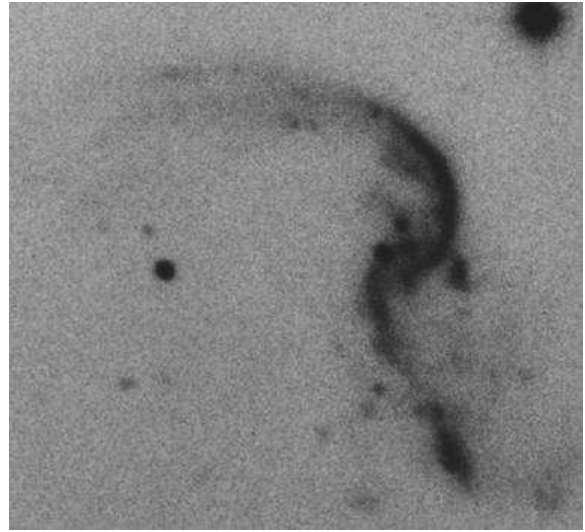}
\caption{
A close-up view of the \citet{arp66} Atlas image of Arp 256N.
The northern tail appears double.
The `X' shape in the center of the galaxy is due to star formation
outside of the main disk of the galaxy (see Figure 25).
}
\end{figure}

\begin{figure}
\plotone{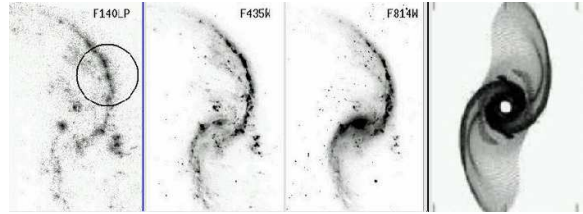}
\caption{
A comparison of the HST UV/optical images of Arp 256N (left three panels)
with an analytical model from \citet{struck12}.  See text for more details
on the model.
}
\end{figure}

\begin{figure}
\plotone{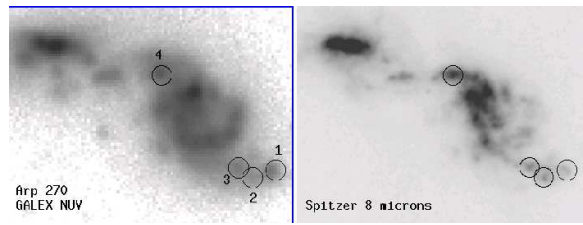}
\caption{
Arp 270. 
Left: GALEX NUV image.   Right: Spitzer 8 $\mu$m
image.  The field of view is 1\farcm9 $\times$ 2\farcm5.
North is up and east to the left. 
The four targeted clumps are marked and labeled.
NGC 3396 is the galaxy to the east (left), while NGC 3395
is to the west.
}
\end{figure}

\begin{figure}
\plotone{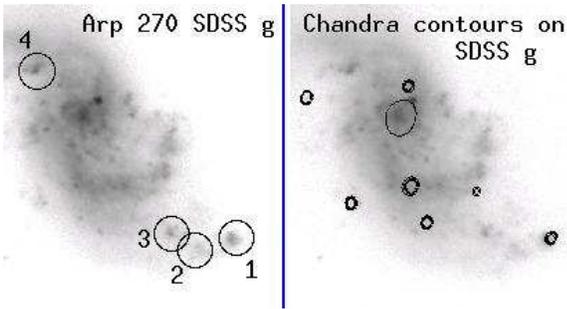}
\caption{
Left:
A close-up view of the SDSS g image of Arp 270,
with the targeted clumps marked and labeled. 
Right: The SDSS g image with 
the smoothed 0.3 $-$ 8 keV Chandra image 
superimposed as contours.
The field of view is 1\farcm1 $\times$ 1\farcm2.
}
\end{figure}

\begin{figure}
\plotone{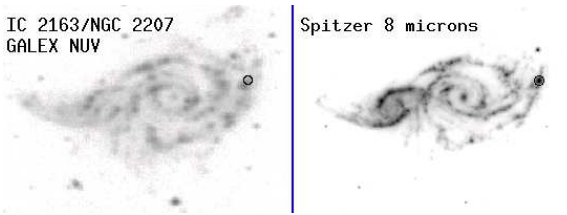}
\caption{
The 
IC 2163/NGC 2207 interacting galaxy pair.
Left: The GALEX NUV image.  Right: The Spitzer 8 micron
image.
North is up and east to the left.
The field of view is 4\farcm0 $\times$ 6\farcm7.
NGC 2207 is the galaxy to the west (right) in this picture.
The hinge clump is marked by a circle with a 5$''$ radius.
}
\end{figure}

\begin{figure}
\plotone{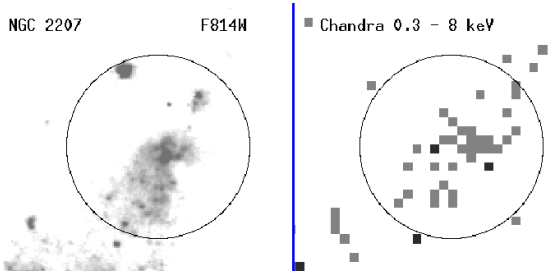}
\caption{
A closer view of the NGC 2207 hinge clump.
Left: The HST F814W image.
Right: The unsmoothed
Chandra 0.3 $-$ 8 keV image.
The circle has a 5$''$ radius.
North is up and east to the left.
}
\end{figure}

\begin{figure}
\plotone{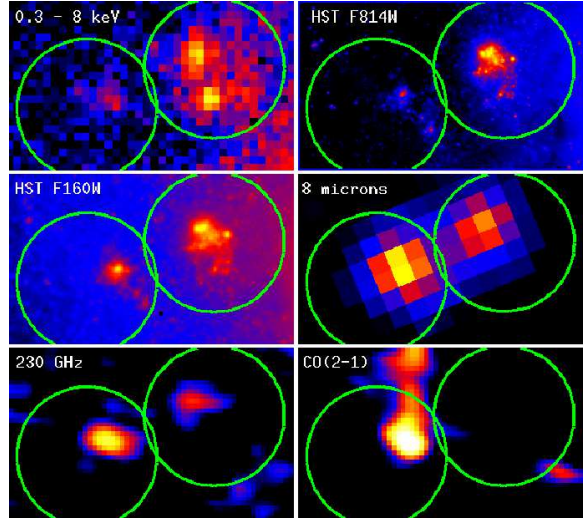}
\caption{
A multi-wavelength montage of the overlap region of the Antennae.
Upper left: the co-added 0.3 $-$ 8 keV Chandra map.
Upper right: an archival HST ACS F814W optical image.
Middle left: an archival HST WFC3 F160W near-infrared image.
Middle right: the 8 $\mu$m Spitzer image.
Lower left: an ALMA 230 GHz continuum map.  
Lower right: an ALMA CO(2-1) map.
North is up and east to the left.
The left and right circles are \citet{zhang10} regions 4 and 3, respectively,
with radii of 4\farcs5. 
The brightest cluster in region 4 is known as WS80 (source 80 from
\citealp{whitmore95}), while the bright complex of star clusters in
region 3 is called Knot B \citep{rubin70, whitmore10}.
The spatial resolution of the ALMA maps is 1\farcs68 $\times$ 0\farcs85.
}
\end{figure}

\clearpage

\begin{deluxetable}{cccrcc}
\tabletypesize{\scriptsize}
\setlength{\tabcolsep}{0.03in}
\def\et#1#2#3{${#1}^{+#2}_{-#3}$}
\tablewidth{0pt}
\tablecaption{Hinge Clump Sample}
\tablehead{
\multicolumn{1}{c}{Clump} &
\multicolumn{1}{c}{RA} &
\multicolumn{1}{c}{Dec.} &
\multicolumn{1}{c}{Distance$^*$} &
\multicolumn{1}{c}{GALEX?} &
\multicolumn{1}{c}{SDSS?}\\ 
\\ 
\multicolumn{1}{c}{} &
\multicolumn{1}{c}{(J2000)} &
\multicolumn{1}{c}{(J2000)} &
\multicolumn{1}{c}{(Mpc)} &
\multicolumn{1}{c}{(NUV/FUV)} &
\multicolumn{1}{c}{} \\
\\ 
}
\startdata

Arp 82-1  &  08 11 13.9  &  +25 13 09.6  &  59.2  &  yy  &  y\\
Arp 240-1  &  13 39 52.3  &  +00 50 22.8  &  101.7  &  yn  &  y\\
Arp 240-2  &  13 39 52.5  &  +00 50 17.3  &  101.7  &  yn  &  y\\
Arp 240-3  &  13 39 53.0 &  +00 50 12.9  &  101.7  &  yn  &  y\\
Arp 240-4  &  13 39 53.6  &  +00 50 28.6  &  101.7  &  yn  &  y\\
Arp 240-5  &  13 39 57.2  &  +00 49 46.8  &  101.7  &  yn  &  y\\
Arp 256-1  &  00 18 49.8  &  $-$10 21 33.8  &  109.6  &  yy  &  y\\
Arp 270-1  &  10 49 46.6  &  +32 58 23.9  &  29.0  &  yy  &  y\\
Arp 270-2  &  10 49 47.5  &  +32 58 20.3  &  29.0  &  yy  &  y\\
Arp 270-3  &  10 49 48.1  &  +32 58 25.2  &  29.0  &  yy  &  y\\
Arp 270-4  &  10 49 51.2  &  +32 59 11.7  &  29.0  &  yy  &  y\\
NGC 2207-1  &  06 16 15.9  &  $-$21 22 02.6  &  38.0  &  yn  &  n\\
\enddata
\tablenotetext{*}{
Distances from the NASA Extragalactic Database, assuming H$_0$ = 73 km~s$^{-1}$~Mpc$^{-1}$,
and accounting for peculiar velocities due to the Virgo Cluster, the Great
Attractor, and the Shapley Supercluster.
}
\end{deluxetable}

\clearpage
\begin{deluxetable}{cccr}
\tablewidth{0pt}
\tablecaption{Hubble Space Telescope Data }
\tablehead{
\multicolumn{1}{c}{Galaxy} &
\multicolumn{1}{c}{Instrument(s)} &
\multicolumn{1}{c}{Filters} &
\multicolumn{1}{c}{Observing} \\
\multicolumn{1}{c}{} &
\multicolumn{1}{c}{} &
\multicolumn{1}{c}{} &
\multicolumn{1}{c}{Time} \\
\multicolumn{1}{c}{} &
\multicolumn{1}{c}{} &
\multicolumn{1}{c}{} &
\multicolumn{1}{c}{(Seconds)} \\
}
\startdata
Arp 82 & WFPC2 & F606W & 1900 \\
       &       & F814W & 1900 \\
Arp 240 & ACS & F140LP & 2516 \\
   & & F435W & 1260 \\
   & & F814W & 720 \\
Arp 240E&  WFC3 & F160W &2395 \\
Arp 240W&  NICMOS & F160W & 2303\\
Arp 256  & ACS & F140LP & 2520 \\
  &  & F435W &1260 \\
  &  & F814W & 720\\
                 & NICMOS & F160W &2495 \\
NGC 2207&WFPC2 & F439W & 2000 \\
        &      & F555W & 660 \\
        &      & F814W & 720 \\
\enddata
\end{deluxetable}

\begin{deluxetable}{crr}
\tablewidth{0pt}
\tablecaption{Chandra Telescope Data }
\tablehead{
\multicolumn{1}{c}{Galaxy} &
\multicolumn{1}{c}{Dataset} &
\multicolumn{1}{c}{Exposure} \\
\multicolumn{1}{c}{} &
\multicolumn{1}{c}{} &
\multicolumn{1}{c}{Time} \\
\multicolumn{1}{c}{} &
\multicolumn{1}{c}{} &
\multicolumn{1}{c}{(ksec)} \\
\\
}
\startdata
Arp 240 & 10565 & 19.9\\
Arp 256 & 13823 & 29.5\\
Arp 270 & 2042 & 19.3\\
NGC 2207 & 14914 & 12.9\\
\enddata
\end{deluxetable}

\begin{deluxetable}{ccccc}
\tabletypesize{\scriptsize}
\setlength{\tabcolsep}{0.03in}
\def\et#1#2#3{${#1}^{+#2}_{-#3}$}
\tablewidth{0pt}
\tablecaption{GALEX Aperture Corrections}
\tablehead{
\multicolumn{1}{c}{Galaxy} &
\multicolumn{1}{c}{Filter} &
\multicolumn{1}{c}{Aperture Correction} &
\multicolumn{1}{c}{Filter} &
\multicolumn{1}{c}{Aperture Correction}
\\ 
\multicolumn{1}{c}{} &
\multicolumn{1}{c}{} &
\multicolumn{1}{c}{} &
\multicolumn{1}{c}{} &
\multicolumn{1}{c}{} 
\\
}
\startdata

   Arp 82  &  NUV  &   1.19  $\pm$  0.04 
       &  FUV  &   1.23  $\pm$  0.15   \\
   Arp 240  &  NUV  &   1.23  $\pm$  0.08   \\
   Arp 256  &  NUV  &   1.30  $\pm$  0.11 
       &  FUV  &   1.46  $\pm$  0.07   \\
   Arp 270  &  NUV  &   1.16  $\pm$  0.02 
       &  FUV  &   1.41  $\pm$  0.31   \\
  NGC 2207  &  NUV  &   1.24  $\pm$  0.05   \\
\enddata
\tablenotetext{*}{
Multiplicative Aperture Corrections, for a 5$''$ radius aperture.
}
\end{deluxetable}

\begin{deluxetable}{ccccccccccccc}
\rotate
\tabletypesize{\scriptsize}
\setlength{\tabcolsep}{0.03in}
\def\et#1#2#3{${#1}^{+#2}_{-#3}$}
\tablewidth{0pt}
\tablecaption{Large Aperture GALEX/SDSS/Spitzer Magnitudes for Sample Hinge Clumps\label{tab-1}}
\tablehead{
\multicolumn{1}{c}{Clump} &
\multicolumn{1}{c}{FUV} &
\multicolumn{1}{c}{NUV} &
\multicolumn{1}{c}{u} &
\multicolumn{1}{c}{g} &
\multicolumn{1}{c}{r} &
\multicolumn{1}{c}{i} &
\multicolumn{1}{c}{z} &
\multicolumn{1}{c}{3.6 $\mu$m} &
\multicolumn{1}{c}{4.5 $\mu$m} &
\multicolumn{1}{c}{5.8 $\mu$m} &
\multicolumn{1}{c}{8.0 $\mu$m} &
\multicolumn{1}{c}{24 $\mu$m} 
\\ 
\multicolumn{1}{c}{} &
\multicolumn{1}{c}{(mag)} &
\multicolumn{1}{c}{(mag)} &
\multicolumn{1}{c}{(mag)} &
\multicolumn{1}{c}{(mag)} &
\multicolumn{1}{c}{(mag)} &
\multicolumn{1}{c}{(mag)} &
\multicolumn{1}{c}{(mag)} &
\multicolumn{1}{c}{(mag)} &
\multicolumn{1}{c}{(mag)} &
\multicolumn{1}{c}{(mag)} &
\multicolumn{1}{c}{(mag)} &
\multicolumn{1}{c}{(mag)} 
\\ 
}
\startdata

Arp 82-1   &    19.19 $\pm$  0.02   &    19.02 $\pm$  0.02   &    18.37 $\pm$  0.03   &    17.92 $\pm$  0.01   &    17.94 $\pm$  0.03   &    18.29 $\pm$  0.05    &     17.92 $\pm$  0.08   &    14.71 $\pm$  0.01   &    14.43 $\pm$  0.01   &    12.40 $\pm$  0.01   &    10.47 $\pm$  0.01   &     6.44 $\pm$  0.01   \\
Arp 240-1   &      $-$   &    16.82 $\pm$  0.04   &    16.16 $\pm$  0.01   &    15.47 $\pm$  0.02   &    14.97 $\pm$  0.02   &    14.80 $\pm$  0.02    &     14.67 $\pm$  0.03   &    11.92 $\pm$  0.01   &    11.84 $\pm$  0.01   &     9.93 $\pm$  0.01   &     8.02 $\pm$  0.01   &     4.65 $\pm$  0.02   \\
Arp 240-2   &      $-$   &    16.97 $\pm$  0.05   &    16.56 $\pm$  0.03   &    15.59 $\pm$  0.02   &    15.44 $\pm$  0.03   &    15.22 $\pm$  0.03    &     14.88 $\pm$  0.03   &    11.93 $\pm$  0.01   &    11.83 $\pm$  0.01   &     9.90 $\pm$  0.01   &     7.96 $\pm$  0.01   &     4.35 $\pm$  0.02   \\
Arp 240-3   &      $-$   &    17.61 $\pm$  0.12   &    17.35 $\pm$  0.07   &    16.18 $\pm$  0.04   &    16.44 $\pm$  0.08   &    16.04 $\pm$  0.07    &     15.57 $\pm$  0.07   &    12.41 $\pm$  0.01   &    12.21 $\pm$  0.01   &    10.29 $\pm$  0.01   &     8.32 $\pm$  0.01   &     3.98 $\pm$  0.01   \\
Arp 240-4   &      $-$   &    17.04 $\pm$  0.06   &    16.48 $\pm$  0.02   &    15.74 $\pm$  0.02   &    15.49 $\pm$  0.03   &    15.35 $\pm$  0.04    &     15.08 $\pm$  0.04   &    12.00 $\pm$  0.01   &    11.90 $\pm$  0.01   &    10.08 $\pm$  0.01   &     8.12 $\pm$  0.01   &     5.02 $\pm$  0.01   \\
Arp 240-5   &      $-$   &    18.14 $\pm$  0.06   &    17.11 $\pm$  0.03   &    15.97 $\pm$  0.03   &    15.37 $\pm$  0.03   &    14.93 $\pm$  0.03    &     14.61 $\pm$  0.03   &    11.27 $\pm$  0.01   &    11.08 $\pm$  0.01   &     9.05 $\pm$  0.01   &     7.09 $\pm$  0.01   &     3.39 $\pm$  0.01   \\
Arp 256-1   &    17.71 $\pm$  0.04   &    17.44 $\pm$  0.04   &    17.18 $\pm$  0.02   &    16.49 $\pm$  0.02   &    15.80 $\pm$  0.02   &    15.76 $\pm$  0.02    &     15.78 $\pm$  0.03   &    12.88 $\pm$  0.01   &    12.72 $\pm$  0.01   &    10.84 $\pm$  0.01   &     8.85 $\pm$  0.01   &     5.53 $\pm$  0.01   \\
Arp 270-1   &    18.18 $\pm$  0.04   &    18.31 $\pm$  0.06   &    17.90 $\pm$  0.03   &    17.72 $\pm$  0.02   &    17.57 $\pm$  0.03   &    17.80 $\pm$  0.04    &     17.77 $\pm$  0.08   &    14.71 $\pm$  0.01   &    14.62 $\pm$  0.01   &    12.57 $\pm$  0.01   &    10.89 $\pm$  0.01   &     7.88 $\pm$  0.02   \\
Arp 270-2   &    18.54 $\pm$  0.09   &    18.57 $\pm$  0.11   &    18.39 $\pm$  0.06   &    17.80 $\pm$  0.04   &    17.60 $\pm$  0.04   &    17.77 $\pm$  0.05    &     17.85 $\pm$  0.10   &    14.48 $\pm$  0.02   &    14.32 $\pm$  0.02   &    12.13 $\pm$  0.01   &    10.33 $\pm$  0.01   &     7.33 $\pm$  0.03   \\
Arp 270-3   &    18.16 $\pm$  0.14   &    18.29 $\pm$  0.18   &    17.95 $\pm$  0.06   &    17.57 $\pm$  0.06   &    17.42 $\pm$  0.07   &    17.50 $\pm$  0.08    &     17.73 $\pm$  0.11   &    14.37 $\pm$  0.02   &    14.15 $\pm$  0.02   &    12.04 $\pm$  0.02   &    10.27 $\pm$  0.01   &     7.18 $\pm$  0.03   \\
Arp 270-4   &    17.11 $\pm$  0.09   &    17.08 $\pm$  0.13   &    16.89 $\pm$  0.04   &    16.47 $\pm$  0.06   &    16.56 $\pm$  0.05   &    16.68 $\pm$  0.07    &     16.56 $\pm$  0.09   &    13.12 $\pm$  0.01   &    12.98 $\pm$  0.01   &    10.71 $\pm$  0.01   &     8.92 $\pm$  0.01   &     5.83 $\pm$  0.01   \\
NGC 2207-1   &      $-$   &    18.10 $\pm$  0.07   &      $-$   &      $-$   &      $-$   &      $-$    &       $-$   &    12.31 $\pm$  0.01   &    11.84 $\pm$  0.01   &     9.69 $\pm$  0.01   &     7.79 $\pm$  0.01   &     3.03 $\pm$  0.01   \\
\enddata
\tablenotetext{*}{Zero magnitude flux densities are  
3631 Jy for the GALEX and SDSS bands, and 
277.5 Jy,
179.5 Jy,
116.6 Jy,
63.1 Jy, and
7.3 Jy for 3.6 $\mu$m, 4.5 $\mu$m, 5.8 $\mu$m, 8.0 $\mu$m, and 
24 $\mu$m, respectively.
}
\end{deluxetable}

\begin{deluxetable}{cccccc}
\tabletypesize{\scriptsize}
\setlength{\tabcolsep}{0.03in}
\def\et#1#2#3{${#1}^{+#2}_{-#3}$}
\tablewidth{0pt}
\tablecaption{H$\alpha$ and 2MASS Photometry}
\tablehead{
\multicolumn{1}{c}{Clump} &
\multicolumn{1}{c}{L$_{H\alpha}$}&
\multicolumn{1}{c}{H$\alpha$} &
\multicolumn{1}{c}{J} &
\multicolumn{1}{c}{H}&
\multicolumn{1}{c}{K}
\\ 
\multicolumn{1}{c}{} &
\multicolumn{1}{c}{(erg/s)} &
\multicolumn{1}{c}{EW} &
\multicolumn{1}{c}{(mag)} &
\multicolumn{1}{c}{(mag)} &
\multicolumn{1}{c}{(mag)} 
\\
\multicolumn{1}{c}{} &
\multicolumn{1}{c}{(observed)} &
\multicolumn{1}{c}{(\AA)} &
\multicolumn{1}{c}{} &
\multicolumn{1}{c}{} &
\multicolumn{1}{c}{} 
\\
}
\startdata
Arp 82-1 & 2.2 $\pm$ 0.7  $\times$ 10$^{40}$  & 342.2 $\pm$ 102.7 &      16.62 $\pm$ 0.08 &      17.18 $\pm$ 0.3 &      15.54 $\pm$ 0.11 \\
Arp 240-1 & 25.4 $\pm$ 7.6  $\times$ 10$^{40}$  & 75.2 $\pm$ 22.6 &      13.46 $\pm$ 0.06 &      12.68 $\pm$ 0.07 &      12.34 $\pm$ 0.07 \\
Arp 240-2 & 23.8 $\pm$ 7.2  $\times$ 10$^{40}$  & 111.8 $\pm$ 33.6 &      13.57 $\pm$ 0.07 &      12.71 $\pm$ 0.07 &      12.38 $\pm$ 0.07 \\
Arp 240-3 & 21.4 $\pm$ 6.4  $\times$ 10$^{40}$  & 281.8 $\pm$ 84.6 &      13.99 $\pm$ 0.11 &      13.33 $\pm$ 0.13 &      12.72 $\pm$ 0.1 \\
Arp 240-4 & 16 $\pm$ 4.8  $\times$ 10$^{40}$  & 76.4 $\pm$ 22.9 &      13.6 $\pm$ 0.08 &      12.88 $\pm$ 0.08 &      12.37 $\pm$ 0.07 \\
Arp 240-5 & 17.4 $\pm$ 5.2  $\times$ 10$^{40}$  & 74.6 $\pm$ 22.4 &      13.2 $\pm$ 0.06 &      12.44 $\pm$ 0.07 &      12.0 $\pm$ 0.06 \\
Arp 256-1 & 11.8 $\pm$ 3.5  $\times$ 10$^{40}$  & 63 $\pm$ 18.9 &      14.45 $\pm$ 0.06 &      13.79 $\pm$ 0.07 &      13.66 $\pm$ 0.08 \\
Arp 270-1 & 0.2 $\pm$ 0.1  $\times$ 10$^{40}$  & 92.3 $\pm$ 27.7 &      16.59 $\pm$ 0.13 &      15.98 $\pm$ 0.17 &      15.66 $\pm$ 0.17 \\
Arp 270-2 & 0.2 $\pm$ 0.1  $\times$ 10$^{40}$  & 69.3 $\pm$ 20.8 &      16.59 $\pm$ 0.13 &      15.36 $\pm$ 0.1 &      16.1 $\pm$ 0.28 \\
Arp 270-3 & 0.4 $\pm$ 0.1  $\times$ 10$^{40}$  & 159.9 $\pm$ 48 &      16.05 $\pm$ 0.12 &      15.55 $\pm$ 0.16 &      14.88 $\pm$ 0.11 \\
Arp 270-4 & 2.2 $\pm$ 0.7  $\times$ 10$^{40}$  & 443.3 $\pm$ 133 &      14.65 $\pm$ 0.13 &      14.04 $\pm$ 0.13 &      13.66 $\pm$ 0.12 \\
NGC 2207-1 & 1.7 $\pm$ 0.5  $\times$ 10$^{40}$  & 121.1 $\pm$ 36.3 &      14.77 $\pm$ 0.13 &      14.35 $\pm$ 0.13 &      13.92 $\pm$ 0.1 \\
\enddata
\end{deluxetable}

\begin{deluxetable}{clllllll}
\tabletypesize{\scriptsize}
\setlength{\tabcolsep}{0.03in}
\def\et#1#2#3{${#1}^{+#2}_{-#3}$}
\tablewidth{0pt}
\tablecaption{Small Aperture HST Photometry for Sample Hinge Clumps$^a$\label{tab-7}}
\tablehead{
\multicolumn{1}{c}{Clump} &
\multicolumn{1}{c}{F140LP} &
\multicolumn{1}{c}{F435W} &
\multicolumn{1}{c}{F439W} &
\multicolumn{1}{c}{F555W} &
\multicolumn{1}{c}{F606W} &
\multicolumn{1}{c}{F814W} &
\multicolumn{1}{c}{F160W} \\
\\ 
}
\startdata

Arp 82-1   &    \nodata   &    \nodata   &    \nodata   &    \nodata   &    21.13 $\pm$  0.70   &     7.84 $\pm$  0.21    &     \nodata    \\
Arp 240-1   &    295.89 $\pm$  2.36   &    45.63 $\pm$  0.49   &    \nodata   &    \nodata   &    \nodata   &     8.61 $\pm$  0.14    &      5.88 $\pm$  0.85    \\
Arp 240-2   &    273.69 $\pm$  3.22   &    68.16 $\pm$  1.48   &    \nodata   &    \nodata   &    \nodata   &    20.47 $\pm$  0.40    &     42.55 $\pm$  1.13    \\
Arp 240-3   &    229.81 $\pm$  0.94   &    62.98 $\pm$  0.42   &    \nodata   &    \nodata   &    \nodata   &    16.76 $\pm$  0.19    &     19.49 $\pm$  1.01    \\
Arp 240-4   &    360.35 $\pm$  1.89   &    37.57 $\pm$  0.58   &    \nodata   &    \nodata   &    \nodata   &     7.26 $\pm$  0.19    &      8.11 $\pm$  0.37    \\
Arp 240-5   &    149.54 $\pm$  0.58   &    158.76 $\pm$  1.09   &    \nodata   &    \nodata   &    \nodata   &    50.06 $\pm$  0.78    &     14.57 $\pm$  0.82    \\
Arp 256-1   &    267.53 $\pm$  1.99   &    42.61 $\pm$  0.84   &    \nodata   &    \nodata   &    \nodata   &     8.38 $\pm$  0.24    &      6.30 $\pm$  0.39    \\
NGC 2207-1   &    \nodata   &    \nodata   &    12.69 $\pm$  0.55   &     8.49 $\pm$  0.55   &    \nodata   &     7.49 $\pm$  0.56    &     \nodata    \\
\enddata
\tablenotetext{a}{
These fluxes were obtained using a 0\farcs15 radius, and aperture
corrections have been applied (see text).
All fluxes are in units of 10$^{-18}$ erg~s$^{-1}$~cm$^{-2}$~\AA$^{-1}$,
and have been corrected for Galactic extinction as described in the text.
}
\end{deluxetable}

\begin{deluxetable}{clllllll}
\tabletypesize{\scriptsize}
\setlength{\tabcolsep}{0.03in}
\def\et#1#2#3{${#1}^{+#2}_{-#3}$}
\tablewidth{0pt}
\tablecaption{Large Aperture HST Photometry for Sample Hinge Clumps$^a$\label{tab-8}}
\tablehead{
\multicolumn{1}{c}{Clump} &
\multicolumn{1}{c}{F140LP} &
\multicolumn{1}{c}{F435W} &
\multicolumn{1}{c}{F439W} &
\multicolumn{1}{c}{F555W} &
\multicolumn{1}{c}{F606W} &
\multicolumn{1}{c}{F814W} &
\multicolumn{1}{c}{F160W} \\
\\ 
}
\startdata

Arp 82-1   &    \nodata   &    \nodata   &    \nodata   &    \nodata   &     2.65 $\pm$  0.01   &     1.34 $\pm$  0.01    &     \nodata    \\
Arp 240-1   &    73.86 $\pm$  0.01   &    33.13 $\pm$  0.01   &    \nodata   &    \nodata   &    \nodata   &    17.29 $\pm$  0.01    &     20.09 $\pm$  0.07    \\
Arp 240-2   &    61.92 $\pm$  0.01   &    26.90 $\pm$  0.01   &    \nodata   &    \nodata   &    \nodata   &    14.35 $\pm$  0.01    &     14.55 $\pm$  0.06    \\
Arp 240-3   &    32.22 $\pm$  0.01   &    14.19 $\pm$  0.01   &    \nodata   &    \nodata   &    \nodata   &     8.07 $\pm$  0.01    &      5.76 $\pm$  0.06    \\
Arp 240-4   &    69.16 $\pm$  0.01   &    25.36 $\pm$  0.01   &    \nodata   &    \nodata   &    \nodata   &    13.93 $\pm$  0.01    &      6.47 $\pm$  0.11    \\
Arp 240-5   &    16.08 $\pm$  0.01   &    23.34 $\pm$  0.02   &    \nodata   &    \nodata   &    \nodata   &    20.34 $\pm$  0.01    &     12.39 $\pm$  0.01    \\
Arp 256-1   &    44.42 $\pm$  0.01   &    13.02 $\pm$  0.01   &    \nodata   &    \nodata   &    \nodata   &     6.99 $\pm$  0.01    &     10.59 $\pm$  0.02    \\
NGC 2207-1   &    \nodata   &    \nodata   &    13.88 $\pm$  0.03   &     9.30 $\pm$  0.01   &    \nodata   &     7.27 $\pm$  0.01    &     \nodata    \\
\enddata
\tablenotetext{a}{
These fluxes are all within a 5$''$ radius.
All fluxes are in units of 10$^{-16}$ erg~s$^{-1}$~cm$^{-2}$~\AA$^{-1}$,
and have been corrected for Galactic extinction as described in the text.
}
\end{deluxetable}

\begin{deluxetable}{cccrcccc}
\tabletypesize{\scriptsize}
\setlength{\tabcolsep}{0.03in}
\def\et#1#2#3{${#1}^{+#2}_{-#3}$}
\tablewidth{0pt}
\tablecaption{Chandra Results}
\tablehead{
\multicolumn{1}{c}{Clump} &
\multicolumn{1}{c}{R.A.$^{\dagger}$}&
\multicolumn{1}{c}{Dec.$^{\dagger}$}&
\multicolumn{1}{c}{Background-}&
\multicolumn{1}{c}{Flux$^{*}$} &
\multicolumn{1}{c}{size$^{**}$}&
\multicolumn{1}{c}{L$_{\rm X}$$^{***}$}&
\multicolumn{1}{c}{L$_{\rm X}$$^{\ddag}$}
\\ 
\multicolumn{1}{c}{} &
\multicolumn{1}{c}{(J2000)} &
\multicolumn{1}{c}{(J2000)} &
\multicolumn{1}{c}{Subtracted} &
\multicolumn{1}{c}{(erg/s/cm$^2$)} &
\multicolumn{1}{c}{($''$)} &
\multicolumn{1}{c}{(erg/s)} &
\multicolumn{1}{c}{(erg/s)} 
\\
\multicolumn{1}{c}{} &
\multicolumn{1}{c}{} &
\multicolumn{1}{c}{} &
\multicolumn{1}{c}{Counts} &
\multicolumn{1}{c}{(0.3 $-$ 8 keV)} &
\multicolumn{1}{c}{} &
\multicolumn{1}{c}{(0.3 $-$ 8 keV)} &
\multicolumn{1}{c}{(0.3 $-$ 8 keV)} 
\\
}
\startdata
Arp 240-1  &  13$^{\rm h}$39$^{\rm m}$52.28$^{\rm s}$& +00$^{\circ}$50$'$21\farcs1 & 29 & 3.2 $\times$ 10$^{-15}$  & 1.5 $\times$ 3.5  &   1.5 $\times$ 10$^{40}$  &   2.2 $\times$ 10$^{40}$ \\
Arp 240-2  &  13$^{\rm h}$39$^{\rm m}$52.59$^{\rm s}$& +00$^{\circ}$50$'$15\farcs6 & 18 & 3.2 $\times$ 10$^{-15}$  & 1.5 $\times$ 1.5  &   1.8 $\times$ 10$^{40}$  &   5.4 $\times$ 10$^{40}$ \\
Arp 240-3  &  13$^{\rm h}$39$^{\rm m}$52.95$^{\rm s}$& +00$^{\circ}$50$'$12\farcs5 & 17 & 4 $\times$ 10$^{-15}$  & $\le$0.75 $\times$ 0.75  &   3.3 $\times$ 10$^{40}$  &  5.0 $\times$ 10$^{40}$ \\
Arp 240-4  &  13$^{\rm h}$39$^{\rm m}$53.51$^{\rm s}$& +00$^{\circ}$50$'$30\farcs0 & 99 & 39 $\times$ 10$^{-15}$  & $\le$0.65 $\times$ 0.65  &   17.8 $\times$ 10$^{40}$  &   21.9 $\times$ 10$^{40}$   \\
Arp 240-5  &  13$^{\rm h}$39$^{\rm m}$57.14$^{\rm s}$& +00$^{\circ}$49$'$46\farcs6 & 86 & 12.1 $\times$ 10$^{-15}$  & 4.8 $\times$ 4.8  &   14.2 $\times$ 10$^{40}$  &   14.2 $\times$ 10$^{40}$ \\
Arp 256-1  &  00$^{\rm h}$18$^{\rm m}$49.84$^{\rm s}$& $-$10$^{\circ}$21$'$34\farcs5 & 42 & 6.1 $\times$ 10$^{-15}$  & 2.0 $\times$ 1.0  &   3.2 $\times$ 10$^{40}$  &   3.2 $\times$ 10$^{40}$   \\
Arp 270-1  &  10$^{\rm h}$49$^{\rm m}$46.60$^{\rm s}$& +32$^{\circ}$58$'$23\farcs7 & 50 & 15 $\times$ 10$^{-15}$  & $\le$0.85 $\times$ 0.85  &   0.4 $\times$ 10$^{40}$  &   0.4 $\times$ 10$^{40}$   \\
Arp 270-2  &  \nodata & \nodata & $\le$6 & $\le$1.2 $\times$ 10$^{-15}$  & \nodata    &   $\le$0.4 $\times$ 10$^{39}$  &  $\le$0.4 $\times$ 10$^{39}$   \\
Arp 270-3  &  \nodata & \nodata & $\le$6 & $\le$1.5 $\times$ 10$^{-15}$  & \nodata    &   $\le$0.4 $\times$ 10$^{39}$  &  $\le$0.3 $\times$ 10$^{39}$   \\
Arp 270-4  &  \nodata & \nodata & $\le$6 & $\le$1.4 $\times$ 10$^{-15}$  & \nodata    &   $\le$0.3 $\times$ 10$^{39}$  &  $\le$0.3 $\times$ 10$^{39}$   \\
NGC 2207-1  &  16$^{\rm h}$16$^{\rm m}$15.5$^{\rm s}$& $-$21$^{\circ}$22$'$02\farcs6 & 26 & 11 $\times$ 10$^{-15}$  & 10 $\times$ 6  &   2.5 $\times$ 10$^{40}$  &   2.5 $\times$ 10$^{40}$   \\
\enddata
\tablenotetext{\dagger}{
Coordinates of central Chandra source.
}
\tablenotetext{*}{
Observed flux from source in center of clump.
}
\tablenotetext{**}{
Approximate total angular extent of X-ray emission from central
source within clump.
}
\tablenotetext{***}{
X-ray luminosity of the central source within the clump; 
corrected for internal extinction using estimates made from
the L$_{H\alpha}$/L$_{24}$ ratio as described in the text.
}
\tablenotetext{\ddag}{
X-ray luminosity within a 5$''$ radius aperture.
Corrected for internal extinction using estimates made from
the L$_{H\alpha}$/L$_{24}$ ratio as described in the text.
}
\end{deluxetable}

\begin{deluxetable}{ccccccc}
\tabletypesize{\scriptsize}
\setlength{\tabcolsep}{0.03in}
\def\et#1#2#3{${#1}^{+#2}_{-#3}$}
\tablewidth{0pt}
\tablecaption{Star Formation Rates of the Hinge Clumps }
\tablehead{
\multicolumn{1}{c}{Clump} &
\multicolumn{1}{c}{L$_{NUV}$$^a$}&
\multicolumn{1}{c}{L$_{24}$$^b$}&
\multicolumn{1}{c}{SFR$^c$} &
\multicolumn{1}{c}{percent$^{c,d}$} &
\multicolumn{1}{c}{SFR$^e$} &
\multicolumn{1}{c}{percent$^{d,e}$} 
\\ 
\multicolumn{1}{c}{} &
\multicolumn{1}{c}{(erg/s)} &
\multicolumn{1}{c}{(erg/s)} &
\multicolumn{1}{c}{(M$_{\sun}$/yr)} &
\multicolumn{1}{c}{Global} &
\multicolumn{1}{c}{(M$_{\sun}$/yr)} &
\multicolumn{1}{c}{Global} 
\\
\multicolumn{1}{c}{} &
\multicolumn{1}{c}{(observed)} &
\multicolumn{1}{c}{(observed)} &
\multicolumn{1}{c}{} &
\multicolumn{1}{c}{SFR} &
\multicolumn{1}{c}{} &
\multicolumn{1}{c}{SFR} 
\\
}
\startdata
Arp 82-1 & 5.5  $\times$ 10$^{41}$ & 10.2  $\times$ 10$^{41}$  & 0.29 & 7\% & 0.18 & 5\% \\
Arp 240-1 & 109.9  $\times$ 10$^{41}$ & 155.4  $\times$ 10$^{41}$  & 4.04 & 19\% & 3.06 & 18\% \\
Arp 240-2 & 96  $\times$ 10$^{41}$ & 206.5  $\times$ 10$^{41}$  & 4.83 & 22\% & 3.73 & 22\% \\
Arp 240-3 & 53.4  $\times$ 10$^{41}$ & 290.0  $\times$ 10$^{41}$  & 6.11 & 28\% & 4.7 & 28\% \\
Arp 240-4 & 89.8  $\times$ 10$^{41}$ & 110.7  $\times$ 10$^{41}$  & 2.76 & 13\% & 2.25 & 13\% \\
Arp 240-5 & 32.5  $\times$ 10$^{41}$ & 498.3  $\times$ 10$^{41}$  & 9.45 & 56\% & 7.69 & 55\% \\
Arp 256-1 & 76.9  $\times$ 10$^{41}$ & 80.4  $\times$ 10$^{41}$  & 2.01 & 43\% & 1.71 & 40\% \\
Arp 270-1 & 2.2  $\times$ 10$^{41}$ & 0.6  $\times$ 10$^{41}$  & 0.02 & 1\% & 0.02 & 1\% \\
Arp 270-2 & 1.8  $\times$ 10$^{41}$ & 1.1  $\times$ 10$^{41}$  & 0.02 & 1\% & 0.02 & 1\% \\
Arp 270-3 & 2.3  $\times$ 10$^{41}$ & 1.2  $\times$ 10$^{41}$  & 0.04 & 2\% & 0.03 & 1\% \\
Arp 270-4 & 6.9  $\times$ 10$^{41}$ & 4.3  $\times$ 10$^{41}$  & 0.19 & 8\% & 0.11 & 5\% \\
NGC 2207-1 & 7  $\times$ 10$^{41}$ & 97.1  $\times$ 10$^{41}$  & 1.74 & 24\% & 1.5 & 22\% \\
\enddata
\tablenotetext{a}{using L$_{NUV}$ = $\nu$L$_{\nu}$.}
\tablenotetext{b}{using L$_{24}$ = $\nu$L$_{\nu}$.}
\tablenotetext{c}{
Calculated 
using
SFR (M$_{\sun}$/yr)
= 5.5 $\times$ 10$^{-42}$[L$_{H\alpha}$ + 0.031~L$_{24}$] (erg/s)
\citep{kennicutt09}.
}
\tablenotetext{d}{
Percent of total SFR from parent galaxy in the target clump.}
\tablenotetext{e}{Calculated 
using 
L$_{NUV}$(corr) = L$_{NUV}$ + 2.26L$_{24}$
and 
log(SFR) = log(L$_{NUV}$(corr)) - 43.17
\citep{hao11}.
}
\end{deluxetable}

\begin{deluxetable}{ccccrc}
\tabletypesize{\scriptsize}
\tablecaption{Broadband UV/Optical Single-Age Instantaneous Burst Models}
\tablewidth{0pt}
\tablehead{
\colhead{Clump} & \colhead{Age} & \colhead{E(B-V)} & \colhead{Stellar}  & \colhead{Reduced} & \colhead{Colors Used}
\\
 & \colhead{(Myr)} & \colhead{(mag)} & \colhead{Mass} & \colhead{{Chi Squared} } & \colhead{}
\\
 & \colhead{} & \colhead{} & \colhead{(M$_{\sun}$)}  & \colhead{{($\chi$$^2$/(N$-$2))} } & \colhead{}
}
\startdata
   Arp 82-1   & 100  $\pm$ $^{  1 }_{  21 }$ &  0.00  $\pm$ $^{ 0.06 }_{  0.00 }$ &        2.5  $\times$ 10$^7$ &  11.5  &                  FUV-NUV, NUV-g, u-g, g-r, r-i, i-z    \\
   Arp 240-1   &  6  $\pm$ $^{  1 }_{   1 }$ &  0.54  $\pm$ $^{ 0.04 }_{  0.06 }$ &       45.1  $\times$ 10$^7$ &  10.9  &                           NUV-g, u-g, g-r, r-i, i-z    \\
   Arp 240-2   & 150  $\pm$ $^{  1 }_{   1 }$ &  0.14  $\pm$ $^{ 0.06 }_{  0.06 }$ &       203.3  $\times$ 10$^7$ &   6.8  &                           NUV-g, u-g, g-r, r-i, i-z    \\
   Arp 240-3   & 150  $\pm$ $^{ 101 }_{   1 }$ &  0.14  $\pm$ $^{ 0.10 }_{  0.14 }$ &       95.5  $\times$ 10$^7$ &  14.0  &                           NUV-g, u-g, g-r, r-i, i-z    \\
   Arp 240-4   & 45  $\pm$ $^{ 11 }_{   6 }$ &  0.28  $\pm$ $^{ 0.06 }_{  0.06 }$ &       111.0  $\times$ 10$^7$ &   2.3  &                           NUV-g, u-g, g-r, r-i, i-z    \\
   Arp 240-5   & 70  $\pm$ $^{ 31 }_{  26 }$ &  0.56  $\pm$ $^{ 0.12 }_{  0.10 }$ &       338.4  $\times$ 10$^7$ &   4.3  &                           NUV-g, u-g, g-r, r-i, i-z    \\
   Arp 256-1   &  6  $\pm$ $^{  1 }_{   2 }$ &  0.54  $\pm$ $^{ 0.22 }_{  0.06 }$ &       21.8  $\times$ 10$^7$ &  32.8  &                  FUV-NUV, NUV-g, u-g, g-r, r-i, i-z    \\
   Arp 270-1   &  5  $\pm$ $^{  1 }_{   2 }$ &  0.28  $\pm$ $^{ 0.12 }_{  0.06 }$ &        0.2  $\times$ 10$^7$ &   7.1  &                  FUV-NUV, NUV-g, u-g, g-r, r-i, i-z    \\
   Arp 270-2   & 70  $\pm$ $^{ 31 }_{  67 }$ &  0.00  $\pm$ $^{ 0.58 }_{  0.00 }$ &        0.7  $\times$ 10$^7$ &   1.0  &                  FUV-NUV, NUV-g, u-g, g-r, r-i, i-z    \\
   Arp 270-3   &  6  $\pm$ $^{ 90 }_{   2 }$ &  0.26  $\pm$ $^{ 0.20 }_{  0.26 }$ &        0.2  $\times$ 10$^7$ &   1.3  &                  FUV-NUV, NUV-g, u-g, g-r, r-i, i-z    \\
   Arp 270-4   & 50  $\pm$ $^{ 21 }_{  45 }$ &  0.00  $\pm$ $^{ 0.28 }_{  0.00 }$ &        1.7  $\times$ 10$^7$ &   0.6  &                  FUV-NUV, NUV-g, u-g, g-r, r-i, i-z    \\
\enddata
\end{deluxetable}

\begin{deluxetable}{crrrrrcrrr}
\tabletypesize{\scriptsize}
\setlength{\tabcolsep}{0.03in}
\def\et#1#2#3{${#1}^{+#2}_{-#3}$}
\tablewidth{0pt}
\tablecaption{Reddening and Ages from the H$\alpha$ Data, and Related Quantities}
\tablehead{
\multicolumn{1}{c}{clump} &
\multicolumn{1}{c}{A$_{\rm V}^a$} &
\multicolumn{1}{c}{E(B$-V$)$^b$} &
\multicolumn{1}{c}{N$_{\rm H}^c$} &
\multicolumn{1}{c}{Age$^d$}&
\multicolumn{1}{c}{Stellar$^e$}&
\multicolumn{1}{c}{L$_{\rm X}$/L$_{H\alpha}$$^f$}&
\multicolumn{1}{c}{L$_{\rm X}$/SFR$^g$}&
\multicolumn{1}{c}{n$_{\rm e}$$\sqrt{f}$$^h$}&
\multicolumn{1}{c}{XPE$^i$}
\\ 
\multicolumn{1}{c}{} &
\multicolumn{1}{c}{(mag)} &
\multicolumn{1}{c}{(mag)} &
\multicolumn{1}{c}{(10$^{21}$ cm$^{-2}$)} &
\multicolumn{1}{c}{(Myrs)} &
\multicolumn{1}{c}{Mass} &
\multicolumn{1}{c}{(corrected)} &
\multicolumn{1}{c}{((erg/s)/(M$_{\sun}$/yr))} &
\multicolumn{1}{c}{(cm$^{-3}$)} &
\multicolumn{1}{c}{} 
\\
\multicolumn{1}{c}{} &
\multicolumn{1}{c}{} &
\multicolumn{1}{c}{} &
\multicolumn{1}{c}{} &
\multicolumn{1}{c}{} &
\multicolumn{1}{c}{(M$_{\sun}$)} &
\multicolumn{1}{c}{} &
\multicolumn{1}{c}{(corrected)} &
\multicolumn{1}{c}{} &
\multicolumn{1}{c}{} 
\\
}
\startdata

 Arp 82-1  &   1.34  $\pm$   0.02     &   0.33  $\pm$  0.01     &   1.93  $\pm$  0.02     &   5.4  $\pm$ $^{ 0.2}_{0.0}$ &         0.7    $\times$ 10$^7$ &   \nodata     &  \nodata     &  \nodata              & \nodata \\
 Arp 240-1  &   1.59  $\pm$   0.02     &   0.39  $\pm$  0.01     &   2.27  $\pm$  0.02     &   6.1  $\pm$ $^{ 0.5}_{0.0}$ &        35.2    $\times$ 10$^7$ &   0.026     &    4.1     $\times$ 10$^{39}$ &        0.035   & 0.0022      \\
 Arp 240-2  &   1.93  $\pm$   0.02     &   0.48  $\pm$  0.01     &   2.76  $\pm$  0.02     &   5.8  $\pm$ $^{ 0.3}_{0.0}$ &        27.5    $\times$ 10$^7$ &   0.053     &    4.2     $\times$ 10$^{39}$ &        0.077   & 0.0053      \\
 Arp 240-3  &   2.41  $\pm$   0.02     &   0.59  $\pm$  0.01     &   3.45  $\pm$  0.02     &   5.5  $\pm$ $^{ 0.1}_{0.0}$ &        25.7    $\times$ 10$^7$ &   0.038     &    6.2     $\times$ 10$^{39}$ &        0.297   & 0.0083      \\
 Arp 240-4  &   1.71  $\pm$   0.02     &   0.42  $\pm$  0.01     &   2.45  $\pm$  0.02     &   6.1  $\pm$ $^{ 0.5}_{0.0}$ &        22.0    $\times$ 10$^7$ &   0.378     &   70.9     $\times$ 10$^{39}$ &        0.850   & 0.0319      \\
 Arp 240-5  &   3.27  $\pm$   0.02     &   0.81  $\pm$  0.01     &   4.69  $\pm$  0.02     &   6.1  $\pm$ $^{ 0.5}_{0.0}$ &        63.0    $\times$ 10$^7$ &   0.069     &   16.5     $\times$ 10$^{39}$ &        0.038   & 0.0058      \\
 Arp 256-1  &   1.69  $\pm$   0.02     &   0.42  $\pm$  0.01     &   2.42  $\pm$  0.02     &   6.3  $\pm$ $^{ 0.5}_{0.0}$ &        17.6    $\times$ 10$^7$ &   0.076     &   17.3     $\times$ 10$^{39}$ &        0.098   & 0.0065      \\
 Arp 270-1  &   0.97  $\pm$   0.02     &   0.24  $\pm$  0.01     &   1.39  $\pm$  0.02     &   5.9  $\pm$ $^{ 0.4}_{0.0}$ &         0.1    $\times$ 10$^7$ &   0.893     &  147.2     $\times$ 10$^{39}$ &        0.526   & 0.0795      \\
 Arp 270-2  &   1.63  $\pm$   0.02     &   0.40  $\pm$  0.01     &   2.33  $\pm$  0.02     &   6.2  $\pm$ $^{ 0.5}_{0.0}$ &         0.2    $\times$ 10$^7$ &   $\le$ 0.076   &   $\le$15.9  $\times$ 10$^{39}$ & \nodata    & $\le$0.0060         \\
 Arp 270-3  &   0.98  $\pm$   0.02     &   0.24  $\pm$  0.01     &   1.40  $\pm$  0.02     &   5.7  $\pm$ $^{ 0.1}_{0.0}$ &         0.2    $\times$ 10$^7$ &   $\le$ 0.031   &   $\le$ 8.9  $\times$ 10$^{39}$ & \nodata    & $\le$0.0045         \\
 Arp 270-4  &   0.73  $\pm$   0.02     &   0.18  $\pm$  0.01     &   1.04  $\pm$  0.02     &   5.3  $\pm$ $^{ 0.2}_{0.0}$ &         0.5    $\times$ 10$^7$ &   $\le$ 0.007   &   $\le$ 2.1  $\times$ 10$^{39}$ & \nodata    & $\le$0.0024         \\
NGC 2207-1  &   4.12  $\pm$   0.02     &   1.02  $\pm$  0.01     &   5.90  $\pm$  0.02     &   5.7  $\pm$ $^{ 0.2}_{0.0}$ &         4.3    $\times$ 10$^7$ &   0.064     &   15.1     $\times$ 10$^{39}$ &        0.033   & 0.0074      \\
\enddata
\tablenotetext{a}{From
A$_{H\alpha}$ = 2.5~log[1 + 0.038L$_{24}$/L$_{H\alpha}$] \citep{kennicutt07}.
}
\tablenotetext{b}{Using A$_{\rm V}$/E(B$-$V) = 4.05 
\citep{calzetti00}.}
\tablenotetext{c}{Using 
N$_{\rm H}$(cm$^{-2}$) = 5.8 $\times$ 10$^{21}$E(B$-$V)
from
\citet{bohlin78}.}
\tablenotetext{d}{From the H$\alpha$
equivalent width, assuming an instantaneous burst.
}
\tablenotetext{e}{
Obtained by
scaling the 
observed i band flux 
(or HST F814W flux, if no SDSS i image available)
to Starburst99 models
assuming an instantaneous burst,
using the age determined from the H$\alpha$
equivalent width and the absorption determined from the H$\alpha$/24 $\mu$m
flux ratio.
}
\tablenotetext{f}{The ratio of the 0.3 $-$ 8 keV X-ray luminosity L$_{\rm X}$
to the H$\alpha$ luminosity, corrected for extinction using the 
H$\alpha$/24 $\mu$m flux ratio as described in the text.}
\tablenotetext{g}{The ratio of the extinction-corrected 0.3 $-$ 8 keV luminosity to the 
SFR.  When both estimates of SFR are available, the average value is used.}
\tablenotetext{h}{Electron number density n$_{\rm e}$ $\times$ $\sqrt{f}$,
where f is the volume filling factor (see text for details).  Note that these are not relevant if the X-ray source is a ULX, which is likely for 
Arp 270-1 and Arp 240-4 (see text).}
\tablenotetext{i}{X-Ray Production Efficiency = L$_X$/L$_{mech}$.}
\end{deluxetable}

\end{document}